\documentclass[preprint2]{aastex631}

\newcommand{\Ha}{H$\alpha$}

\usepackage{amsmath}
\usepackage{bm}
\usepackage{xcolor}
\usepackage{hyperref}

\received{October 5, 2021}
\revised{November 23, December 14, 2021}
\accepted{December 17, 2021}

\shorttitle{The ``super-Chandrasekhar'' SN~Ia 2020esm}
\shortauthors{Dimitriadis et al.}
\graphicspath{{./}{figures/}}

\begin{document}

\title{A Carbon/Oxygen-dominated Atmosphere Days After Explosion for the ``Super-Chandrasekhar'' Type Ia SN~2020esm}

\author[0000-0001-9494-179X]{Georgios Dimitriadis}
\affiliation{Department of Astronomy and Astrophysics, University of California, Santa Cruz, California, 95064, USA}
\affiliation{School of Physics, Trinity College Dublin, The University of Dublin, Dublin 2, Ireland}

\author[0000-0002-2445-5275]{Ryan~J.~Foley}
\affiliation{Department of Astronomy and Astrophysics, University of California, Santa Cruz, California, 95064, USA}

\author[0000-0001-5409-6480]{Nikki~Arendse}
\affiliation{DARK, Niels Bohr Institute, University of Copenhagen, Jagtvej 128, 2200 Copenhagen, Denmark}

\author[0000-0003-4263-2228]{David~A.~Coulter}
\affiliation{Department of Astronomy and Astrophysics, University of California, Santa Cruz, California, 95064, USA}

\author[0000-0002-3934-2644]{Wynn~V.~Jacobson-Gal\'an}
\affiliation{Center for Interdisciplinary Exploration and Research in Astrophysics (CIERA) and Department of Physics and Astronomy, Northwestern University, Evanston, IL 60208, USA}
\affiliation{Department of Astronomy and Astrophysics, University of California, Berkeley, CA 94720, USA}

\author[0000-0003-2445-3891]{Matthew~R.~Siebert}
\affiliation{Department of Astronomy and Astrophysics, University of California, Santa Cruz, California, 95064, USA}

\author[0000-0001-9695-8472]{Luca~Izzo}
\affiliation{DARK, Niels Bohr Institute, University of Copenhagen, Jagtvej 128, 2200 Copenhagen, Denmark}

\author[0000-0002-6230-0151]{David~O.~Jones}
\affiliation{Department of Astronomy and Astrophysics, University of California, Santa Cruz, California, 95064, USA}

\author[0000-0002-5740-7747]{Charles~D.~Kilpatrick}
\affiliation{Center for Interdisciplinary Exploration and Research in Astrophysics (CIERA) and Department of Physics and Astronomy, Northwestern University, Evanston, IL 60208, USA}

\author[0000-0001-8415-6720]{Yen-Chen~Pan}
\affiliation{Institute of Astronomy, National Central University, 300 Jhongda Road, 32001 Jhongli, Taiwan}

\author[0000-0002-5748-4558]{Kirsty~Taggart}
\affiliation{Department of Astronomy and Astrophysics, University of California, Santa Cruz, California, 95064, USA}

\author[0000-0002-4449-9152]{Katie~Auchettl}
\affiliation{Department of Astronomy and Astrophysics, University of California, Santa Cruz, California, 95064, USA}
\affiliation{DARK, Niels Bohr Institute, University of Copenhagen, Jagtvej 128, 2200 Copenhagen, Denmark}
\affiliation{School of Physics, The University of Melbourne, VIC 3010, Australia}
\affiliation{ARC Centre of Excellence for All Sky Astrophysics in 3 Dimensions (ASTRO 3D)}

\author[0000-0002-8526-3963]{Christa~Gall}
\affiliation{DARK, Niels Bohr Institute, University of Copenhagen, Jagtvej 128, 2200 Copenhagen, Denmark}

\author[0000-0002-4571-2306]{Jens~Hjorth}
\affiliation{DARK, Niels Bohr Institute, University of Copenhagen, Jagtvej 128, 2200 Copenhagen, Denmark}

\author{Daniel~Kasen}
\affiliation{Department of Astronomy and Astrophysics, University of California, Berkeley, CA 94720, USA}
\affiliation{Nuclear Science Division, Lawrence Berkeley National Laboratory, 1 Cyclotron Road, Berkeley, CA, 94720, USA}

\author[0000-0001-6806-0673]{Anthony~L.~Piro}
\affiliation{The Observatories of the Carnegie Institution for Science, 813 Santa Barbara St., Pasadena, CA 91101, USA}

\author[0000-0002-6248-398X]{Sandra~I.~Raimundo}
\affiliation{DARK, Niels Bohr Institute, University of Copenhagen, Jagtvej 128, 2200 Copenhagen, Denmark}
\affiliation{Department of Physics and Astronomy, University of Southampton, Highfield, Southampton SO17 1BJ, UK}
\affiliation{Department of Physics and Astronomy, University of California, Los Angeles, CA 90095, USA}

\author[0000-0003-2558-3102]{Enrico~Ramirez-Ruiz}
\affiliation{Department of Astronomy and Astrophysics, University of California, Santa Cruz, California, 95064, USA}

\author[0000-0002-4410-5387]{Armin~Rest}
\affiliation{Department of Physics and Astronomy, Johns Hopkins University, 3400 North Charles Street, Baltimore, MD 21218, USA}
\affiliation{Space Telescope Science Institute, 3700 San Martin Drive, Baltimore, MD 21218, USA}

\author[0000-0002-9486-818X]{Jonathan~J.~Swift}
\affiliation{The Thacher School, 5025 Thacher Rd, Ojai, CA 93023, USA}

\author[0000-0002-3352-7437]{Stan~E.~Woosley}
\affiliation{Department of Astronomy and Astrophysics, University of California, Santa Cruz, California, 95064, USA}

\correspondingauthor{Georgios Dimitriadis}
\email{dimitrig@tcd.ie}



\begin{abstract}

Seeing pristine material from the donor star in a Type Ia supernova (SN~Ia) explosion  can reveal the nature of the binary system. In this paper, we present photometric and spectroscopic observations of SN\,2020esm, one of the best-studied SNe of the class of  ``super-Chandrasekhar'' SNe~Ia (SC SNe~Ia), with data obtained $-12$ to +360 days relative to peak brightness, obtained from a variety of ground- and space-based telescopes. Initially misclassified as a Type II supernova, SN\,2020esm peaked at $M_{B}=-19.9$~mag, declined slowly ($\Delta m_{15}(B)=0.92$~mag), and had particularly blue UV and optical colors at early times.  Photometrically and spectroscopically, SN~2020esm evolved similarly to other SC SNe~Ia, showing the usual low ejecta velocities, weak intermediate mass elements (IMEs), and the enhanced fading at late times, but its early spectra are unique. Our first few spectra (corresponding to a phase of $\gtrsim$10~days before peak) reveal a nearly-pure carbon/oxygen atmosphere during the first days after explosion.  This composition can only be produced by pristine material, relatively unaffected by nuclear burning.  The lack of H and He may further indicate that SN\,2020esm is the outcome of the merger of two carbon/oxygen white dwarfs (WDs). Modeling its bolometric light curve, we find a $^{56}$Ni mass of $1.23^{+0.14}_{-0.14}$~M$_{\sun}$ and an ejecta mass of $1.75^{+0.32}_{-0.20}$~M$_{\sun}$, in excess of the Chandrasekhar mass. Finally, we discuss possible progenitor systems and explosion mechanisms of SN\,2020esm and, in general, the SC SNe~Ia class.

\end{abstract}

\keywords{supernovae: general --- supernovae: individual (SN\,2020esm) --- white dwarfs}


\section{Introduction} \label{sec:intro}

Observations of Type Ia supernovae (SNe~Ia) first showed that the expansion of the Universe is accelerating \citep{Riess98:Lambda, Perlmutter99}. SNe~Ia are also key to measuring the local expansion rate \citep{Riess16,Freedman2019ApJ}, and those measurements differ from inferences from early-Universe probes that may indicate unaccounted physics in the current cosmological model \citep{Freedman21ApJ}. While there is strong observational evidence that SNe~Ia result from the thermonuclear explosion of a degenerate carbon/oxygen white dwarf (WD) star in a binary system \citep{Bloom12}, details of the progenitor system and explosion are poorly constrained \citep{Maoz14}.  

The peak luminosity of most SNe~Ia correlates strongly with their decline rate (or light-curve width, parametrized with their magnitude decline from peak to 15 days after, $\Delta m_{15}$, \citealt{Phillips93}) and color \citep{Riess96}. By observing the brightness, decline rate, and color of a SN~Ia, one can infer its relative distance, which in turn, can be used to constrain cosmological parameters \citep[e.g.,][]{Scolnic18:ps1,Jones19}. The width-luminosity relation (WLR) can be explained as all SNe~Ia having a similar ejecta mass with varying amounts of radioactive $^{56}$Ni \citep{Kasen07:wlr}, which sets the peak luminosity. Alternatively, the total ejecta mass may be the primary factor that causes differences in $^{56}$Ni and luminosity \citep{Goldstein18ApJ}. Moreover, SNe~Ia are characterized by maximum-light spectra that lack hydrogen and helium emission features, but have prominent absorption features from intermediate-mass  (e.g., Ca, S, Si) and iron-group elements \citep{Filippenko97,Parrent14}.

A small fraction of SNe~Ia are outliers in the WLR \citep{Taubenberger17}, perhaps because of deviations in ejecta mass or $^{56}$Ni, and not identifying such events in a distance-independent way could bias cosmological results \citep{Rubin2015ApJ}.

One of the most intriguing peculiar sub-classes of SNe~Ia is the ``super-Chandrasekhar'' SNe Ia (SC SNe~Ia). The moniker for this subclass comes from modeling that suggests a total mass that is in excess of the Chandrasekhar mass ($\mathrm{M_{Ch}}$; \citealt{Chandrasekhar31}), the theoretical maximum mass for a non-rotating, zero-temperature WD. Many (but not all) SC SNe~Ia do not conform to the WLR, generally having peak luminosities higher than expected for their decline rate. SC SNe Ia have slowly declining light curves, high early-time UV flux, an i-band primary maximum several days before the B-band maximum, a lack of a secondary maximum in the $i$ and NIR bands, relatively hot photospheres, abnormally low ejecta velocities, a weak (or delayed) NIR H-band break, and relatively strong carbon features in peak-light spectra  \citep{Ashall2021arXiv}. Modeling their light curves in the same way as typical SNe~Ia \citep{Arnett82,Jeffery1999}, one finds that their total ejecta mass, and occasionally the $^{56}$Ni mass alone, is in excess of the Chandrasekhar mass, $M_{\rm Ch} \approx 1.4$~M$_{\sun}$ \citep{Scalzo10}.

Prime examples of SC SNe~Ia include SNe\,2003fg \citep{Howell06}, 2006gz \citep{Hicken07ApJ,Maeda09ApJ}, 2007if \citep{Scalzo10,Yuan10ApJ}, 2009dc \citep{Yamanaka09ApJ,Silverman2011MNRAS,Taubenberger11,Hachinger12MNRAS}, 2012dn \citep{Chakradhari14MNRAS,Yamanaka2016PASJ,Parrent16MNRAS,Taubenberger19}, LSQ14fmg \citep{Hsiao20ApJ}, ASASSN-15pz \citep{Chen2019ApJ} and ASASSN-15hy \citep{Lu2021ApJ}. Of these events, the best-studied ones are SNe\,2009dc and 2012dn, coincidentally representing the significant diversity within SC SNe~Ia: While SN\,2009dc shows all the observational characteristics mentioned above, SN\,2012dn is noticeably fainter at peak (matching the luminosity of normal SNe~Ia) with higher ejecta velocities compared to its sub-class.

Late-time observations of SC SNe~Ia are invaluable tools in order to shed light on the peculiar nature of these objects. \citet{Taubenberger2013MNRAS} performed the first systematic study of SC SNe~Ia at late epochs, finding low ionization states based on the weak [\ion{Fe}{3}] emission lines due to high central densities, and a sudden rapid fading of their light curves (with SN\,2007if being the exception, at least at the epoch of its observation), interpreted as flux redistribution into infrared wavelengths, with the source of this effect (CO formation, dust formation, $\gamma$-ray and/or positron escape, IR catastrophe, discontinued circumstellar interaction) still being debated. Moreover, \citet{Yamanaka2016PASJ} present a late-time NIR excess of SN\,2012dn, coinciding with its optical fading, while its nebular spectra show a prominent emission feature at $\sim$6,300~\AA, which \citet{Taubenberger19} identified as [\ion{O}{1}] $\lambda\lambda$6300, 6364.

From a theoretical perspective, a definitive explosion mechanism and progenitor scenario for SC SNe~Ia still do not exist. The earliest suggestion of a rapidly differentially rotating massive WD \citep{Yoon05AA} is challenged by more recent numerical simulations \citep{Pfannes10AA2,Pfannes10AA,Fink18AA}. A merger of two WDs \citep{Iben84, Webbink84} that exceed $\mathrm{M_{Ch}}$ can naturally explain the high ejecta mass \citep{Moll14}, while increased luminosity at early times can be achieved with hydrogen-free circumstellar medium (CSM) of material ejected during the merger in the close vicinity of the system \citep{Raskin13,Raskin14,Noebauer16MNRAS}, however, an open question remains whether this binary configuration can lead to a type Ia explosion or an accretion induced collapse  \citep[see][for relevant discussions]{Marsh2004MNRAS,Saio2004ApJ,Dan2011ApJ,Shen2012ApJ}. On the other hand, \citet{Hsiao20ApJ}, \citet{Ashall2021arXiv} and \citet{Lu2021ApJ} favor an explosion of a C-O degenerate core inside a carbon-rich envelope \citep{Hoeflich96ApJ}, possibly under the ``core-degenerate'' scenario \citep{Kashi11MNRAS}.

In this paper, we present observations of the ``super-Chandrasekhar'' SN~Ia 2020esm, with excellent photometric and spectroscopic coverage, from $\sim$12 days before to $\sim$360 days after maximum light. We show that SN\,2020esm has a nearly-pure carbon/oxygen atmosphere for the first days after explosion, an observation in accordance with the merger of two carbon/oxygen WDs, providing the strongest evidence yet that WD mergers produce SNe~Ia. We additionally detect early blue UV/optical colors, indicating interaction between the SN and a significant fraction of the disrupted WD that was ejected into the circumstellar environment. Finally, we confirm previous late-time observations of SC SNe~Ia that show an enchanced fading at the optical light curves. We present the discovery of SN\,2020esm, discuss its initial misclassification and summarize our observing campaign in Section~\ref{sec:obs_data_red}. The host-galaxy properties, its distance and extinction on the line of sight, alongside its photometric and spectroscopic evolution are presented in Section~\ref{sec:analysis}. We discuss our findings in the context of the general SC SNe~Ia population and their proposed progenitor systems in Section~\ref{sec:discussion}. Finally, we conclude in Section~\ref{sec:conclusion}.

Throughout this paper, we will use the moniker SC SNe~Ia to describe the members of the ``super-Chandrasekhar'' SNe~Ia class. We note that, in the literature, different monikers have been proposed, such as 09dc-like in \citet{Taubenberger19} and 03fg-like in \citet{Ashall2021arXiv}, however we choose to use the historical designation. Moreover, we adopt the AB magnitude system, unless where noted, and a Hubble constant of $H_0 = 73$ km s$^{-1}$ Mpc$^{-1}$.

\section{Observations and Data Reduction} \label{sec:obs_data_red}

In this Section, we present the discovery of SN\,2020esm, its initial classification and our subsequent photometric and spectroscopic followup campaign.

\subsection{Discovery and classification} \label{subsec:discovery}

SN\,2020esm was discovered on UT 2020 March 22.37 by ASAS-SN \citep{Shappee14}, with an internal name of ASASSN-20dl \citep{Brimacombe20ATel,Stanek20TNSTR}, at $\textit{g}=17.7\pm0.04$ mag, with the last non-detection from ASAS-SN on UT 2020 March 18.90, to a limiting magnitude of 18.5\footnote{\url{https://asas-sn.osu.edu/}}. The last non-detection of the transient with ZTF was on UT 2020 March 6.48 to a limiting magnitude of 19.67 in \textit{g}-band, while the last non-detection with ATLAS was on UT 2020 March 16.66 to a limiting magnitude of 17.73 in \textit{o}-band.

\begin{figure*}[t]
\plottwo{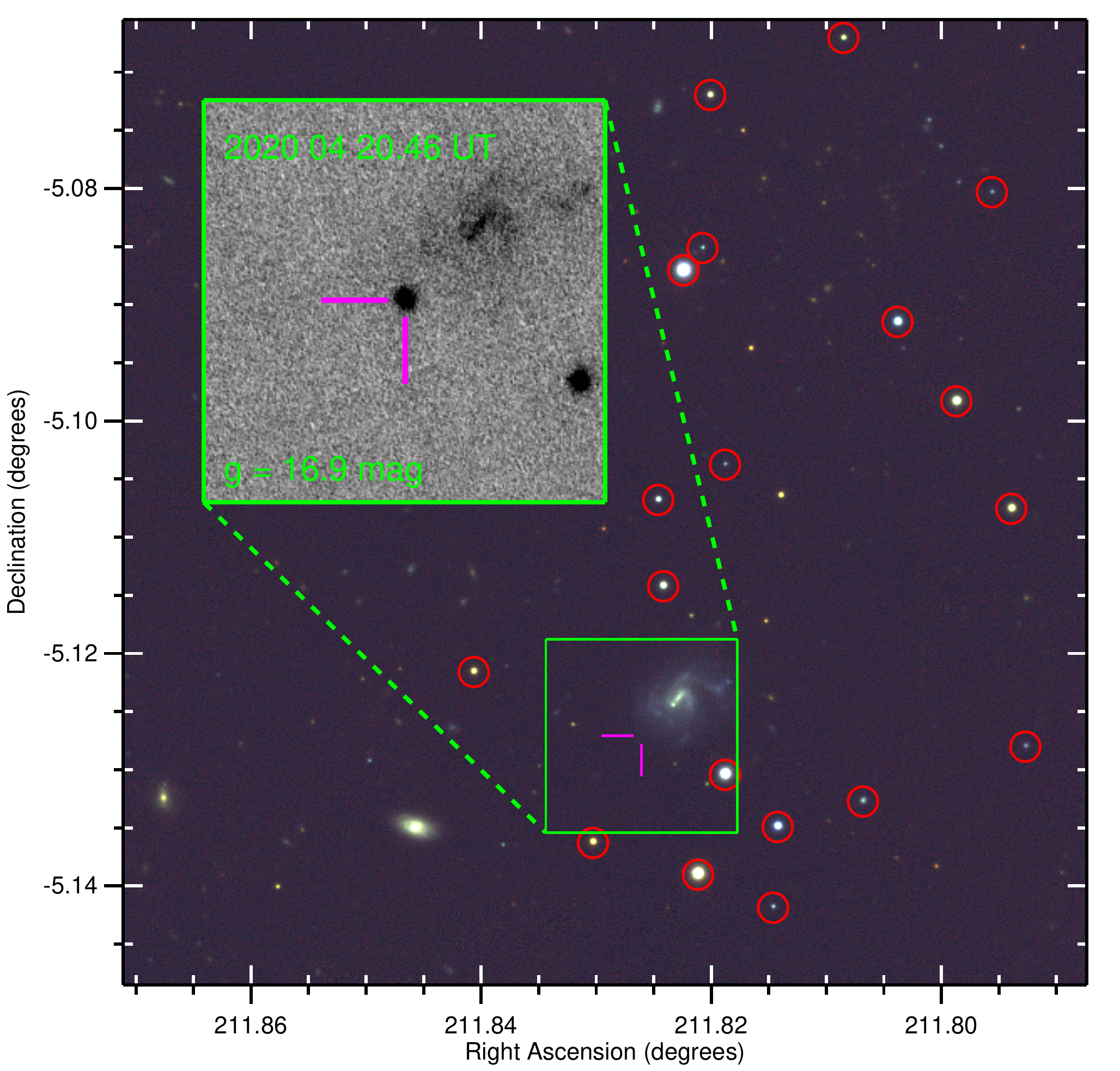}{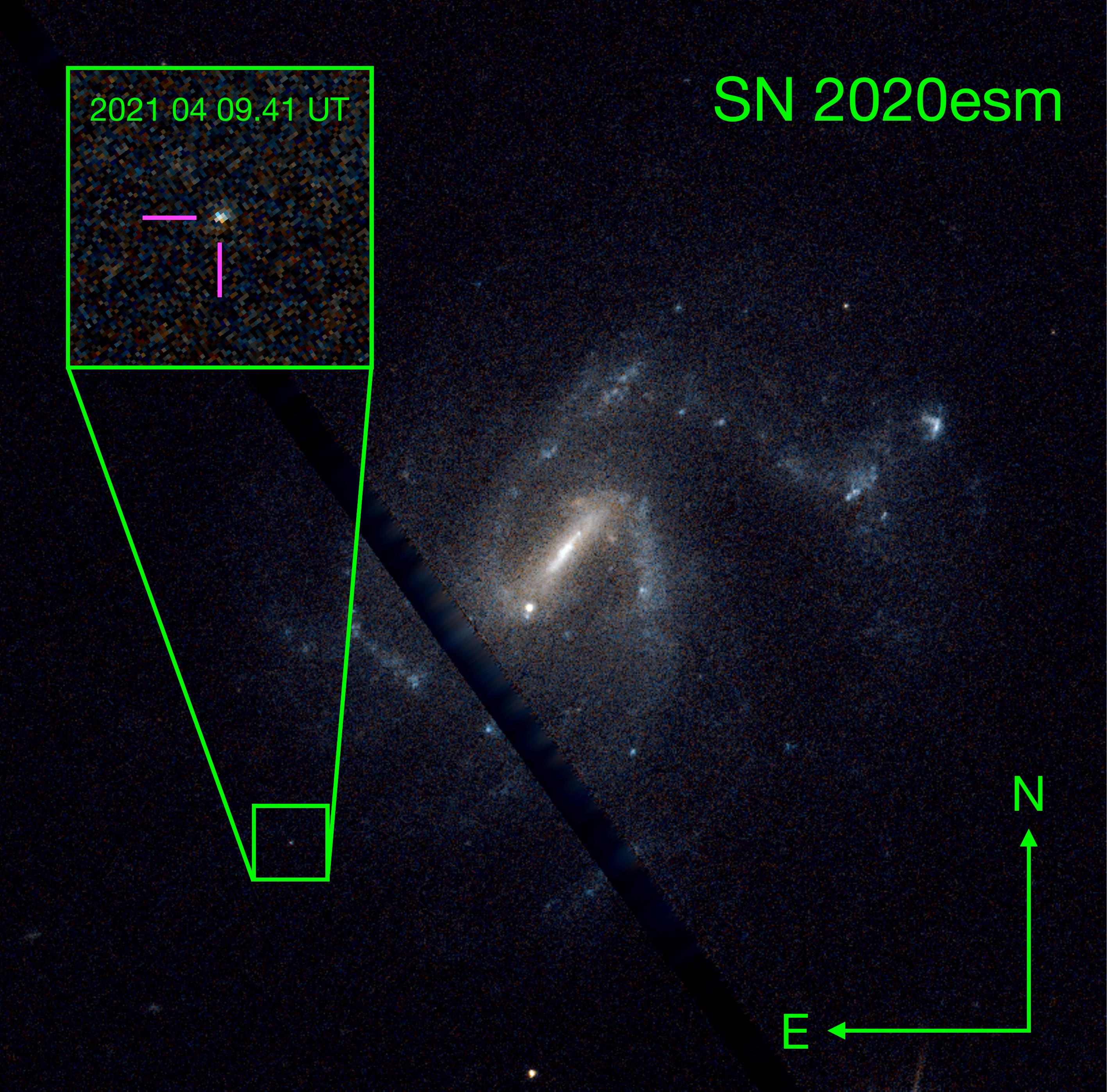}
\caption{Pan-STARRS 5\arcmin{}$\times$5\arcmin{} color composite (\textit{g}/\textit{i}/\textit{y}, left) and HST/WFC3 45\arcsec{}$\times$45\arcsec{} color composite (\textit{F555W}/\textit{F814W}, right) image stamp of the field of WISEA J140717.48-050726.1, the host of SN\,2020esm. The location of the SN is indicated with the magenta tick-marks. PS1 standard stars in the field are marked with red circles. The green insets show zoomed-in regions, centered on the SN location ($1\arcmin\times1\arcmin$, left and 3\arcsec{}$\times$3\arcsec{}, right), taken at a phase of $14.35$ and $356.02$ rest-frame days from $\textit{B}$-band maximum. 
\label{fig:image}}
\end{figure*}

The supernova is located at $\alpha=14^{\rm{h}}07^{\rm{m}}18^{\rm{s}}.260$, $\delta=-05^{\rm{o}}07\arcmin37\arcsec.67$ (J2000.0), $10.9$\arcsec\ East and $11.8$\arcsec\ South of WISEA J140717.48-050726.1, a star-forming irregular galaxy at a redshift of $z = 0.03619 \pm 0.00015$ \citep{6df_2}. In Figure~\ref{fig:image}, left panel, we present a deep pre-explosion color composite (\textit{g}/\textit{i}/\textit{y}) image stamp of WISEA J140717.48-050726.1, with the location of SN\,2020esm marked with magenta tick-marks, while the green inset shows a zoomed-in region of a \textit{g}-band Pan-STARRS image stamp of the supernova, at a phase of $14.35$ rest-frame days from \textit{B}-band maximum.

Originally, SN\,2020esm was classified as a young Type II supernova \citep{Tucker20TNSCR}, based on a spectrum taken on UT 2020 March 23.45 ($\sim$1.08 days after discovery) with the University of Hawaii 2.2-m telescope (UH88) Supernova Integral Field Spectrograph \citep[SNIFS;][]{Lantz2004SPIE}. That classification was the result of misclassifying the \ion{C}{2} absorption at $\sim$6,300~\AA\ and $\sim$7,000~\AA\ as a P-Cygni \Ha\ feature. Moreover, based on its distance and minimal extinction on the line of sight (Section~\ref{subsec:host_dist_extinction}), SN\,2020esm was discovered at an absolute magnitude of $-18.4\pm0.15$, indicating a luminous event, generally not consistent with core-collapse SNe. Examination of spectra taken at later phases (Figure~\ref{fig:spec_series}) clearly shows that SN\,2020esm is a SN~Ia, and particularly of the ``super-Chandrasekhar'' (SC) subclass. The discovery of a candidate of one of the most rare sub-type of SNe~Ia led us to initiate an extensive multi-wavelength observational campaign.

\subsection{Photometry} \label{subsec:phot}

We obtained optical photometric observations of SN\,2020esm with various telescopes/instruments. Our main photometry was performed with the SINISTRO cameras of the LCOGT \citep{LCOGT13PASP} network of 1-m telescopes (NOAO2020A-012 and NOAO2020B-011, PI: Foley). Images were obtained in \textit{ugri}, from 2020-03-25 UT (approximately 3 days after discovery and 11 days before peak brightness) until 2020-07-22 UT (107 days after maximum). Additional \textit{griz} photometry was obtained through the Young Supernova Experiment \citep[YSE;][]{Jones21ApJ} with the Pan-STARRS1 telescope (PS1) between 2020-03-25 UT and 2020-07-24 UT, \textit{BVgri} photometry with the 1-m telescope at the Lulin observatory in Taiwan, and \textit{griz} photometry with the  0.7-m telescope of the Thacher observatory in California (Swift et~al., in prep.). Late-time \textit{griz} imaging (at 280 and 305 days from maximum brightness) was performed with the Gemini Multi-Object Spectrograph (GMOS) on the 8.1-m Gemini North telescope in Mauna Kea (GN-2020B-Q-324 and GN-2021A-DD-102, PI: Foley) and $F555W$ and $F814W$ imaging with \textit{HST}/WFC3 at 356 days from maximum brightness (SNAP-16239, PI: Foley).

All ground-based images were reduced, resampled, and calibrated using \textsc{photpipe} \citep{Rest14} with absolute flux calibration performed using PS1 standard stars in the SN field. For the \textit{HST}/WFC3 photometry, reduced images were downloaded from the Mikulski Archive for Space Telescopes and drizzled following the techniques of \citet{Kilpatrick2018}. Aperture photometry was performed with \texttt{photutils} \citep{Bradley2020} using a 0.2 arcsec aperture, and instrumental magnitudes were calibrated using AB zeropoint conversions in the fits headers. No difference imaging was performed, as the background light from the host galaxy is minimal.

UV photometric observations were performed with the Ultraviolet Optical Telescope \citep[UVOT;][]{Roming05} onboard the Neil Gehrels Swift Observatory \citep{Gehrels04} from 2020-03-24 UT until 2020-04-30 UT, with template images obtained on 2020-12-21 UT. Aperture photometry on the difference images was performed in a 5\arcsec\ region on the SN location, following the standard guidelines from \citet{Brown09}, using the most recent calibration database (CALDB, version 20201008).

\begin{figure}[t]
\begin{center}
\includegraphics[width=0.47\textwidth]{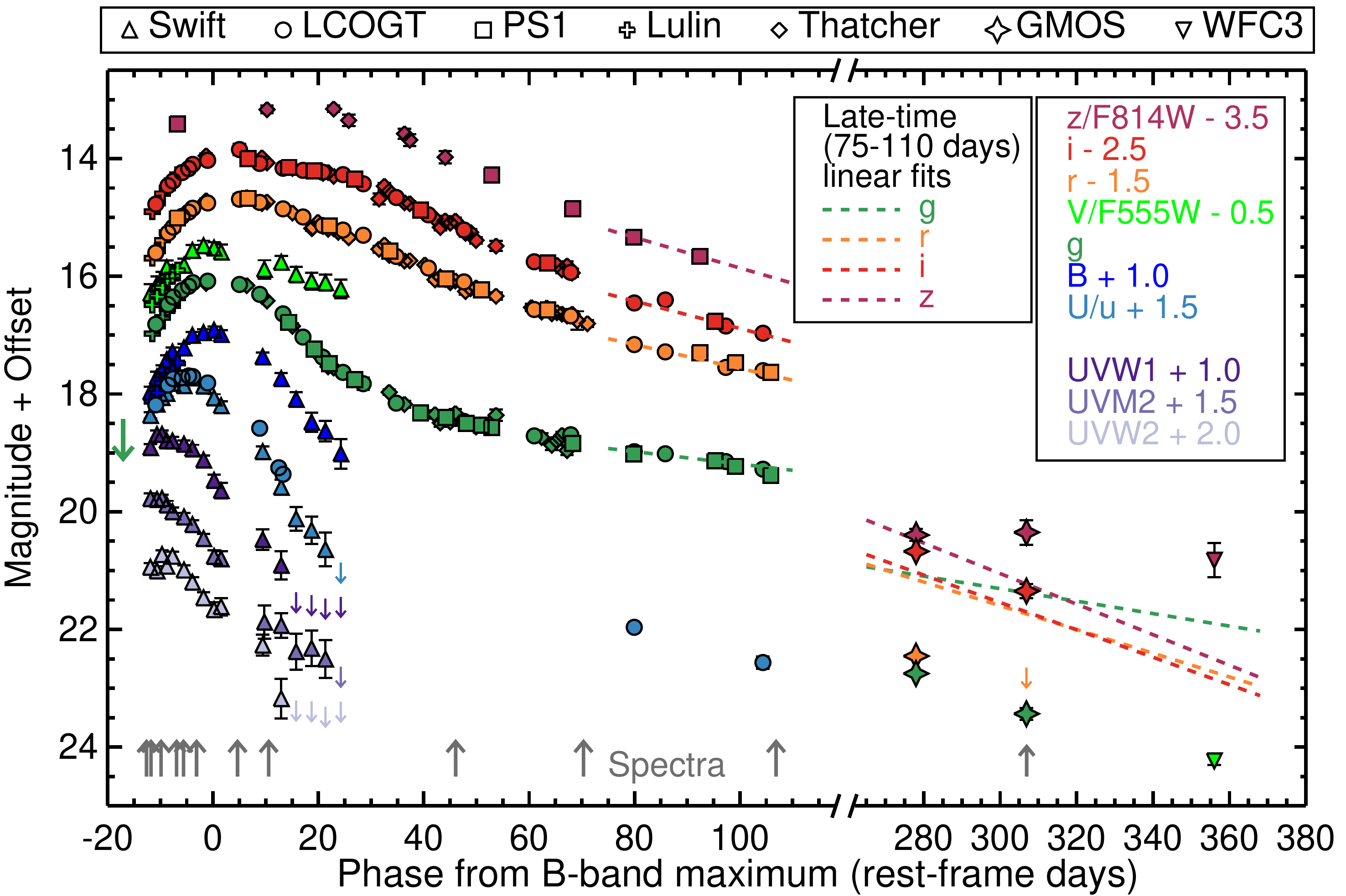}
\caption{Multicolor light curves (\textit{UV}+\textit{uBgVriz}+WFC3) of SN\,2020esm in rest-frame days, with respect to \textit{B}-band maximum. The light curves are plotted with offsets, as described in the legend. Downward arrows mark non-detections at the location of the SN. The dashed lines shown in the late-time regime are linear fits to the 75 to 110 days \textit{griz} light curves, extended at the phase of our late-time observations. We additionally show, with upward arrows, the phases corresponding to our spectral series (Table~\ref{table:spec_log}). \label{fig:lc_apparent}}
\end{center}
\end{figure}

We present the complete SN\,2020esm light curves, corrected for MW extinction, in Figure~\ref{fig:lc_apparent}, spanning from $-11.89$ to $+356.02$ days relative to peak $B$ brightness, which we estimate by fitting a polynomial to the \textit{B}-band Swift light curve from MJD $58935.0$ to $58958.0$, to be at $\mathrm{MJD}_{max}^{B}=58944.585$. Our complete photometric dataset is reported in Table~\ref{table:phot_data}. We note that no attempt for cross-filter corrections and K-corrections was made.

\subsection{Spectroscopy} \label{subsec:spec}

Spectroscopic coverage of SN\,2020esm ranges from roughly $-12$ to $+307$ days relative to $B$-band maximum. We obtained a total of seven spectroscopic observations with the FLOYDS spectrograph on the Faulkes 2-m telescopes of the Las Cumbres Observatory Global Telescope Network \citep[LCOGTN;][]{LCOGT13PASP} robotic network (NOAO2020A-012, PI: Foley), one spectrum with the Kast spectrograph \citep{KAST} on the Lick Shane telescope (2020A-S011, PI: Foley), two spectra with the Low-Resolution Imaging Spectrometer \citep[LRIS;][]{LRIS} on the Keck I telescope (2020A-U209, PI: Foley) and one spectrum with Gemini Multi-Object Spectrograph \citep[GMOS;][]{GMOS} on the Gemini-North telescope (GN-2021A-DD-102, PI: Foley). 

The spectra were reduced using standard \textsc{iraf/pyraf}\footnote{IRAF is distributed by the National Optical Astronomy Observatory, which is operated by the Association of Universities for Research in Astronomy (AURA) under a cooperative agreement with the National Science Foundation.} and python routines for bias/overscan subtractions and flat fielding. The wavelength solution was derived using arc lamps while the final flux calibration and telluric lines removal were performed using spectro-photometric standard star spectra, obtained the same night \citep{Silverman2012MNRAS}. 

\begin{figure}[t]
\begin{center}
\includegraphics[width=0.47\textwidth]{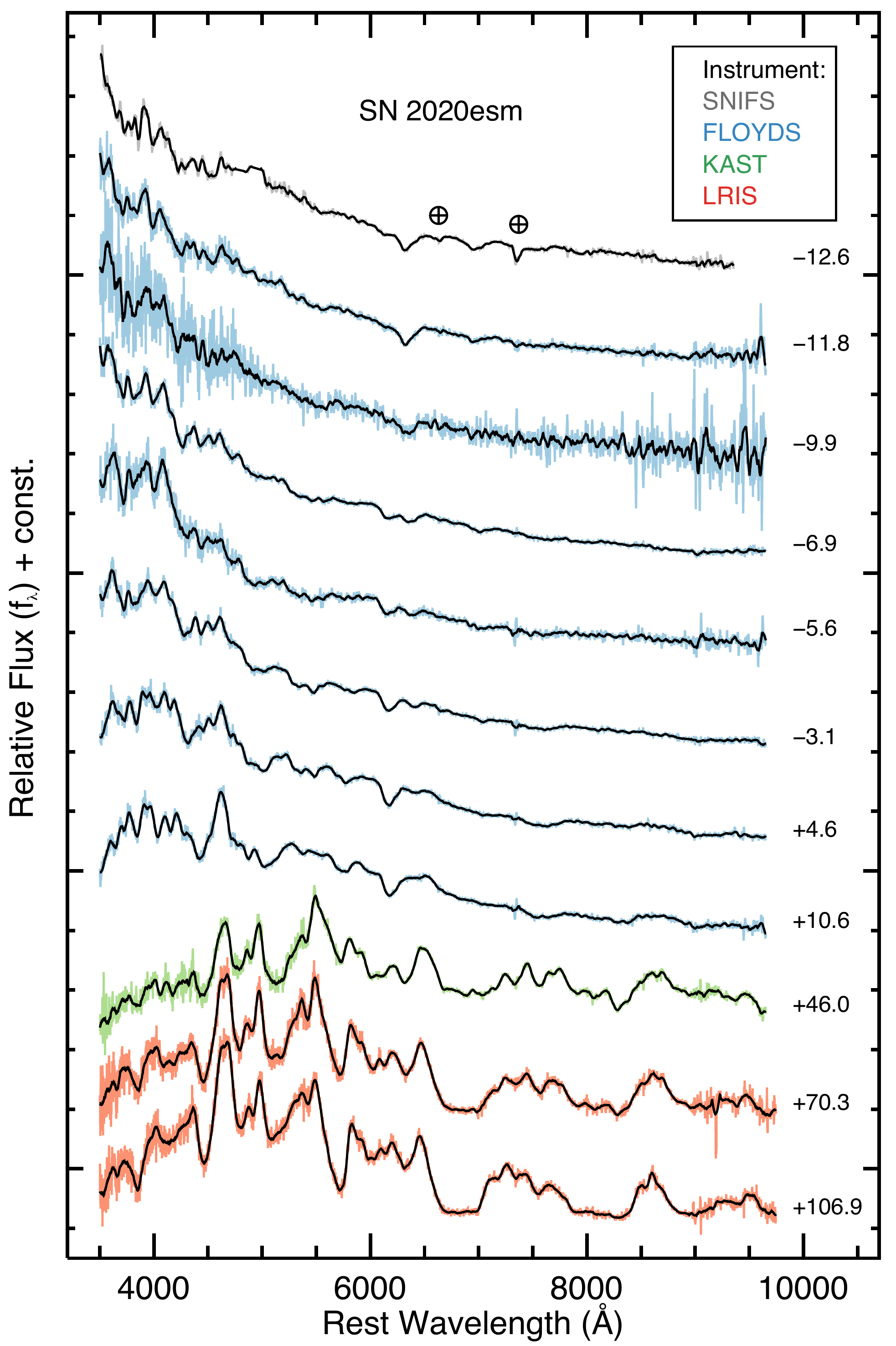}
\caption{The spectroscopic series of SN\,2020esm, spanning from $-13$ days to $+107$ days from \textit{B}-band peak brightness. Each spectrum is color-coded according to each source, indicated in the legend. The black lines correspond to the smoothed versions of the spectra, using a Savitsky–Golay filter, with each spectrum's phase additionally labeled. The spectra have been corrected for Milky Way extinction ($E(B-V)_{\rm MW}=0.0207$) and placed in rest-frame wavelength ($z = 0.03619$). \label{fig:spec_series}}
\end{center}
\end{figure}

Figure~\ref{fig:spec_series} shows the $-13$ days to $+107$ days from $B$-band peak brightness spectral series (the spectrum at $+307$ days is presented and analysed at Section~\ref{subsec:spec_analysis}). Detailed information of each observation is provided in Table~\ref{table:spec_log} and the complete spectroscopic data set is available in the electronic edition.

\section{Analysis} \label{sec:analysis}

In this Section, we discuss the host galaxy of SN\,2020esm, its distance and extinction along the line of sight, and we present our analysis of its photometric and spectroscopic data set. Throughout our analysis, we will use data for SN\,2009dc from \citet{Taubenberger11}, \citet{Brown2014}, \citet{Silverman2011MNRAS} and \citet{Friedman2015ApJS}, for SN\,2012dn from \citet{Taubenberger19}, \citet{Brown2014} and \citet{Yamanaka2016PASJ} and for SN\,2011fe from \citet{Pereira13}, \citet{Zhang16}, \citet{Brown2012ApJ} and \citet{Matheson2012ApJ}. We note that, while all of these SNe were photometrically observed primarily in \textit{UBVRI}-bands, they all have excellent spectrophotometric coverage, and using these spectra we can estimate \textit{ugri} light curves, and directly compare with the light curves of SN\,2020esm.

\subsection{Host galaxy, distance and extinction} \label{subsec:host_dist_extinction}

The host galaxy of SN\,2020esm is the star-forming irregular galaxy WISEA J140717.48-050726.1, at a redshift of $z = 0.03619 \pm 0.00015$ \citep{6df_2}. An estimate of the distance based on fitting the SN light curves (a usual method for determining distances) is not possible, since SN\,2020esm is not a normal SN~Ia, as it is brighter for its light curve shape. With no redshift-independent distances available, we choose to use the cosmological distance of $D = 156.8 \pm 11.0$~Mpc (H$_{0} = 73.0 \pm 5.0$~km~s$^{-1}$~Mpc$^{-1}$ and correcting for peculiar motions related to the Virgo cluster and Great Attractor, \citealt{Mould00}). At this distance, SN\,2020esm is located 12.2 kpc from the galaxy's core.

The Milky Way extinction on the line of sight is $E(B-V)_{\rm MW}=0.0207\pm0.0007$~mag, using the \citet{Schlafly11} dust maps. Regarding host galaxy extinction, the large (projected) distance of the SN to its host galaxy indicates minimal extinction. Visual inspection of our high-resolution Keck/LRIS spectra show no obvious narrow \ion{Na}{1} D absorption at the host redshift. Using these spectra, we are able to provide a 3-$\sigma$ upper limit of the \ion{Na}{1} D absorption lines equivalent width of $EW_{\mathrm{upper}}=0.072$~\AA, and assuming a \citet{Fitzpatrick99} reddening law with $R_{V}=3.1$, this corresponds to an upper limit of $E(B-V)_{\rm host}=0.017\pm0.012$~mag \citep[][with updated uncertainty as per \citealt{Phillips2013ApJ}]{Poznanski12MNRAS}. Thus, we use only the Milky Way reddening on the line of sight and we adopt this value to correct all of our photometry and spectra.

SC SNe~Ia tend to explode in relatively low-mass galaxies, with high specific star formation rate (sSFR), with events hosted by more massive galaxies tending to explode in remote locations \citep{Taubenberger11}. In order to place SN\,2020esm in the context of SC SN~Ia host galaxies, we derive a stellar mass and SFR for WISEA J140717.48-050726.1 using the SED fitting package \textsc{Le PHARE} \citep{Arnouts1999MNRAS,Ilbert2006}. The code uses the population-synthesis templates of \citet{Bruzual2003MNRAS}, summed according to an exponentially declining burst of star formation. We assume a \citet{Chabrier2003PASP} initial mass function (IMF) and a stellar metallicity set between 0.2--1.0 Z$_{\sun}$.  Dust attenuation in the galaxy is applied to the SED models using the \citet{Calzetti2000ApJ} template. As an input, we performed elliptical aperture photometry on PS1 \citep{PS1} images in \textit{grizy} bands, using PS1 field stars for calibration, and corrected all photometry for foreground extinction. We find a stellar mass of $\mathrm{log}(M/\mathrm{M_{\sun})}=9.72^{+0.05}_{-0.03}$ and a SFR of $\mathrm{log}(M/\mathrm{M_{\sun}\:yr^{-1})}=-0.07^{+0.51}_{-0.67}$ ($\mathrm{log\:sSFR (yr^{-1})}=-9.79$). For consistency, we performed the same analysis for the host galaxies of SNe\,2009dc and 2012dn, finding stellar masses of $\mathrm{log}(M/\mathrm{M_{\sun})}=10.67^{+0.04}_{-0.06}$ and $\mathrm{log}(M/\mathrm{M_{\sun})}=9.65^{+0.05}_{-0.08}$, with SFRs $\mathrm{log}(M/\mathrm{M_{\sun}\:yr^{-1})}=-0.50^{+0.52}_{-0.89}$ and $\mathrm{log}(M/\mathrm{M_{\sun}\:yr^{-1})}=0.11^{+0.32}_{-0.06}$ ($\mathrm{log\:sSFR (yr^{-1})}=-11.17$ and $-9.54$), respectively.

\subsection{Photometric evolution} \label{subsec:phot_analysis}

\begin{figure}[t]
\begin{center}
\includegraphics[width=0.47\textwidth]{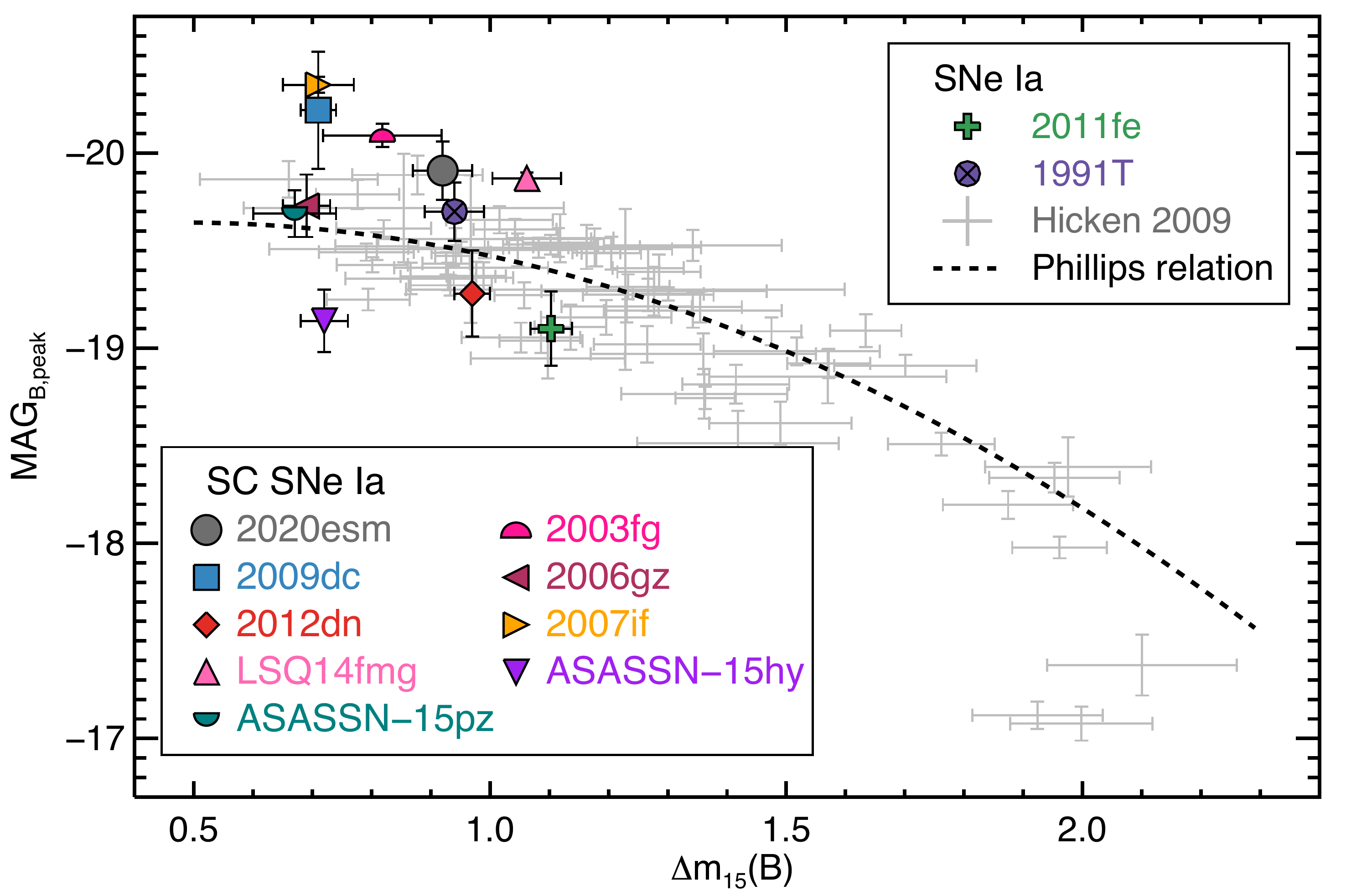}
\caption{The absolute $B$-band magnitude vs. $\Delta m_{15}(B)$ of SNe~Ia. Grey crosses correspond to the CfA SNe Ia sample from \citet{Hicken09ApJ}, with the dashed line showing the Phillips relation. We overplot SN\,2020esm with a grey circle, and various other SC SNe Ia as described in the legend. The normal SN\,2011fe is shown with an green cross and the overluminous SN\,1991T with a purple circle with X. \label{fig:phillips_rel}}
\end{center}
\end{figure}

SN\,2020esm peaked in \textit{B} band on 2020-04-05 UT, at $B = 16.16\pm0.03$~mag and taking into account the distance to WISEA J140717.48-050726.1 and the extinction on the line of sight (Section~\ref{subsec:host_dist_extinction}), the peak absolute \textit{B}-band magnitude was $-19.91\pm0.15$~mag. The magnitude decline in \textit{B}-band after 15 days was $\Delta m_{15}(B) = 0.92\pm0.05$~mag, comparable to other high-luminosity/slowky declining SC SNe~Ia ($\Delta m_{15}(B)=0.97$ and $0.71$~mag for SN\,2012dn and SN\,2009dc, respectively, as opposed to $\Delta m_{15}(B)=1.1$~mag for the normal SN\,2011fe, Figure~\ref{fig:phillips_rel}).

\begin{figure}[t]
\begin{center}
\includegraphics[width=0.47\textwidth]{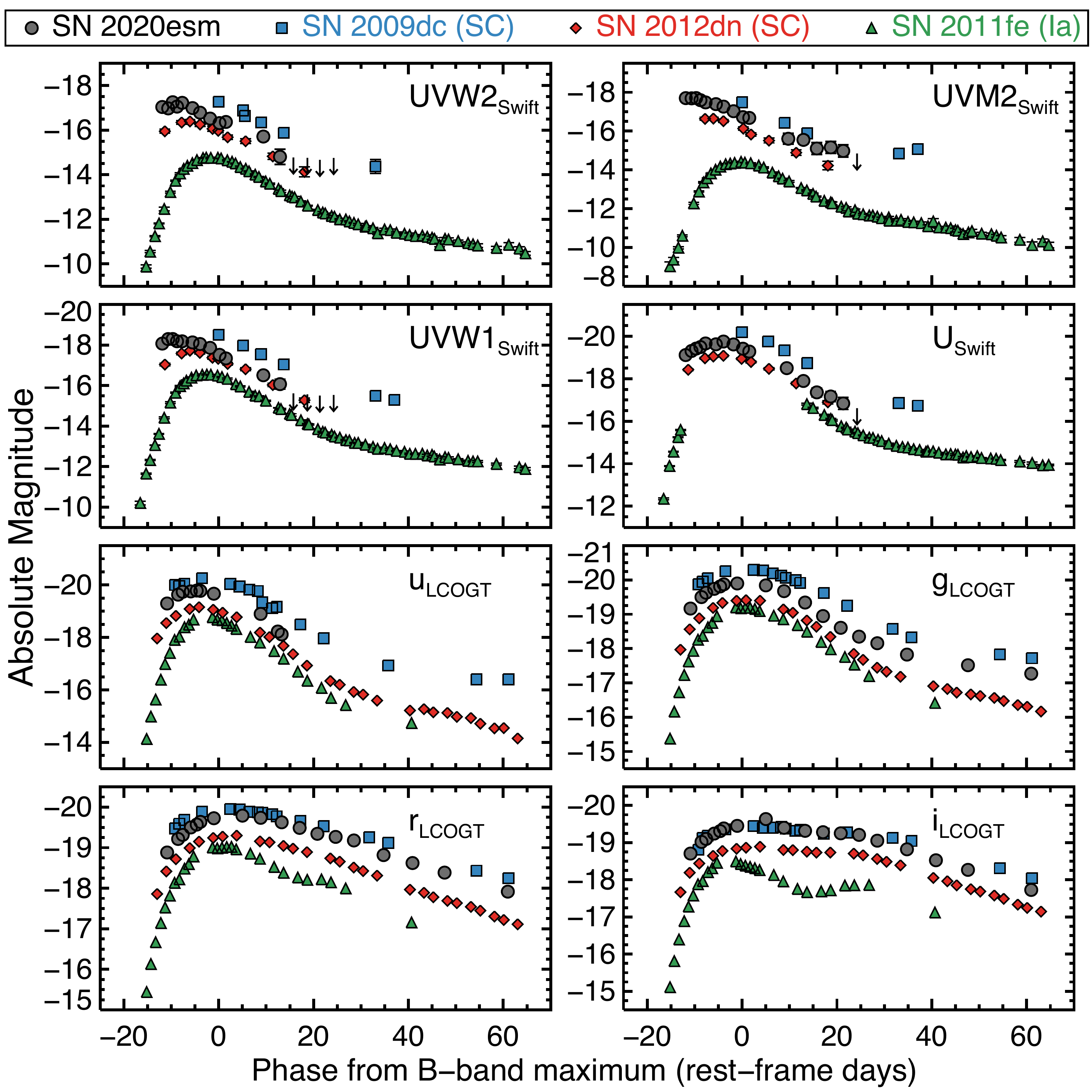}
\caption{The UV (Swift \textit{UVW2}, \textit{UVM2}, \textit{UVW1} and \textit{U}) and optical (\textit{ugri}) light curves of SN\,2020esm (grey circles) are presented in absolute magnitude. The light curves are compared with the light curves of SN\,2009dc (blue squares), SN\,2012dn (red diamonds) and SN\,2011fe (green upward triangles).
\label{fig:abs_lcs}}
\end{center}
\end{figure}

We show the absolute magnitude light curves of SN\,2020esm in Figure~\ref{fig:abs_lcs}. SN\,2020esm displays several characteristics associated with SC SNe~Ia in contrast to normal SNe~Ia: It is substantially brighter at peak ($\sim$2-3 mags brighter in the UV and $\sim$1 mag in the optical), shows a slower evolution of the light curve in all wavelengths and lacks the distinctive secondary maximum in the redder bands. Of particular interest are the UV light curves: While SN\,2011fe showed a steep rise from around $-15$~days and peaked at $-1.7$ (\textit{UVW2}), $-0.2$ (\textit{UVM2}) and $-2.3$ (\textit{UVW1}) days from \textit{B}-band maximum, SN\,2020esm is close to peak or fading from our first observations ($-12$ days from \textit{B}-band maximum). The general evolution of the light curves match the SC SNe~Ia, closely resembling SN\,2009dc, although somewhat fainter, while being brighter than SN\,2012dn.

\begin{figure}[t]
\begin{center}
\includegraphics[width=0.47\textwidth]{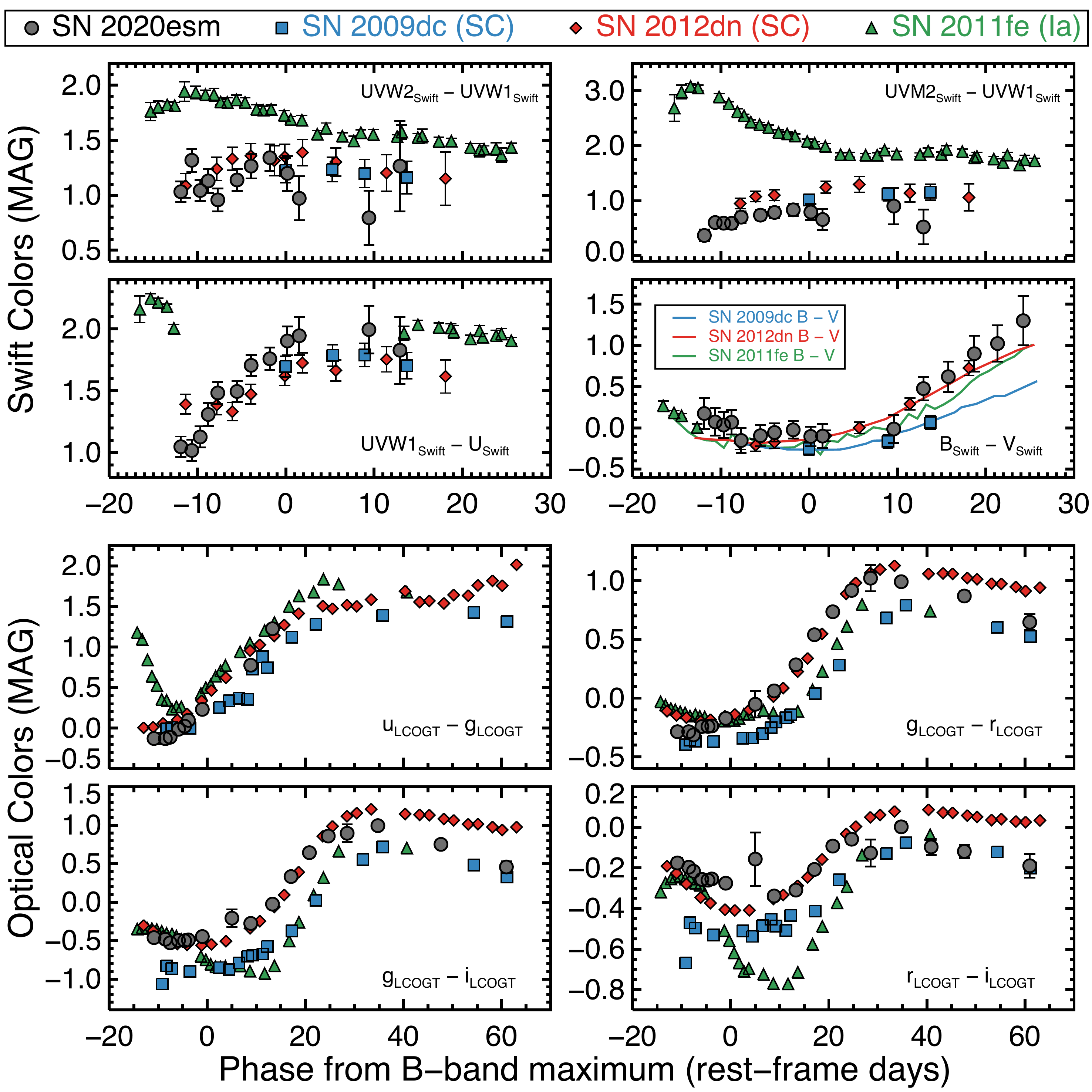}
\caption{The Swift (\textit{UVW2-UVW1}, \textit{UVM2-UVW1}, \textit{UVW1-U}, \textit{B-V}) and optical (\textit{u-g}, \textit{g-r}, \textit{g-i} and \textit{r-i}) color curves of SN\,2020esm (grey circles), compared with SN\,2009dc (blue squares), SN\,2012dn (red diamonds) and SN\,2011fe (green upward triangles). For the Swift \textit{B-V} panel, we additionally plot ground-based \textit{B-V} color curves for each SN, as solid lines.\label{fig:colors_all}}
\end{center}
\end{figure}

Figure~\ref{fig:colors_all} shows the {\it Swift} and optical color evolution of SN\,2020esm. The observed early colors are considerably bluer than those of normal SNe~Ia, particularly in the UV, due to the small amount of line blanketing and the weak \ion{Ca}{2} H\&K absorption feature (Figure~\ref{fig:spec_series}). Another characteristic of SC SNe~Ia, is the optical color evolution in \textit{r}-\textit{i} from peak up to about +30 days, when normal SNe~Ia show their bluest colors, while SC SNe~Ia continuously evolve to redder colors, a behavior seen in SN\,2020esm.

The evolution of the light curve changes dramatically at later times (Figure~\ref{fig:lc_apparent}). Our \textit{griz} observations at $\sim$280 and $\sim$305 days from \textit{B}-band maximum show that the \textit{g}- and \textit{r}-band luminosities are substantially fainter than expected from linear fits at the 75--110 day regime ($\sim$1.7--2.0 and $\sim$1.3~mag, respectively), while \textit{i} and \textit{z} bands have faded less (only $\sim$0.35 and $\sim$0.1--0.9 mag, respectively), with the \textit{z} band showing a (nearly) constant evolution. Similar evolution is seen in the \textit{HST} photometry (at $\sim$356 days, Figure~\ref{fig:image}, right), where the $F555W$ (\textit{V}-band equivalent) is substantially faint, while $F814W$ (wide \textit{I}) remains bright.

This photometric behaviour is remarkably similar to SN\,2009dc \citep{Taubenberger11}: its \textit{R}-band light curve continues its earlier phase linear decline, while the \textit{B} band at $\sim$260 days from maximum is $\sim$0.7 mag fainter and the \textit{I} band at $\sim$295 days from maximum is $\sim$0.9 mag brighter, relative to their linear declines, with an analogous behaviour (although observed with fewer data) seen in the SC SN\,2006gz \citep{Maeda09ApJ}. A more drastic change is observed in LSQ14fmg and SN\,2012dn even earlier: the optical light curves start to rapidly fade after $\sim$30 and 60 days from maximum, respectively, with a simultaneous increase of the NIR luminosity, which has been attributed to either pre-existing dust in a NIR echo scenario \citep{Yamanaka2016PASJ} or the CO formation/dust formation in the SN ejecta \citep{Hsiao20ApJ,Taubenberger19}. We note that, while we have not acquired NIR photometry for SN\,2020esm at these epochs, there is no indication of a dramatic shift of the emission to longer wavelengths, as our \textit{z}-band observations, although brighter than expected, do not show an increase in brightness, ruling out the formation of a significant amount of dust.
 
This change in decline rates (seen also in the bolometric light curves, Section~\ref{subsec:bol_lc}) is in contrast with normal SNe~Ia (such as SN\,2011fe, \citealt{Zhang16}) that have consistent late-time declines until $\sim$500 days \citep{Dimitriadis17MNRAS}. Moreover, the rapid fading is inconsistent with the canonical SN~Ia radioactive decay evolution, where the (bolometric) decline rate is expected to slow down after $\sim$250--300 days from maximum, and resemble the $^{56}$Co decay rate. This shared characteristic of SC SNe~Ia indicates a common explosion mechanism for this subclass of thermonuclear explosions, potentially different from normal SNe~Ia. We defer to Section~\ref{sec:discussion} for further discussion on the physical mechanisms that may lead to this unexpected behaviour.

\subsection{Spectroscopic evolution} \label{subsec:spec_analysis}

\begin{figure*}[t]
\begin{center}
\includegraphics[width=0.97\textwidth]{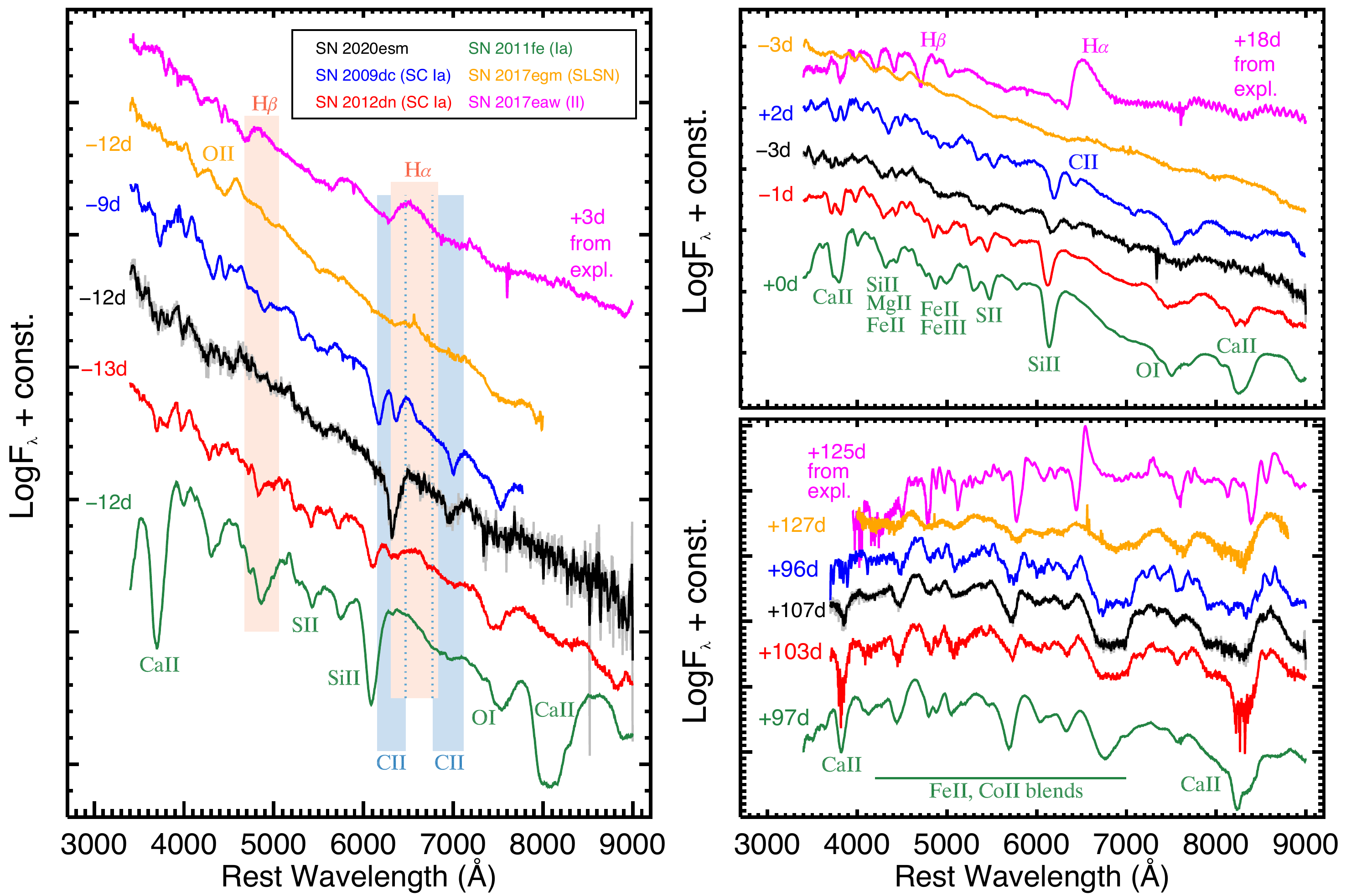}
\caption{Pre-maximum spectra ($<-$10 days from peak brightness, left), maximum brightness (top right) and post-maximum ($>$90 days from peak brightness, bottom right) of SN\,2020esm, compared to the normal type Ia SN\,2011fe, two SC SNe~Ia (SN\,2009dc and SN\,2012dn), the SLSN SN\,2017egm \citep{Nicholl2017ApJ} and the Type II SN\,2017eaw \citep{Szalai19}. All spectra are in flux density per unit wavelength, $F_{\lambda}$, and presented in logarithmic scale with additional offsets, for presentation purposes. Orange-shaded regions represent the H$\alpha$ and H$\beta$ at $\pm$12,000 $\mathrm{km s^{-1}}$ (typical ejecta velocities of early hydrogen-rich core-collapse SNe), while the blue-shaded ones correspond to \ion{C}{2} at $-20,000$ to $-5,000$ $\mathrm{km s^{-1}}$ (typical ejecta velocities of unburnt material, seen in early thermonuclear SNe). Other intermediate-mass and iron-peak elements, present in normal SNe~Ia spectra, are marked with green labels, while the \ion{O}{2} line blend, identified in SL SNe is marked in orange. The phases relative to each SN's maximum are additionally labeled. \label{fig:spec_comparisons}}
\end{center}
\end{figure*}

Figure~\ref{fig:spec_comparisons} shows early (left), peak (top right) and late (bottom right) spectra of SN\,2020esm, compared to the SC SNe~Ia 2009dc and 2012dn, the normal SNe~Ia 2011fe, the SLSN 2017egm and the SN II 2017eaw.

\begin{figure}[t]
\begin{center}
\includegraphics[width=0.47\textwidth]{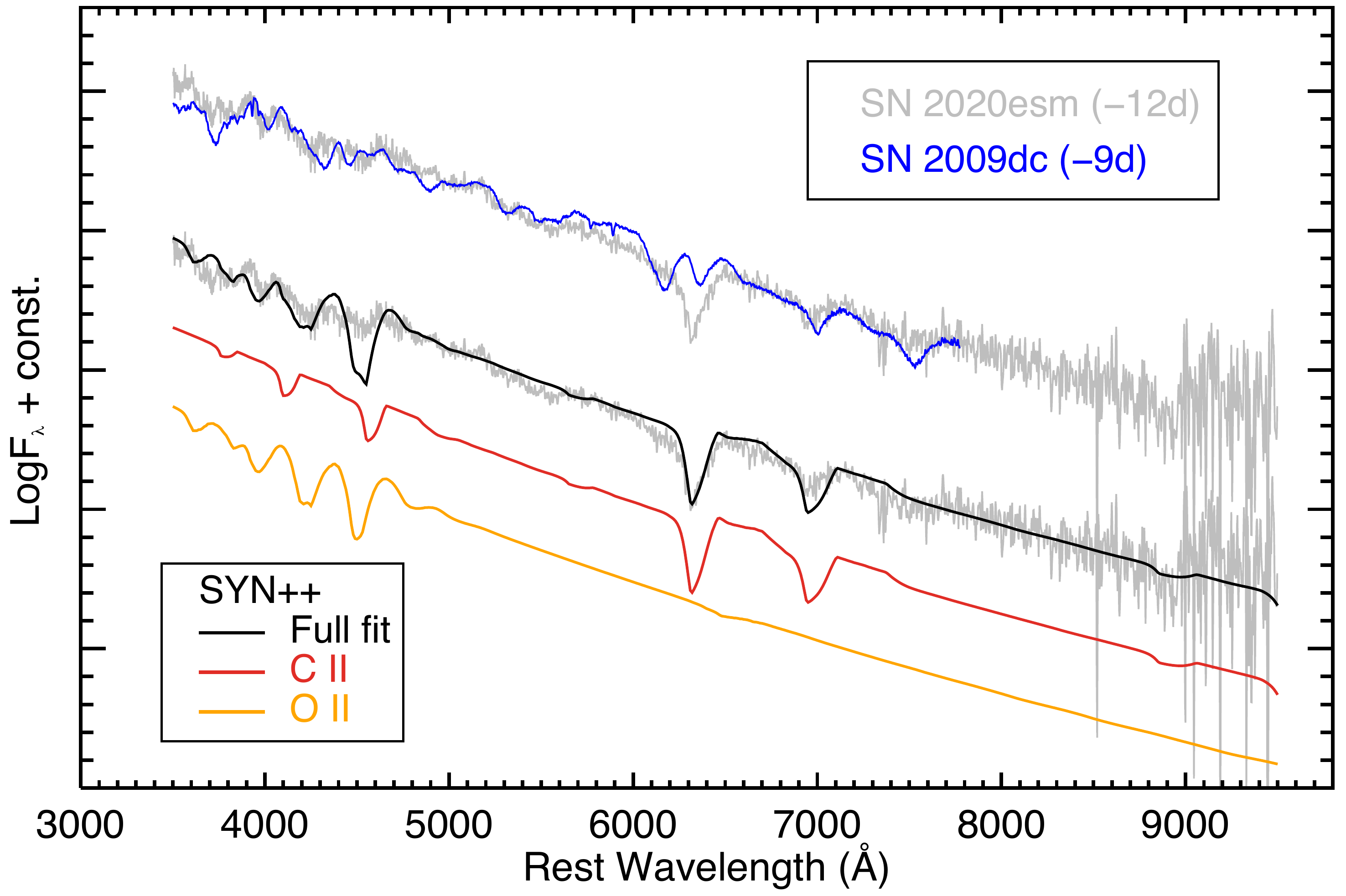}
\caption{The early spectrum of SN\,2020esm is shown as a grey line. Our best fit is shown as a black line, which consists of a 15,500 K black body, with \ion{C}{2} and \ion{O}{2}, with the decomposition of the atomic species shown as red and orange lines respectively. We additionally show the early ($\sim$-9 day) spectrum of SN\,2009dc as a blue line.\label{fig:spec_synpp}}
\end{center}
\end{figure}

The early spectra of SN\,2020esm (Figure~\ref{fig:spec_series} and Figure~\ref{fig:spec_comparisons}, left) are unique among known thermonuclear WD SNe. The spectra are dominated by a blue continuum with a strong absorption feature at $\sim$6,300~\AA\ and a somewhat weaker one at $\sim$7,000~\AA. We model the $-11.8$-day spectrum with \textsc{syn++} \citep{SYNAPPS}, a parameterized SN spectral synthesis software. \textsc{syn++} treats the SN as an optically thick, continuum-emitting pseudo-photosphere surrounded by a line-forming region of various atomic elements, and assumes spherical symmetry, local thermal equilibrium and homologous expansion of the ejecta. While this generalised approach might not accurately describe the physical conditions in the SN, it is useful for inferring the presence (or absence) of specific atomic species at various ejecta velocities and ionization states. Our main results are shown in Figure~\ref{fig:spec_synpp}, where the grey and blue lines correspond to SNe\,2020esm and 2009dc, respectively, at similar phases. Our full fit for SN\,2020esm is shown as a black solid line.

The continuum is well described with a black body of $T=15,500$ K, on which we add \ion{C}{2} at $\sim$12,800~km~s${-1}$ (red line), detached from the photosphere ($\sim$11,500~km~s$^{-1}$), in order to reproduce the flat emission peak and the sharp red edge of the $\sim$6,300~\AA\ P-Cygni absorption feature. There is an indication of a weak absorption feature at $\sim$6,000~\AA, which could be \ion{Si}{2} at the photospheric velocity. However, as it can be seen by comparing SN\,2020esm with SN\,2009dc, other IMEs (such as \ion{S}{2} and \ion{Ca}{2}) are absent or extremely weak. This suggests that silicon is either absent or extremely weak at these early phases (substantially weaker that SN\,2009dc). On the contrary, the absorption features at 3500--4800~\AA\ are adequately matched with \ion{O}{2} (orange line). While other IMEs might be present, although extremely weak, the spectrum can be generally reproduced with only \ion{C}{2} and \ion{O}{2}, giving additional evidence of a nearly pure carbon/oxygen atmosphere above the expanding photosphere. This spectrum is similar to the earliest spectra of some superluminous SNe (thought to have core-collapse progenitors, \citealt{Quimby18}), but later spectra of SN\,2020esm are distinct from those of that class. With only a single early spectrum, one might incorrectly classify a SN similar to SN\,2020esm as a SN\,II (as was done for SN\,2020esm) or a SLSN, potentially contaminating samples of those classes (see \citealt{Jerkstrand2020Sci} for a similar case of a potential misclassification of SN~2006gy).

\begin{figure}[t]
\begin{center}
\includegraphics[width=0.47\textwidth]{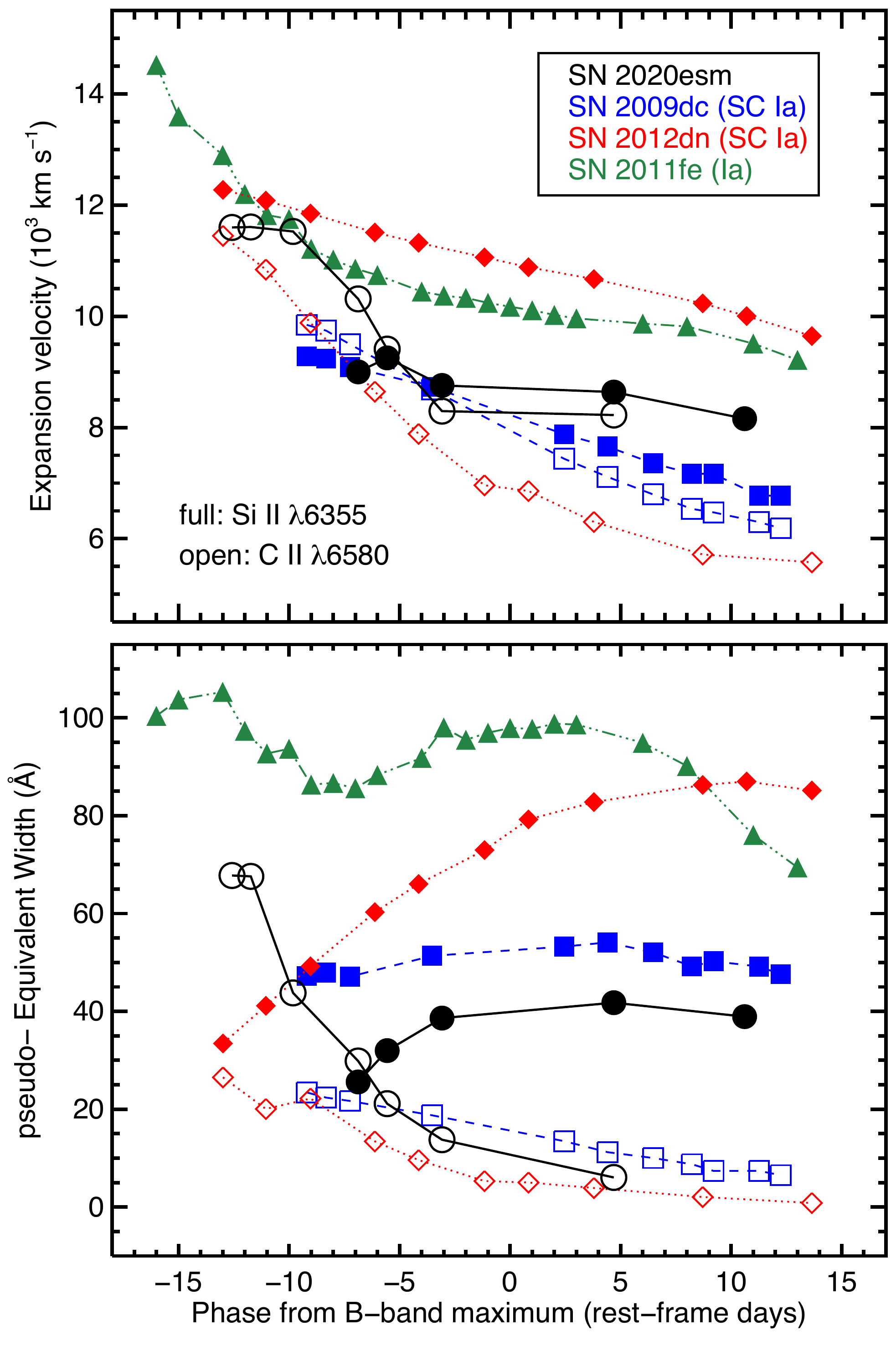}
\caption{The expansion velocity (top) and pseudo- equivalent width (bottom) of \ion{Si}{2} $\lambda$6355 (full symbols) and \ion{C}{2} $\lambda$6580 (open symbols) as a function of phase for SN\,2020esm (black circles), SN\,2009dc (blue squares), SN\,2012dn (red diamonds) and SN\,2011fe (green upward triangles). Typical uncertainties are $\sim$200 km s$^{-1}$ and $\sim$ 5\AA, respectively. \label{fig:spec_vel_ew}}
\end{center}
\end{figure}

As the SN evolves towards peak brightness, the usual IMEs seen in thermonuclear SNe (e.g., \ion{O}{1}, \ion{Mg}{2}, \ion{Ca}{2}, \ion{Si}{2}, \ion{S}{2}, \ion{Fe}{2}, \ion{Fe}{3}) start to appear, although considerably weaker, in accordance with SC SNe~Ia (Figure~\ref{fig:spec_comparisons}, top right). At this phase, the SN has evolved to a clear thermonuclear one, confirming the misclassification of it as a core collapse. The ejecta velocities and strengths of the unburned and synthesized products, probed by \ion{C}{2} and \ion{Si}{2}, respectively, are estimated by fitting these two features with Gaussian profiles across our spectral series, with our results shown in Figure~\ref{fig:spec_vel_ew}.

The ejecta velocities, probed by the IME \ion{Si}{2}, remain approximately constant throughout SN\,2020esm's evolution, with a velocity similar to SN\,2009dc (and substantially lower than SN\,2011fe). However, the \ion{C}{2} velocity is higher than both SN\,2009dc and SN\,2012dn, up to $\sim${}$-5$~days, indicating that unburnt material at early times is located above the photosphere. Moreover, the equivalent width of the \ion{C}{2} line is particularly high, and becomes similar in strength to that of SN\,2009dc around maximum light, after which the \ion{C}{2} and \ion{Si}{2} velocities become comparable, suggesting that the unburnt material is mixed with IMEs in lower layers of the ejecta.

\begin{figure}[t]
\begin{center}
\includegraphics[width=0.47\textwidth]{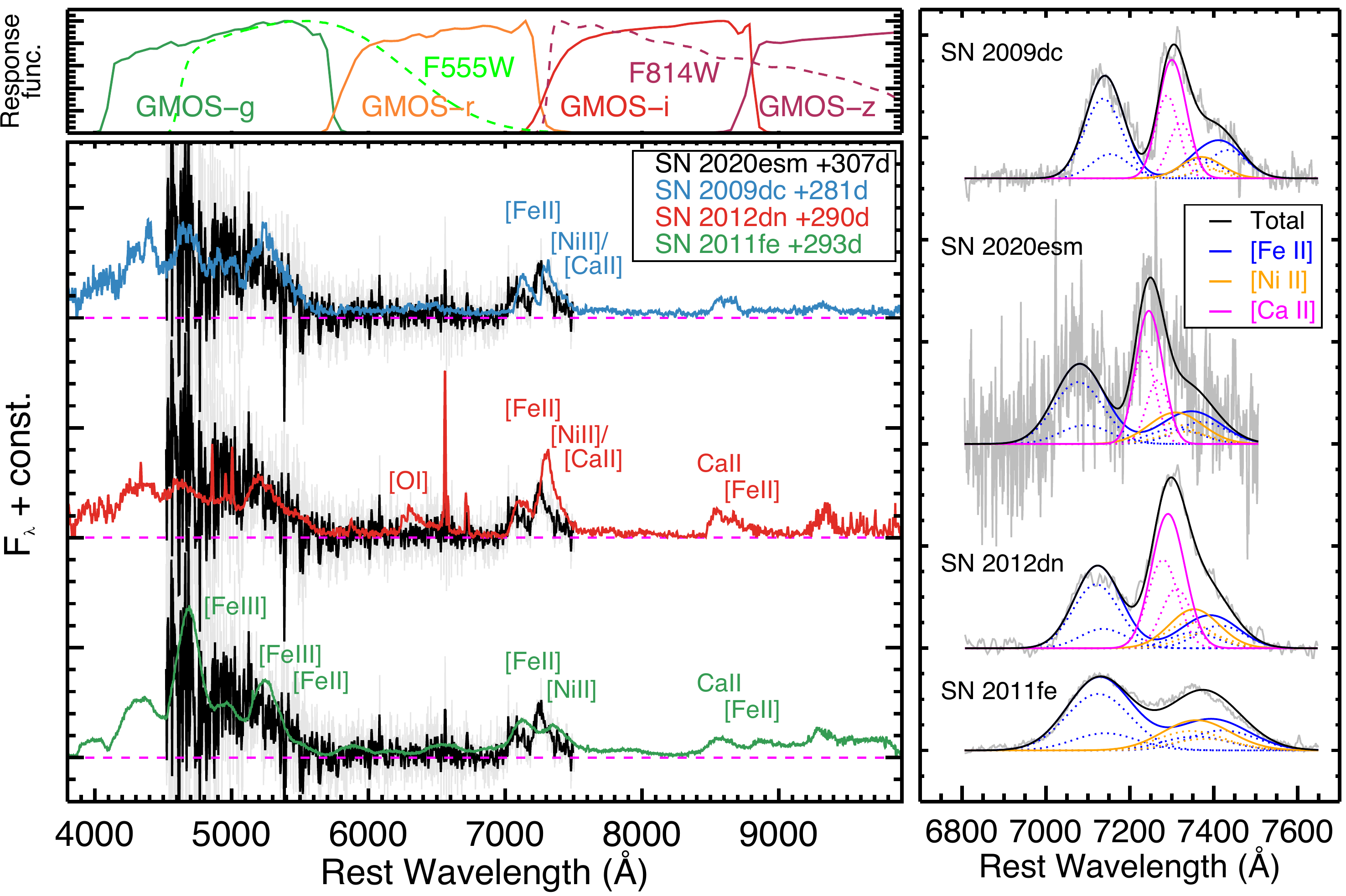}
\caption{The late-time ($\sim$307 days from peak brightness) raw (grey) and binned (black) spectrum of SN\,2020esm is shown in the left panel. Late-time spectra at similar phases are shown for SN\,2009dc (blue), SN\,2021dn (red) and SN\,2011fe (green). Various spectral features of normal and SC SNe~Ia present in nebular epochs, are marked with green and blue/red labels, respectively. In the upper panel, we show the response functions of the GMOS \textit{griz} and HST-WFC3 $F555W$ and $F814W$ filters, with solid and dashed lines, respectively. On the right panel, the $\sim$7,000-7,500 \AA\ line region of each SN is shown, with the model fits overplotted, as in the legend. \label{fig:spec_late}}
\end{center}
\end{figure}

Figure~\ref{fig:spec_late}, left panel, shows our +307 days spectrum of SN\,2020esm (black), compared with nebular spectra of SN\,2009dc (blue), SN\,2012dn (red) and SN\,2011fe (green) at similar phases. While the signal-to-noise of the spectrum, particularly at bluer wavelengths, is not high, common nebular SC SNe~Ia features, such as [\ion{Fe}{3}], [\ion{Fe}{2}] and [\ion{Ni}{2}], can be seen. A defining characteristic of SC SNe~Ia is the relatively low [\ion{Fe}{3}] flux relative to [\ion{Fe}{2}] as probed by the $\lambda$4700/$\lambda$5200 line ratio. The lower ratio of these lines in SC SNe~Ia is attributed to a lower ionization state at these epochs and higher central ejecta densities, compared to normal SNe~Ia \citep{Mazzali2011MNRAS,Taubenberger2013MNRAS}.

The line complex of SC SNe~Ia at $\sim$7,000-7,500 \AA\ differs from normal SNe~Ia, which are dominated by forbidden transitions of [\ion{Fe}{2}] and [\ion{Ni}{2}]. SN\,2020esm is more similar to SN\,2009dc and SN\,2012dn, with two sharp emission peaks in this wavelength range. We attempt to fit the $\sim$7,000-7,500 \AA\ line complex for the nebular spectra of SNe\,2020esm, 2009dc, 2012dn, and 2011fe with a similar approach as in \citealt{Graham2017MNRAS,Maguire2018MNRAS,Siebert2020ApJ}, with the results shown in Figure~\ref{fig:spec_late}, right panel. While for the normal SN\,2011fe, the line complex can be adequately fit with only [\ion{Fe}{2}] and [\ion{Ni}{2}] emission, the SC SNe~Ia require an additional [\ion{Ca}{2}] component. This likely detection of [\ion{Ca}{2}] is consistent with the lower ionization state of SC SNe~Ia, compared to normal SNe~Ia, as [\ion{Ca}{2}] is a very efficient cooling line and dominates the emission features, if calcium exists in the same ejecta region as other Fe-group elements. Moreover, as it can be seen in the upper panel, the \textit{i}-, \textit{z}- and $F814W$-bands probe the line complexes of $\sim$7,000--7,500 and $\sim$8,500--9,000~\AA\, corresponding to regions of the spectrum where [\ion{Ca}{2}] and the calcium NIR triplet emission is expected, respectively, while the \textit{r} band is free of strong emission lines, probing the bolometric decline due to the radioactive decay of $^{56}$Ni. The photometric evolution of this bands at late phases (Section~\ref{subsec:phot_analysis}), with the substantial fading in \textit{gr}- and the constant evolution of \textit{iz}-bands, in conjunction with the +307-day spectrum, confirms this line identification.

Interestingly, SN\,2020esm shows much higher (blue-shifted) iron/nickel and calcium velocities (-3,300 and -2,400~km~s$^{-1}$) compared to SN\,2009dc (-800 and -115~km~s$^{-1}$) and SN\,2012dn (-1,500 and -520~km~s$^{-1}$). This substantial net blueshift of the lines can be interpreted as formation of dust in the ejecta, as the newly formed dust can obscure more emission from the far side of the SN, causing a suppression of the redshifted emission, and an asymmetric line profile \citep{Smith2008ApJ}. However, the line ratio of [\ion{Fe}{2}]/[\ion{Ni}{2}] of SN\,2020esm is similar to the one of the dust-free SN\,2011fe ($3.02\pm0.54$ and $2.91\pm0.02$, respectively), indicating that this blueshift is not an effect of a line asymmetry. This spectroscopic observation, in combination with the late time photometry (see Section~\ref{subsec:phot_analysis}), may rule out the formation of dust. Unlike SN\,2012dn, we do not detect [\ion{O}{1}] $\lambda\lambda$6300,6364 and we place an upper limit of the [\ion{O}{1}] luminosity at $2.5\times10^{36}$~erg~s$^{-1}$. Finally, we also do not detect H$\alpha$, with an upper limit of the luminosity at $1.7\times10^{36}$~erg~s$^{-1}$.

\subsection{Bolometric light curve} \label{subsec:bol_lc}

\begin{figure*}[t]
\begin{center}
\includegraphics[width=0.97\textwidth]{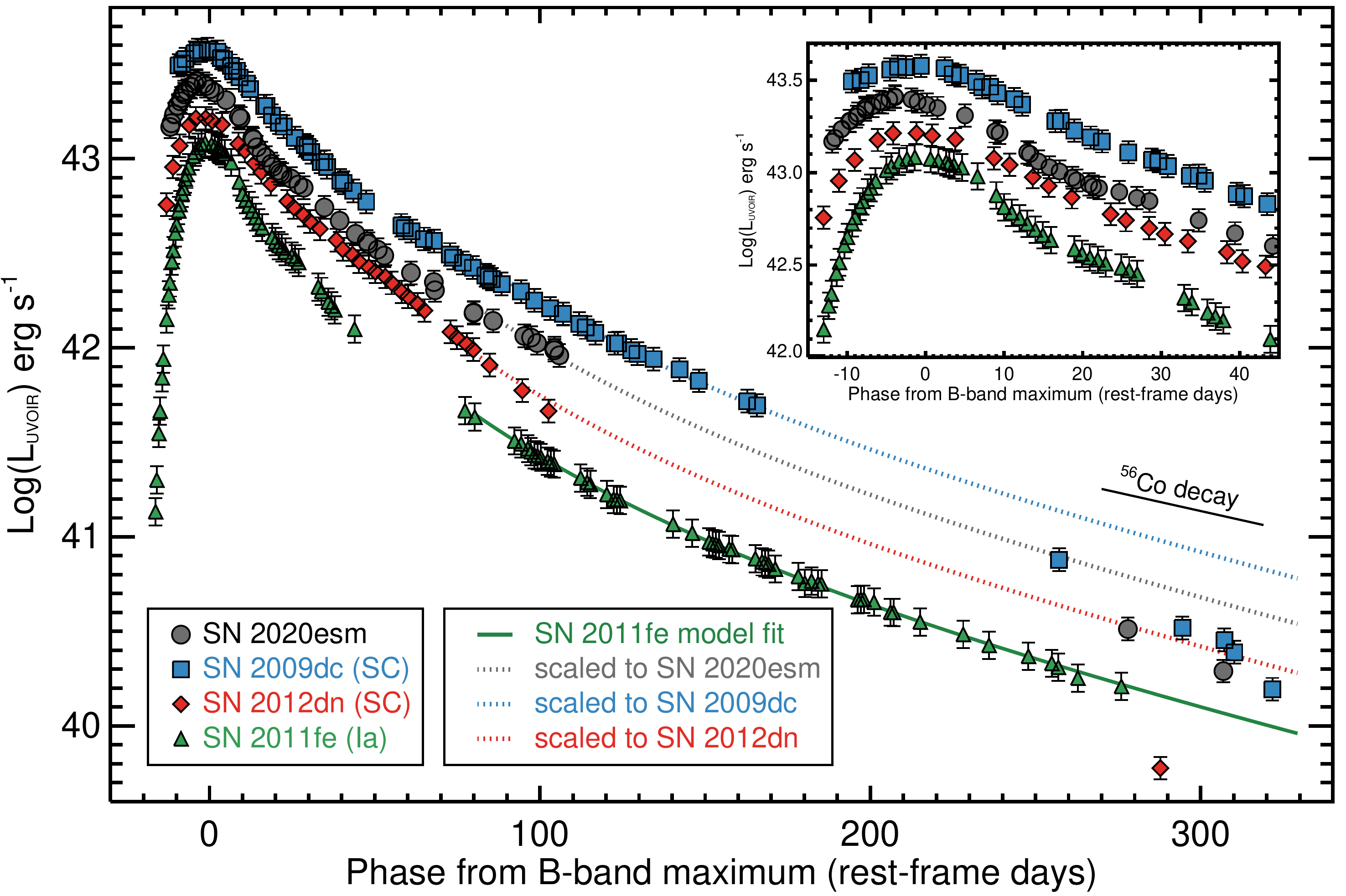}
\caption{The constructed bolometric light curve of SN\,2020esm (grey circles), compared to the bolometric light curves of the normal SN\,2011fe (green upward triangles) and the SC SNe 2009dc (blue squares) and 2012dn (red diamonds), in units of $\mathrm{erg\,s^{-1}}$, with the inset showing the zoomed-in region around peak luminosity. At late times, the solid  green line is a fit to the bolometric light curve of SN\,2011fe, as described in the text, with dotted grey, blue and red lines the fit, scaled to SN\,2020esm, SN\,2009dc and SN\,2012dn, respectively. The slope of the $\mathrm{^{56}Co}$ radioactive decline rate is shown as a solid black line. \label{fig:bol_lc}}
\end{center}
\end{figure*}

We constructed a pseudo-bolometric light curve from our UV optical photometry as follows: After correcting our UV and optical photometric magnitudes for Milky Way extinction, we converted them to monochromatic fluxes. For the case of missing observations in any of the optical bands on a given epoch, an estimation was made with linear interpolation on adjacent epochs, while for the UV light curves after 2020-04-30UT, we assumed zero flux. The final spectral energy distribution was interpolated linearly and integrated with respect to wavelength. We assumed zero flux at the blue end of \textit{UVW2}-band (1,600 Angstroms) and at the red end of \textit{i}-band (10,000 Angstroms). Finally, we used the luminosity distance to the SN to convert the integrated flux to luminosity. For the NIR wavelengths regime (10,000--24,000 Angstroms), we used bolometric corrections derived from SN\,2009dc, which shows similar color evolution in the UV and optical bands (however, see \citet{Ashall2021arXiv} for the variations at early NIR light curves of SC SNe Ia). Adding this NIR luminosity to our UV and optical one, we derive a total pseudo-bolometric UVOIR light curve, as a function of phase from $B$-band maximum, shown in Figure~\ref{fig:bol_lc}, with grey circles. We additionally show the bolometric light curves of SN\,2009dc (blue squares), SN\,2012dn (red diamonds) and SN\,2011fe (green upward triangles), which were constructed using a similar procedure (note that SN\,2009dc, SN\,2012dn and SN\,2011fe have extensive coverage in the NIR, thus the NIR luminosity is directly computed, contrary to SN\,2020esm).

\begin{deluxetable}{ccc}
\tablenum{1}
\tablecaption{The UVOIR bolometric light of SN\,2020esm. \label{table:bol_lc}}
\tablewidth{0.45\textwidth}
\tablehead{
\colhead{Phase\tablenotemark{a}} & \colhead{L} & \colhead{L error}\\
\colhead{(rest-frame days)} & \colhead{($10^{43}\:\mathrm{erg\:s^{-1}}$)} & \colhead{($10^{43}\:\mathrm{erg\:s^{-1}}$)} 
}
\startdata
$-11.90$ &  $1.472$  & $0.203$ \\
$-11.54$ &  $1.541$  & $0.213$ \\
$-10.74$ &  $1.717$  & $0.237$ \\
$-10.65$ &  $1.726$  & $0.238$
\enddata
\tablenotetext{a}{Relative to $B$-band maximum (MJD 58944.58)}
\tablecomments{The complete bolometric light curve is available in the online edition.}
\end{deluxetable}

\begin{figure}[t]
\begin{center}
\includegraphics[width=0.47\textwidth]{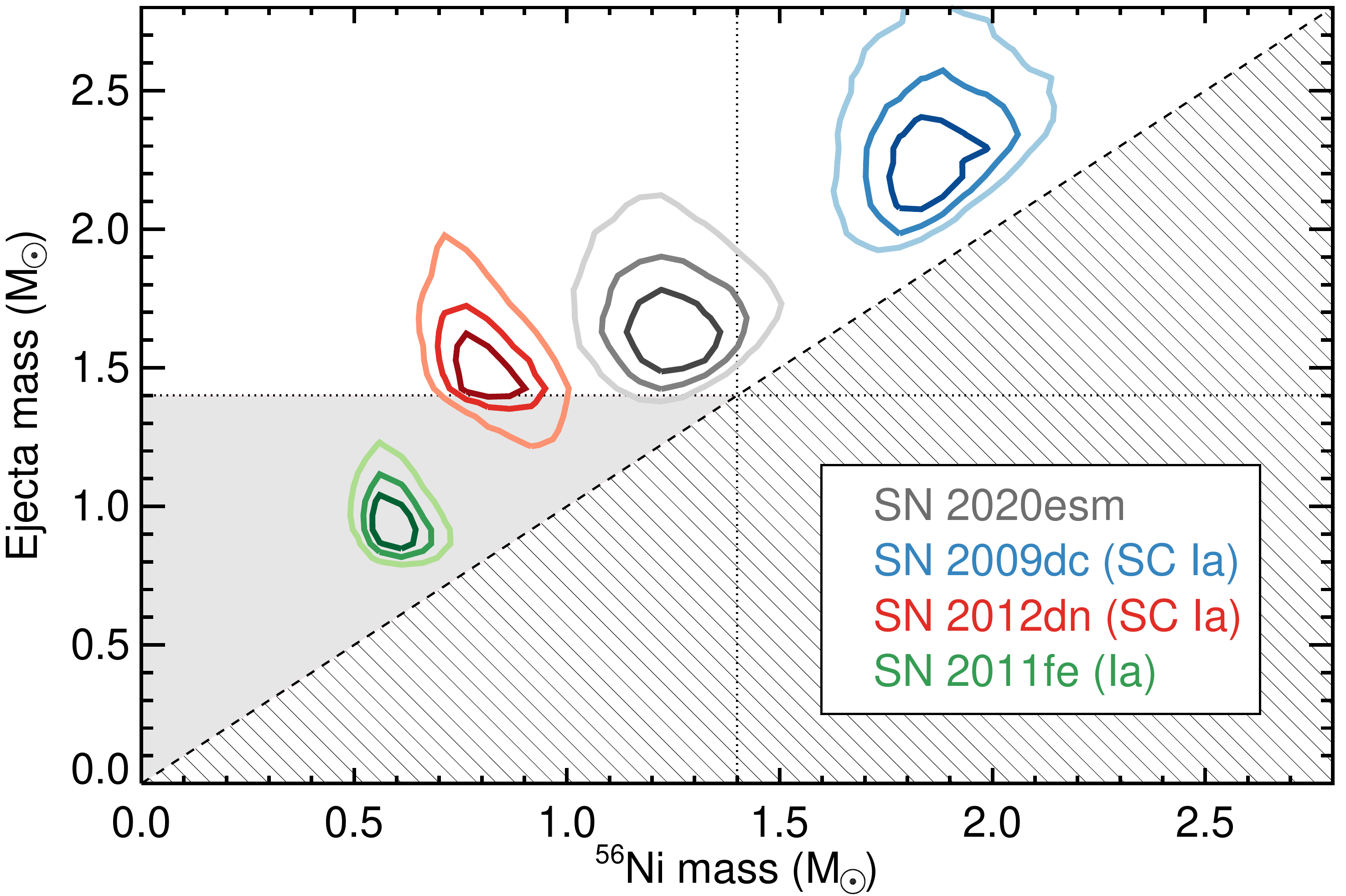}
\caption{Posteriors of the $M_{\rm ^{56}Ni}$ and $M_{\rm ej}$ inferred using \textsc{pyBoloSN} \citep{Scalzo14}. The contours for each SN correspond to the 25, 50 and 75 percentile. Dotted horizontal and vertical lines mark the classical Chandrasekhar mass limit (1.4 $\mathrm{M_{\odot}}$), with the grey-shaded region corresponding to the sub-Chandrasekhar region. The forbidden $M_{\rm ^{56}Ni} > M_{\rm ej}$ region is crossed out. \label{fig:ni_ejecta}}
\end{center}
\end{figure}

In order to estimate the $^{56}$Ni and the ejecta mass of SN\,2020esm, we use the Markov chain Monte Carlo (MCMC) code \textsc{pyBoloSN} \citep{Scalzo14}. The code fits the late-time bolometric light curve, with the $^{56}$Ni and the ejecta mass as free parameters, and the rise time of the bolometric light curve as a prior. In this calculation, we assume that the pseudo-bolometric luminosity of the SN is only due to the radioactive decay of $^{56}$Ni. We performed the same analysis for SN\,2011fe, SN\,2012dn and SN\,2009dc, using rise time priors that are reported in their relevant studies: $t_\mathrm{rise}^{\mathrm{11fe}}=16.6\pm0.15$, $t_\mathrm{rise}^{\mathrm{12dn}}=20.0\pm0.5$ and $t_\mathrm{rise}^{\mathrm{09dc}}=22.0\pm0.5$ days. Due to the uncertain time of explosion of SN\,2020esm, but motivated by its similar light curve evolution with the one of SN\,2009dc, we choose $t_\mathrm{rise}^{\mathrm{20esm}}=22.0\pm2.0$ days. The results are shown in Figure~\ref{fig:ni_ejecta}, with our final estimates reported in Table~\ref{table:ni_ejecta}. Our results place SN\,2020esm comfortably in the SC SNe~Ia regime, particularly in terms of the ejecta mass.

\begin{deluxetable}{lcc}
\tablenum{2}
\tablecaption{Estimates of $\mathrm{^{56}Ni}$ and ejecta mass of SN\,2020esm, SN\,2009dc, SN\,2012dn and SN\,2011fe. \label{table:ni_ejecta}}
\tablewidth{0.45\textwidth}
\tablehead{
\colhead{SN name} & \colhead{M$_{\mathrm{^{56}Ni}}$} & \colhead{$M_{\mathrm{ej}}$}\\
\nocolhead{SN name} & \colhead{(M$_{\sun}$)} & \colhead{(M$_{\sun}$)} 
}
\startdata
SN\,2020esm &  $1.23^{+0.14}_{-0.14}$  & $1.75^{+0.32}_{-0.20}$ \\
SN\,2009dc &  $1.84^{+0.17}_{-0.14}$  & $2.34^{+0.26}_{-0.22}$ \\
SN\,2012dn &  $0.82^{+0.12}_{-0.10}$  & $1.56^{+0.33}_{-0.18}$ \\
SN\,2011fe &  $0.60^{+0.07}_{-0.06}$  & $1.00^{+0.19}_{-0.11}$
\enddata
\tablecomments{Uncertainties are reported as 1-$\sigma$ percentiles.}
\end{deluxetable}

As it can be seen in Figure~\ref{fig:bol_lc}, the decline rate of the bolometric light curve of SN\,2020esm changes dramatically at later times. SN\,2020esm shows a similar decline rate as SN\,2009dc, and is substantially brighter than SN\,2011fe. The decline rate is significantly faster than the $^{56}$Co decay (shown with the solid black line in Figure~\ref{fig:bol_lc}), and is attributed to the increasing escape fraction of the $\gamma$-rays, as the opacity of the ejecta decreases. We do not detect the increasing fading seen in SN\,2012dn after $\sim60$ days from peak which indicates that any of the proposed mechanisms that led to this rapid decline in SN\,2012dn's luminosity (as proposed in \citealt{Taubenberger19}) has not occurred yet. However, at much later phases ($\sim280$ days), we observe a decline in the bolometric luminosity, similar to SN\,2009dc.

We fit the late-time (50--300 days) bolometric light curve of SN\,2011fe with a simple $^{56}$Co radioactive decay model that takes into account the decreasing optical depth to the $\gamma$-rays, while the positrons and X-rays entirely deposit their energy in the ejecta (i.e. we consider the positrons and X-rays fully trapped). Under these assumptions, the bolometric luminosity from the $^{56}$Co decay chain is:

\begin{multline*}
L=\frac{ 2.221\times10^{43}}{56} \frac{\lambda_{56}}{\mathrm{days^{-1}}} \frac{M_{56}}{\mathrm{M_{\odot}}} \frac{q_{56}^{\gamma}f_{56}^{\gamma}(t)+q_{56}^{l}+q_{56}^{X}}{\mathrm{keV}} \\
\times\exp{(-\lambda_{56}t)}\:\mathrm{erg\:s^{-1}}
\end{multline*}

where $M_{56}$ is the initial mass synthesised, $\lambda_{56}$ is the decay constant, $q_{56}^{\gamma}$, $q_{56}^{l}$ and $q_{56}^{X}$ are the average energies per decay carried by $\gamma$-rays, positrons and X-rays and $f_{56}^{\gamma}$ is the fraction of the $\gamma$-rays that contribute to the luminosity, and is given by $f_{56}^{\gamma}(t)=1-\exp{(-(t^{\gamma}_{56}/t)^{2})}$, where $t^{\gamma}_{56}$ corresponds to the time when the optical depth to $\gamma$-rays becomes unity. The values for the decay energies and constants were obtained from the National Nuclear Data Center\footnote{\url{https://www.nndc.bnl.gov/}}. This simple toy model can reasonably well reproduce the late-time emission of normal SNe Ia, as seen for the case of SN\,2011fe (green solid line) up to $\sim$300 days. We find $M_{56}=0.29\pm0.04\:\mathrm{M_{\odot}}$ and $t^{\gamma}_{56}=38.9\pm4.7$ days, in tension with the estimate from the peak luminosity ($M_{56}=0.6\:\mathrm{M_{\odot}}$), a result that has also been confirmed in other studies \citep{Zhang16,Shappee17ApJ,Dimitriadis17MNRAS}, with possible explanations including positron escape, additional flux beyond the UVOIR wavelength regime or an Infrared catastrophe (IRC).

We scale the SN\,2011fe fit to match the  50--100 days bolometric light curves of the SC SNe~Ia, shown in Figure~\ref{fig:bol_lc}) as grey, blue and red dotted lines, for SN\,2020esm, SN\,2009dc and SN\,2012dn, respectively. While the model fairly matches the light curves at these epochs, the steeper decline rate at later times is obvious, notably occurring earlier for SN\,2012dn. The luminosity of SN\,2020esm evolves similarly to SN\,2009dc, with a decrease of flux at $\sim$278 and $\sim$307 days, when we find that SN\,2020esm has 53\% and 44\% of the scaled SN\,2011fe model, while for SN\,2009dc, at $\sim295-320$ days, we find 37\% to 24\% and for SN\,2012dn, at $\sim288$ days, 20\%. Assuming similar conditions in the ejecta of the SNe and applying scaling relations for the nickel masses according to SN\,2011fe, we measure nickel masses from the $^{56}$Co radioactive tail of $M_{56}=0.29\pm0.06\:\mathrm{M_{\odot}}$ for both SN\,2020esm and SN\,2009dc, and $M_{56}=0.08\pm0.02\:\mathrm{M_{\odot}}$ for SN\,2012dn. We defer to Section~\ref{sec:discussion} for further discussion on the physical mechanisms that may lead to this unexpected behaviour.

\section{Discussion} \label{sec:discussion}

In this Section, we discuss our findings in the context of the proposed progenitor models of the SC SNe~Ia class. In particular, the main observables of SN\,2020esm that any viable candidate model needs to address are its early spectra, its early blue UV/optical colors, its high luminosity and slow decline rate, and its enhanced optical fading at later times.

As mentioned above (Section~\ref{sec:intro}), the earliest suggestion for the origin of a SC SN~Ia was the explosion of a single C/O WD, with mass $>\mathrm{M_{Ch}}$, that is able to sustain this mass while rapidly rotating or in the presence of high magnetic fields \citep{Yoon05AA}. This scenario was initially proposed for the first SC SN~Ia 2003fg \citep{Howell06}. However, \citealt{Fink18AA}, using state-of-the-art numerical simulations, have shown that, while this set of models can potentially predict the high luminosity and slow decline rate of (most, but not all) SC SNe~Ia (due to the high nickel and ejecta masses involved), they cannot explain their spectral characteristics, as they fail to reproduce the low ejecta velocities at peak and the detonation that unbinds the WD does not leave any unburnt carbon in the ejecta. This is in contrast with spectral observations of SC SNe~Ia, particularly with SN\,2020esm (Figures~\ref{fig:spec_series} and~\ref{fig:spec_synpp}), where strong absorption features of carbon (and potentially oxygen) are identified at early times.

For about a week after explosion, SN\,2020esm continued to have an atmosphere consisting of principally carbon and oxygen. While some normal SNe~Ia have carbon lines early in their evolution, their spectra are dominated by absorption from intermediate-mass and even iron-group elements only a few hours after explosion, consistent with the explosion of a compact WD \citep{Nugent11, Foley12:09ig}. However, the persistent carbon and oxygen absorption features of SN\,2020esm imply a large mass of unburnt material in the outermost layers of the ejecta. The only way to have as much pristine, unburnt material as SN\,2020esm is an extended envelope of material above the exploding WD. This configuration can be achieved by disrupting a C/O WD during the process of merging with another (more massive) WD \citep{Moll14}.  This process should leave a relatively large fraction of the disrupted WD just above the burnt material, which would be immediately swept up by the ejecta to produce the early \ion{C}{2} and \ion{O}{2} absorption \citep{Raskin14}.

Hydrodynamical models of this merger process also expect a sizable fraction of the disrupted WD to be unbound or barely bound to the system, producing a large amount of CSM \citep{Raskin13}. There may be a delay between the disruption and explosion, and the density and radius of the CSM should scale with the delay. That is, prompt explosions should have dense material close to the star, while delayed explosions could have material at large radii. Some  merger systems, in particular those with unequal masses, are also expected to eject material well before they merge and potentially detonate \citep{Dan2011ApJ,Dan2012MNRAS}.

Another natural delay is the viscous timescale for the accretion disk formed from the disrupted WD, ranging from minutes to days \citep{Raskin13}. The CSM would expand at roughly the escape speed, resulting in material out to as much as $10^{14}$~cm.  The SN shock would quickly interact with this material, resulting in additional UV/X-ray photons and the shock-heated material would cool on the diffusion time, creating additional UV photons \citep{Piro12}.

\begin{figure}[t]
\begin{center}
\includegraphics[width=0.47\textwidth]{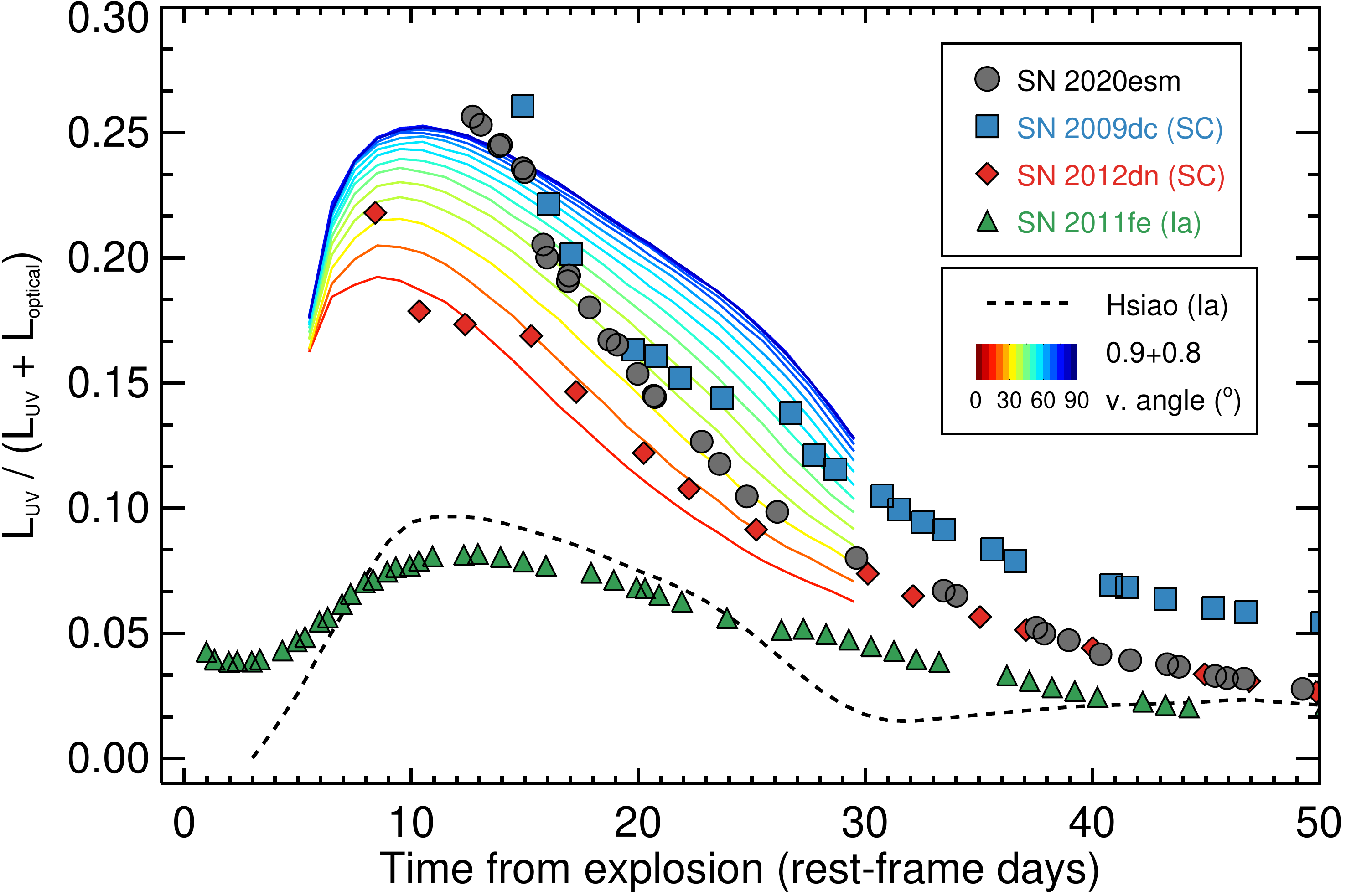}
\caption{The ratio of the UV (1,600--3,000 \AA) to the UV+optical (1,600-10,000 \AA) luminosity of SN\,2020esm (grey circles) compared with SN\,2009dc (blue squares), SN\,2012dn (red diamonds) and SN\,2011fe (green upward triangles). We additionally show, with the black dashed line, the same ratio of the \citealt{Hsiao07} template, representing the normal SNe~Ia, and with the red-to-blue color scheme lines, the merger model of $0.9\mathrm{M_{\odot}}+0.8\mathrm{M_{\odot}}$ WDs from \citet{Raskin14}. The viewing angle dependence is illustrated in the legend. \label{fig:bol_uv_lc}}
\end{center}
\end{figure}

Figure~\ref{fig:bol_uv_lc} shows the UV (1,600--3,000 Angstroms) contribution of SN\,2020esm to the bolometric light curve. SN\,2020esm, along with the SC SNe~Ia 2012dn and 2009dc, emit substantially more in the UV regime at early times, when the UV emission constitutes almost $20-25\%$ of the UV+optical luminosity, compared to the normal SN\,2011fe. We additionally compare with the UV contribution of the $0.9\mathrm{M_{\odot}}+0.8\mathrm{M_{\odot}}$ WDs merger model from \citet{Raskin14}, a violent merger model of two WDs, where the detonation occurs after the secondary WD is disrupted, forming a disk of carbon and oxygen material around the primary WD. The outcome of the simulation is a synthesised spectral series, as a function of the time from explosion and the viewing angle, with 0$^{\circ}$ and 90$^{\circ}$ denoting viewing angles along the pole and along the equator respectively. The viewing angle orientation has strong observational effects in the model spectra, with the UV contribution being higher along the equator, likely due to reduced line blanketing; as the ejecta collide with the disk, $^{56}$Ni is slowed down to low velocities, resulting in the absence of iron group elements above the photosphere and weak line blanketing.

\begin{figure}[t]
\begin{center}
\includegraphics[width=0.47\textwidth]{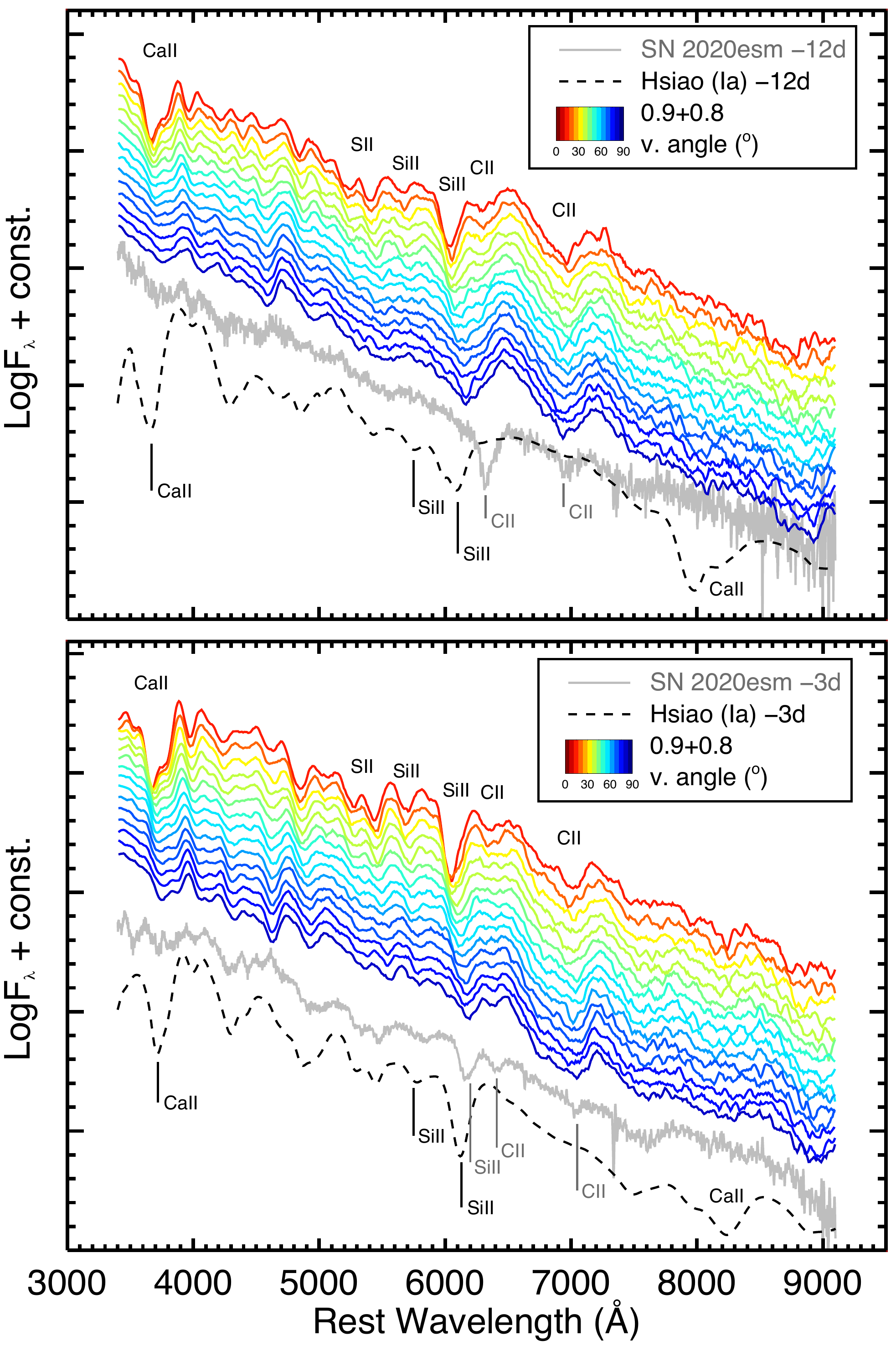}
\caption{Spectra of SN\,2020esm (grey) at early times (top) and peak brightness (bottom), compared with the \citealt{Hsiao07} template and the merger model of $0.9\mathrm{M_{\odot}}+0.8\mathrm{M_{\odot}}$ WDs from \citet{Raskin14}, at similar epochs. The viewing angle dependence is illustrated in the legend. Several spectra features, discussed in the text, are additionally labeled. \label{fig:spec_model_comp}}
\end{center}
\end{figure}

Another strong orientation effect of the \citet{Raskin14} models is the velocity and strength of the unburnt carbon and the IMEs absorption features. For equatorial viewing angles, the surrounding disc decelerates the ejecta, additionally narrowing their velocity range above the photosphere. This can be seen in Figure~\ref{fig:spec_model_comp}, where the model spectra (especially the ones along the equator, for which silicon/carbon is weaker/stronger with lower/higher velocities than the ones along the pole) qualitatively match the spectra of SN 2020esm, showing the strong carbon features (although at larger velocities) and the blue continuum. While the model does not exactly match the observations of SN\,2020esm, it shows distinct observables, such as the weak IMEs, the blue color, and the persistent carbon, that normal SNe~Ia (represented by the \citealt{Hsiao07} template) do not show.

\begin{figure}[t]
\begin{center}
\includegraphics[width=0.47\textwidth]{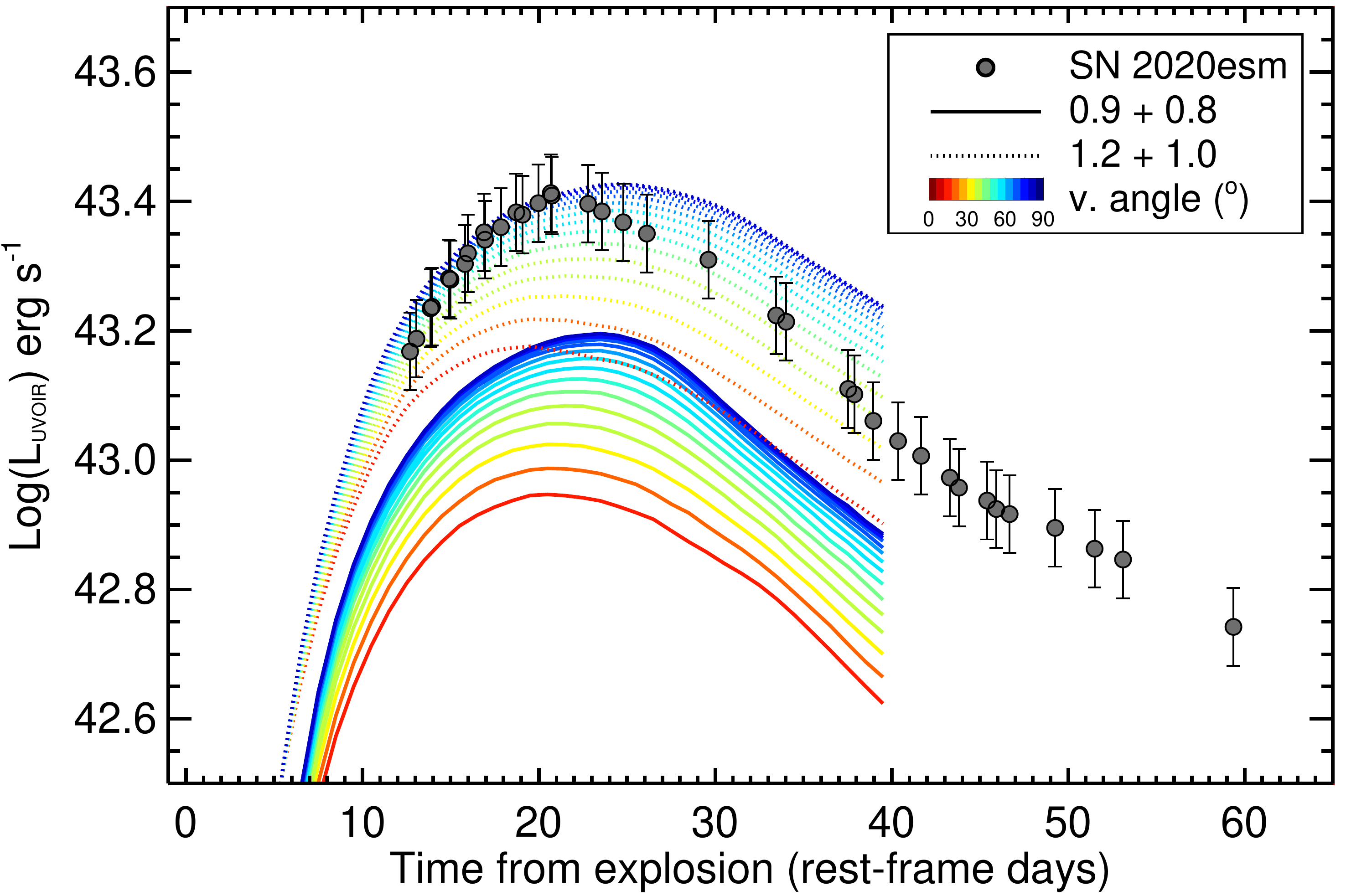}
\caption{The bolometric (1,600-24,000 \AA) bolometric light curve of SN\,2020esm (grey circles) is compared with the $0.9\mathrm{M_{\odot}}+0.8\mathrm{M_{\odot}}$ (solid lines) and $1.2\mathrm{M_{\odot}}+1.0\mathrm{M_{\odot}}$ (dotted lines) WDs merger models from \citet{Raskin14}. The viewing angle dependence is illustrated in the legend. \label{fig:bol_model_comp}}
\end{center}
\end{figure}

While the merger models from \citet{Raskin14} are able to reproduce the general spectroscopic properties of SN\,2020esm, they generally cannot fully reproduce the (bolometric) light curves of SC SNe~Ia. Figure~\ref{fig:bol_model_comp} shows the constructed bolometric light curves of the $0.9\mathrm{M_{\odot}}+0.8\mathrm{M_{\odot}}$ and $1.2\mathrm{M_{\odot}}+1.0\mathrm{M_{\odot}}$ models, compared to SN\,2020esm. This models produce 0.664(1.23) $\mathrm{M_{\odot}}$ of $^{56}$Ni with a total ejecta mass of 1.77(2.20) $\mathrm{M_{\odot}}$, respectively. SN\,2020esm is 1.5--2.7$\times$ brighter at peak, compared to the model at 90$^{\circ}$ (along the equator) and 0$^{\circ}$ (along the pole), respectively. Higher mass models are able to roughly match the peak luminosity (as they produce more $^{56}$Ni), but they predict slower decline rates, as the total ejecta mass increases. However, the \citet{Raskin14} merger models do not include any interaction of the SN ejecta with a carbon/oxygen-rich CSM, which can naturally increase the luminosity of a low-mass merger to the luminosity of SN\,2020esm. Possible origins of the CSM can be material from the disrupted secondary WD \citep{Raskin13} or the (ejected) envelope of an AGB star  \citep{Kashi11MNRAS,Hsiao20ApJ}.

An additional problem with the mergers as paths towards SC SNe~Ia explosions (and in fact for all SNe~Ia) is the asymetric explosions that these models predict \citep{Bulla16MNRAS}. The two SC SNe~Ia that have polarimetric observations, SN\,2007if \citep{Cikota19MNRAS} and SN\,2009dc \citep{Tanaka10ApJ} show low levels of polarization, excluding large assymetries. However, these observations were performed after peak brightness (13, 20, 45 and 46 days and 6 and 90 days after maximum for SN\,2007if and SN\,2009dc, respectively), when the ejecta may have settled in a more spherical geometry. Nevertheless, early-time polarimetry of SC SNe~Ia is crucial to provide an insight to their explosion mechanism.

We note that there is a significant debate about the exact conditions that lead to a detonation during a double C/O WD binary merger and the time delay between the initiation of the merger and the subsequent explosion. In particular, it has been shown \citep{Dan2011ApJ,Dan2012MNRAS} that during detonations at contact, the final merger may take tens to hundreds orbits (with initial periods of hundreds of seconds), during which mass ejection from the system can take place, creating a carbon/oxygen atmosphere and reproducing the observations of SN\,2020esm. These models favor unequal mass systems with high total mass, that are expected to take longer for the dynamical instability to set in and lead to the detonation, while the actual location of the hotspot can alter the synthesised $^{56}$Ni mass. In any case, both the delayed merger models from \citet{Raskin14} and the unequal mass mergers with a detonation triggered close to contact can match the observables of SN\,2020esm, such as the high $^{56}$Ni and ejecta mass, the early carbon/oxygen dominated spectra and the blue early colors.

Finally, another striking observational characteristic of SN\,2020esm is the rapid fading of the optical light curves and the enchanced decline rate of the bolometric light curve at late times (Figure~\ref{fig:bol_lc}), seen in most of SC SNe~Ia. This unique behaviour of SC SNe~Ia has been investigated in previous studies \citep{Silverman2011MNRAS,Taubenberger11,Taubenberger17,Taubenberger19} with no definite conclusions. Initially, it was proposed that this enhanced decline rate was, in fact, a manifestation of the end of the interaction of the ejecta with CSM, a mechanism that could explain the increased peak brightness, since this interaction will result in additional luminosity, that would compensate the normal $^{56}$Ni-powered light curve at early times, and place SC SNe~Ia at the normal SNe~Ia context. However, no indication of hydrogen-rich CSM is evident, due to the absence of narrow hydrogen emission features at early times. On the other hand, hydrogen-free CSM (such as unburnt C/O material, originating from the disrupted secondary WD) is a possible explanation, however, a sustained ejecta-CSM interaction up to $\sim100-200$ days after maximum is not supported by simulations. On a different approach, the enhanced decline rate could be due to a real flux deficit, with possible explanations including a change in the energy deposition rate (such as reduced $\gamma$-ray and/or positron trapping), a redistribution of the emission into longer wavelengths (such as the onset of an early IR catastrophe), or the formation of dust, that results in absorption of the optical light and re-emission of a thermal continuum, determined by the temperature of the dust. In the case of SN\,2020esm, no observations in the mid-to-far IR have been made, thus no conclusive statements regarding the progenitor can be made.

A possible solution to the SC SNe~Ia progenitor problem may be that of thermonuclear explosions inside a dense non-degenerate carbon-rich envelope \citep{Hoeflich96ApJ,Noebauer16MNRAS}, a model suggested in \citet{Hsiao20ApJ} and \citet{Ashall2021arXiv}. This approach can generally explain most of the observables, such as the bright luminosity and slow evolution (as the interaction between the ejecta and the envelope produces a strong reverse shock that converts kinetic to luminous energy) and the low ejecta velocities (as the reverse shock decelerates the ejecta). However, these early models do not explicitly specify the origin of the envelope, while they generally predict very large $^{56}$Ni and envelope masses to explain the slow rise. The authors invoke the ``core-degenerate'' scenario \citep{Kashi11MNRAS}, the explosion of the degenerate C/O core in the center of an AGB star, as a possible progenitor scenario, with additional arguments in favor being the detection of a superwind in observations of LSQ14fmg \citep{Hsiao20ApJ} and the observational correlations presented in \citealt{Ashall2021arXiv} (a more massive envelope would produce brighter explosions, stronger carbon lines and lower Si velocities). However, these models predict significant X-ray luminosity (due to the interaction) and a UV late-time re-brightening (due to interaction with previous superwind episodes of the AGB star), which have not been seen yet. More importantly, the interaction of the ejecta with the AGB’s envelope/wind should produce narrow hydrogen/helium emission lines, an observation that has never been seen in SC SNe~Ia. While this absence may be explained if the AGB star is at the late stage of its evolution, as proposed for LSQ14fmg by \citet{Hsiao20ApJ}, where the dense and non-degenerate envelope does not allow $\gamma$-rays to escape and form narrow emission features, it requires that all SC SNe Ia originate from late-stage AGB stars. At the same time, core-degenerate explosions at earlier stages of the AGB star should be common, but never seen (with a possible exception that of the equally rare subclass of SNe Ia-CSM, \citealt{Silverman2013ApJS}). Nevertheless, this scenario is intriguing, and an exploration of its theoretical and observed rates and complete physical parameters, alongside accurate hydrodynamical, nucleosynthetic and radiative transfer calculations are encouraged.

\section{Conclusion} \label{sec:conclusion}

In this work, we presented extensive UV and optical ground- and space-based observations of the ``super-Chandrasekhar'' SNe Ia candidate 2020esm, spanning from a few days after explosion to its nebular epochs. SN\,2020esm shares all the basic characteristics of the SC SNe~Ia sub-class, such as the high luminosity at peak, the slow evolution of its light curve, the absence of the secondary maximum at the redder photometric bands, the blue early UV and optical colors, the persistent carbon features up to $\sim$ 10 days from maximum, the low ejecta velocities, the weak IMEs absorption lines, the low ionisation in the nebular spectra and the rapid decline of the optical light curves at late-times. Its early spectrum, revealing a nearly-pure carbon/oxygen atmosphere, provides a strong evidence of a WD merger as the progenitor system (ruling out all single WD models). Finally, we discussed theoretical models that have been proposed as viable candidates for these explosions.

We note that SN\,2020esm was initially misclassifed as a core-collapse event, leaving open the possibility of more SC SNe~Ia incorrectly classified in the past as SNe~II or SLSN based only on one spectrum, resulting in contamination of samples of these classes.

As more SC SNe~Ia are discovered, the intrinsic diversity within this sub-class is becoming more prevalent, and the need for high-quality observations, such as early and late-time UV observations, X-ray observations immediately after discovery and high-resolution spectroscopy (especially at the H$\alpha$ and helium wavelength regions) in the optical and NIR, of these events is extremely important, in order to reveal the true nature of these enigmatic events.



\

\textbf{Acknowledgments:}

We thank the anonymous referee for helpful comments that improved the clarity and presentation of this paper.

The UCSC transient team is supported in part by NSF grant AST-1518052, NASA/{\it Swift} grant 80NSSC19K1386, the Gordon \& Betty Moore Foundation, the Heising-Simons Foundation, and by a fellowship from the David and Lucile Packard Foundation to R.J.F.

A major upgrade of the Kast spectrograph on the Shane 3-m telescope at Lick Observatory was made possible through generous gifts from the Heising-Simons Foundation as well as William and Marina Kast. Research at Lick Observatory is partially supported by a generous gift from Google.

Some of the data presented herein were obtained at the W.\ M.\ Keck Observatory, which is operated as a scientific partnership among the California Institute of Technology, the University of California and the National Aeronautics and Space Administration. The Observatory was made possible by the generous financial support of the W.\ M.\ Keck Foundation.  The authors wish to recognize and acknowledge the very significant cultural role and reverence that the summit of Maunakea has always had within the indigenous Hawaiian community.  We are most fortunate to have the opportunity to conduct observations from this mountain.

We acknowledge the use of public data from the {\it Neil Gehrels Swift Observatory} data archive. The work made use of Swift/UVOT data reduced by P. J. Brown and released in the Swift Optical/Ultraviolet Supernova Archive (SOUSA). SOUSA is supported by NASA's Astrophysics Data Analysis Program through grant NNX13AF35G.

The Young Supernova Experiment and its research infrastructure is supported by the European Research Council under the European Union's Horizon 2020 research and innovation programme (ERC Grant Agreement No.\ 101002652, PI K.\ Mandel), the Heising-Simons Foundation (2018-0913, PI R.\ Foley; 2018-0911, PI R.\ Margutti), NASA (NNG17PX03C, PI R.\ Foley), NSF (AST-1720756, AST-1815935, PI R.\ Foley; AST-1909796, AST-1944985, PI R.\ Margutti), the David \& Lucille Packard Foundation (PI R.\ Foley), VILLUM FONDEN (project number 16599, PI J.\ Hjorth), and the Center for AstroPhysical Surveys (CAPS) at the National Center for Supercomputing Applications (NCSA) and the University of Illinois Urbana-Champaign.

The Pan-STARRS1 Surveys (PS1) and the PS1 public science archive have been made possible through contributions by the Institute for Astronomy, the University of Hawaii, the Pan-STARRS Project Office, the Max-Planck Society and its participating institutes, the Max Planck Institute for Astronomy, Heidelberg and the Max Planck Institute for Extraterrestrial Physics, Garching, The Johns Hopkins University, Durham University, the University of Edinburgh, the Queen's University Belfast, the Harvard-Smithsonian Center for Astrophysics, the Las Cumbres Observatory Global Telescope Network Incorporated, the National Central University of Taiwan, the Space Telescope Science Institute, the National Aeronautics and Space Administration under Grant No. NNX08AR22G issued through the Planetary Science Division of the NASA Science Mission Directorate, the National Science Foundation Grant No.\ AST-1238877, the University of Maryland, Eotvos Lorand University (ELTE), the Los Alamos National Laboratory, and the Gordon and Betty Moore Foundation.

Based on observations obtained at the international Gemini Observatory (NOIRLab Prop.\ ID 2020B-0358 and GN-2021A-DD-102), a program of NSF’s NOIRLab, which is managed by the Association of Universities for Research in Astronomy (AURA) under a cooperative agreement with the National Science Foundation on behalf of the Gemini Observatory partnership: the National Science Foundation (United States), National Research Council (Canada), Agencia Nacional de Investigaci\'{o}n y Desarrollo (Chile), Ministerio de Ciencia, Tecnolog\'{i}a e Innovaci\'{o}n (Argentina), Minist\'{e}rio da Ci\^{e}ncia, Tecnologia, Inova\c{c}\~{o}es e Comunica\c{c}\~{o}es (Brazil), and Korea Astronomy and Space Science Institute (Republic of Korea). We thank the director, J.\ Lotz and J.\ Blakeslee for their approval of and assistance with our DD program.

This work makes use of observations from the Las Cumbres Observatory global telescope network (NOIRLab Prop.\ IDs 2020A-0334 and 2020B-0250; PI: R.\ Foley).

This research is based on observations made with the NASA/ESA Hubble Space Telescope obtained from the Space Telescope Science Institute, which is operated by the Association of Universities for Research in Astronomy, Inc., under NASA contract NAS 5–26555. These observations are associated with program SNAP--16239.

This publication has made use of data collected at Lulin Observatory, partly supported by MoST grant 108-2112-M-008-001

D. A. C acknowledges support from the National Science Foundation Graduate Research Fellowship under Grant DGE1339067.

W.J-G is supported by the National Science Foundation Graduate Research Fellowship Program under Grant No.~DGE-1842165 and the IDEAS Fellowship Program at Northwestern University. W.J-G acknowledges support through NASA grants in support of {\it Hubble Space Telescope} program GO-16075.

M.R.S. is supported by the National Science Foundation Graduate Research Fellowship Program Under Grant No. 1842400.

This work was supported by a VILLUM FONDEN Investigator grant to JH (project number 16599) and by a VILLUM FONDEN Young Investigator grant to CG (project number
25501).

Support for D.O.J was provided by NASA through the NASA Hubble Fellowship grant HF2-51462.001 awarded by the Space Telescope Science Institute, which is operated by the Association of Universities for Research in Astronomy, Inc., for NASA, under contract NAS5-26555

Parts of this research were supported by the Australian Research Council Centre of Excellence for All Sky Astrophysics in 3 Dimensions (ASTRO 3D), through project number CE170100013.


%

\vspace{5mm}
\facilities{LCOGT (Sinistro and FLOYDS), PS1, Lulin, Thatcher, Gemini (GMOS), Swift (UVOT), HST (WFC3), Shane (Kast), Keck (LRIS)}


\software{\texttt{\href{https://www.l3harrisgeospatial.com/Software-Technology/IDL}{IDL}} \citep{idl},  \texttt{photpipe} \citep{Rest14}, \texttt{\href{https://iraf-community.github.io/}{iraf/pyraf}} \citep{iraf1,iraf2,iraf3}, \texttt{\href{https://www.cfht.hawaii.edu/~arnouts/LEPHARE/lephare.html}{Le PHARE}} \citep{Arnouts1999MNRAS,Ilbert2006}, \texttt{\href{https://c3.lbl.gov/es/}{SYN++}} \citep{SYNAPPS}, \texttt{\href{https://github.com/rscalzo/pyBoloSN}{pyBoloSN}} \citep{Scalzo14}}



\clearpage
\appendix

\renewcommand\thetable{A\arabic{table}} 
\setcounter{table}{0}

\section{Photometry and Spectroscopy}

\begin{deluxetable}{lcccccc}[h!]
\tablecaption{Observed photometry of SN\,2020esm. \label{table:phot_data}}
\tablehead{
\colhead{MJD} & \colhead{Phase\tablenotemark{a}} & \colhead{Filter} & \colhead{Brightness} & \colhead{Brightness Error} & \colhead{Flux} & \colhead{Flux Error} \\
\nocolhead{MJD} & \colhead{(Rest-frame Days)} & \nocolhead{Filter} & \colhead{(AB mag)} & \colhead{(AB mag)} & \colhead{($10^{-16}\:\mathrm{erg}\:\mathrm{cm}^{-2}\:\mathrm{s}^{-1}\:\mathrm{\AA}^{-1}$)} & \colhead{($10^{-16}\:\mathrm{erg}\:\mathrm{cm}^{-2}\:\mathrm{s}^{-1}\:\mathrm{\AA}^{-1}$)}
}
\startdata
58932.26 & $-11.89$ & B$_{\mathrm{Swift}}$ & 17.056 & 0.098 & 8.7400 & 0.7889 \\
58932.26 & $-11.89$ & UVW2$_{\mathrm{Swift}}$ & 19.126 & 0.070 & 6.0200 & 0.3881 \\
58932.26 & $-11.89$ & U$_{\mathrm{Swift}}$ & 16.967 & 0.055 & 14.9400 & 0.7566 \\
58932.27 & $-11.88$ & UVW1$_{\mathrm{Swift}}$ & 18.053 & 0.062 & 10.3300 & 0.5897 \\
58932.29 & $-11.86$ & V$_{\mathrm{Swift}}$ & 16.860 & 0.159 & 6.6890 & 0.9796 \\
58932.29 & $-11.86$ & UVM2$_{\mathrm{Swift}}$ & 18.471 & 0.090 & 8.9510 & 0.7420 \\
58932.63 & $-11.53$ & g$_{\mathrm{Lulin}}$ & 17.051 & 0.010 & 7.2030 & 0.0663 \\
58932.63 & $-11.53$ & B$_{\mathrm{Lulin}}$ & 17.212 & 0.011 & 7.4280 & 0.0752 \\
58932.63 & $-11.53$ & V$_{\mathrm{Lulin}}$ & 17.049 & 0.009 & 5.4960 & 0.0456 \\
58932.64 & $-11.53$ & i$_{\mathrm{Lulin}}$ & 17.438 & 0.010 & 1.9470 & 0.0179 \\
58932.64 & $-11.53$ & r$_{\mathrm{Lulin}}$ & 17.243 & 0.009 & 3.5580 & 0.0295 \\
58933.32 & $-10.87$ & i$_{\mathrm{LCOGT}}$ & 17.314 & 0.022 & 2.2840 & 0.0463 \\
58933.32 & $-10.87$ & u$_{\mathrm{LCOGT}}$ & 16.786 & 0.024 & 16.9700 & 0.3752 \\
58933.32 & $-10.87$ & g$_{\mathrm{LCOGT}}$ & 16.891 & 0.013 & 8.5270 & 0.1021 \\
58933.32 & $-10.87$ & r$_{\mathrm{LCOGT}}$ & 17.152 & 0.015 & 3.8740 & 0.0535
\enddata
\tablenotetext{a}{Relative to $B$-band maximum (MJD 58944.58)}
\tablecomments{The complete photometric table is available in the online edition.}
\end{deluxetable}

\begin{deluxetable}{lccccc}[h!]
\tablecaption{Observing details of the optical spectra of SN\,2020esm. \label{table:spec_log}}
\tablehead{
\colhead{Obs Date (UT)} & \colhead{Phase\tablenotemark{a}} & \colhead{Telescope + Instrument} & \colhead{Slit Width} & \colhead{Grism/Grating} & \colhead{Exposure time} \\
\nocolhead{Obs Date (UT)} & \colhead{(rest-frame days)} & \nocolhead{Telescope + Instrument} & \nocolhead{Arcseconds} & \nocolhead{Grism/Grating} & \colhead{(s)}
}
\startdata
2020-03-24 &  $-11.8$  & Faulkes North + FLOYDS &  1.6\arcsec\  &  235 l/mm   &   2100 \\
2020-03-26 &  $-9.9$  & Faulkes North + FLOYDS &  1.6\arcsec\  &  235 l/mm   &   1800 \\
2020-03-29 &  $-6.9$  & Faulkes North + FLOYDS &  1.6\arcsec\  &  235 l/mm   &   1500 \\
2020-03-30 &  $-5.6$  & Faulkes South + FLOYDS &  1.6\arcsec\  &  235 l/mm   &   1500 \\
2020-04-02 &  $-3.1$  & Faulkes North + FLOYDS &  1.6\arcsec\  &  235 l/mm   &   1500 \\
2020-04-10 &  4.6  & Faulkes North + FLOYDS &  1.6\arcsec\  &  235 l/mm   &   1500 \\
2020-04-16 &  10.6  & Faulkes North + FLOYDS &  1.6\arcsec\  &  235 l/mm   &   1500 \\
2020-05-23 &  46.0 &  Shane + Kast &  2.0\arcsec\  &   452/3306 + 300/7500  &   1845 (blue), 3$\times$600 (red) \\
2020-06-17 &  70.3 &  Keck + LRIS &  1.0\arcsec\  &   600/4000 + 400/8500  &   610 (blue), 600 (red) \\
2020-07-25 &  106.9 &  Keck + LRIS &  1.0\arcsec\  &   600/4000 + 400/8500  &   1800 (blue), 2$\times$650 (red) \\
2021-02-17 &  306.9  & Gemini North + GMOS &  1.0\arcsec\  &  B600   &  8$\times$900
\enddata
\tablenotetext{a}{Relative to $B$-band maximum (MJD 58944.58)}
\tablecomments{The complete spectroscopic data are available in the online edition.}
\end{deluxetable}


\bibliography{sn2020esm_apj_arxiv}{}

\begin{thebibliography}{}
\expandafter\ifx\csname natexlab\endcsname\relax\def\natexlab#1{#1}\fi
\providecommand{\url}[1]{\href{#1}{#1}}
\providecommand{\dodoi}[1]{doi:~\href{http://doi.org/#1}{\nolinkurl{#1}}}
\providecommand{\doeprint}[1]{\href{http://ascl.net/#1}{\nolinkurl{http://ascl.net/#1}}}
\providecommand{\doarXiv}[1]{\href{https://arxiv.org/abs/#1}{\nolinkurl{https://arxiv.org/abs/#1}}}

\bibitem[{{Arnett}(1982)}]{Arnett82}
{Arnett}, W.~D. 1982, \apj, 253, 785, \dodoi{10.1086/159681}

\bibitem[{{Arnouts} {et~al.}(1999){Arnouts}, {Cristiani}, {Moscardini},
  {Matarrese}, {Lucchin}, {Fontana}, \& {Giallongo}}]{Arnouts1999MNRAS}
{Arnouts}, S., {Cristiani}, S., {Moscardini}, L., {et~al.} 1999, \mnras, 310,
  540, \dodoi{10.1046/j.1365-8711.1999.02978.x}

\bibitem[{{Ashall} {et~al.}(2021){Ashall}, {Lu}, {Hsiao}, {Hoeflich},
  {Phillips}, {Galbany}, {Burns}, {Contreras}, {Krisciunas}, {Morrell},
  {Stritzinger}, {Suntzeff}, {Taddia}, {Anais}, {Baron}, {Brown}, {Busta},
  {Campillay}, {Castell{\'o}n}, {Corco}, {Davis}, {Folatelli}, {Forster},
  {Freedman}, {Gonzal{\'e}z}, {Hamuy}, {Holmbo}, {Kirshner}, {Kumar}, {Marion},
  {Mazzali}, {Morokuma}, {Nugent}, {Persson}, {Piro}, {Roth}, {Salgado},
  {Sand}, {Seron}, {Shahbandeh}, \& {Shappee}}]{Ashall2021arXiv}
{Ashall}, C., {Lu}, J., {Hsiao}, E.~Y., {et~al.} 2021, arXiv e-prints,
  arXiv:2106.12140.
\newblock \doarXiv{2106.12140}

\bibitem[{{Bloom} {et~al.}(2012){Bloom}, {Kasen}, {Shen}, {Nugent}, {Butler},
  {Graham}, {Howell}, {Kolb}, {Holmes}, {Haswell}, {Burwitz}, {Rodriguez}, \&
  {Sullivan}}]{Bloom12}
{Bloom}, J.~S., {Kasen}, D., {Shen}, K.~J., {et~al.} 2012, \apjl, 744, L17,
  \dodoi{10.1088/2041-8205/744/2/L17}

\bibitem[{{Bradley} {et~al.}(2020){Bradley}, {Sip{\H{o}}cz}, {Robitaille},
  {Tollerud}, {Vin{\'\i}cius}, {Deil}, {Barbary}, {Wilson}, {Busko},
  {G{\"u}nther}, {Cara}, {Conseil}, {Bostroem}, {Droettboom}, {Bray}, {Andersen
  Bratholm}, {Lim}, {Barentsen}, {Craig}, {Pascual}, {Perren}, {Greco},
  {Donath}, {De Val-Borro}, {Kerzendorf}, {Bach}, {Weaver}, {D'Eugenio},
  {Souchereau}, \& {Ferreira}}]{Bradley2020}
{Bradley}, L., {Sip{\H{o}}cz}, B., {Robitaille}, T., {et~al.} 2020,
  {astropy/photutils: 1.0.0}, 1.0.0,  Zenodo, \dodoi{10.5281/zenodo.4044744}

\bibitem[{{Brimacombe} {et~al.}(2020){Brimacombe}, {Krannich}, {Cacella},
  {Wiethoff}, {Vallely}, {Stanek}, {Kochanek}, {Basinger}, {Way}, {Bose},
  {Thompson}, {Shappee}, {Holoien}, {Prieto}, {Bersier}, {Dong}, {Chen},
  {Stritzinger}, {Holmbo}, {Bock}, \& {Masi}}]{Brimacombe20ATel}
{Brimacombe}, J., {Krannich}, G., {Cacella}, P., {et~al.} 2020, The
  Astronomer's Telegram, 13666, 1

\bibitem[{{Brown} {et~al.}(2014){Brown}, {Breeveld}, {Holland}, {Kuin}, \&
  {Pritchard}}]{Brown2014}
{Brown}, P.~J., {Breeveld}, A.~A., {Holland}, S., {Kuin}, P., \& {Pritchard},
  T. 2014, \apss, 354, 89, \dodoi{10.1007/s10509-014-2059-8}

\bibitem[{{Brown} {et~al.}(2009){Brown}, {Holland}, {Immler}, {Milne},
  {Roming}, {Gehrels}, {Nousek}, {Panagia}, {Still}, \& {Vanden
  Berk}}]{Brown09}
{Brown}, P.~J., {Holland}, S.~T., {Immler}, S., {et~al.} 2009, \aj, 137, 4517,
  \dodoi{10.1088/0004-6256/137/5/4517}

\bibitem[{{Brown} {et~al.}(2012){Brown}, {Dawson}, {de Pasquale}, {Gronwall},
  {Holland}, {Immler}, {Kuin}, {Mazzali}, {Milne}, {Oates}, \&
  {Siegel}}]{Brown2012ApJ}
{Brown}, P.~J., {Dawson}, K.~S., {de Pasquale}, M., {et~al.} 2012, \apj, 753,
  22, \dodoi{10.1088/0004-637X/753/1/22}

\bibitem[{{Brown} {et~al.}(2013){Brown}, {Baliber}, {Bianco}, {Bowman},
  {Burleson}, {Conway}, {Crellin}, {Depagne}, {De Vera}, {Dilday}, {Dragomir},
  {Dubberley}, {Eastman}, {Elphick}, {Falarski}, {Foale}, {Ford}, {Fulton},
  {Garza}, {Gomez}, {Graham}, {Greene}, {Haldeman}, {Hawkins}, {Haworth},
  {Haynes}, {Hidas}, {Hjelstrom}, {Howell}, {Hygelund}, {Lister}, {Lobdill},
  {Martinez}, {Mullins}, {Norbury}, {Parrent}, {Paulson}, {Petry}, {Pickles},
  {Posner}, {Rosing}, {Ross}, {Sand}, {Saunders}, {Shobbrook}, {Shporer},
  {Street}, {Thomas}, {Tsapras}, {Tufts}, {Valenti}, {Vander Horst}, {Walker},
  {White}, \& {Willis}}]{LCOGT13PASP}
{Brown}, T.~M., {Baliber}, N., {Bianco}, F.~B., {et~al.} 2013, \pasp, 125,
  1031, \dodoi{10.1086/673168}

\bibitem[{{Bruzual} \& {Charlot}(2003)}]{Bruzual2003MNRAS}
{Bruzual}, G., \& {Charlot}, S. 2003, \mnras, 344, 1000,
  \dodoi{10.1046/j.1365-8711.2003.06897.x}

\bibitem[{{Bulla} {et~al.}(2016){Bulla}, {Sim}, {Pakmor}, {Kromer},
  {Taubenberger}, {R{\"o}pke}, {Hillebrandt}, \& {Seitenzahl}}]{Bulla16MNRAS}
{Bulla}, M., {Sim}, S.~A., {Pakmor}, R., {et~al.} 2016, \mnras, 455, 1060,
  \dodoi{10.1093/mnras/stv2402}

\bibitem[{{Calzetti} {et~al.}(2000){Calzetti}, {Armus}, {Bohlin}, {Kinney},
  {Koornneef}, \& {Storchi-Bergmann}}]{Calzetti2000ApJ}
{Calzetti}, D., {Armus}, L., {Bohlin}, R.~C., {et~al.} 2000, \apj, 533, 682,
  \dodoi{10.1086/308692}

\bibitem[{{Chabrier}(2003)}]{Chabrier2003PASP}
{Chabrier}, G. 2003, \pasp, 115, 763, \dodoi{10.1086/376392}

\bibitem[{{Chakradhari} {et~al.}(2014){Chakradhari}, {Sahu}, {Srivastav}, \&
  {Anupama}}]{Chakradhari14MNRAS}
{Chakradhari}, N.~K., {Sahu}, D.~K., {Srivastav}, S., \& {Anupama}, G.~C. 2014,
  \mnras, 443, 1663, \dodoi{10.1093/mnras/stu1258}

\bibitem[{{Chambers} {et~al.}(2016){Chambers}, {Magnier}, {Metcalfe},
  {Flewelling}, {Huber}, {Waters}, {Denneau}, {Draper}, {Farrow}, {Finkbeiner},
  {Holmberg}, {Koppenhoefer}, {Price}, {Rest}, {Saglia}, {Schlafly}, {Smartt},
  {Sweeney}, {Wainscoat}, {Burgett}, {Chastel}, {Grav}, {Heasley}, {Hodapp},
  {Jedicke}, {Kaiser}, {Kudritzki}, {Luppino}, {Lupton}, {Monet}, {Morgan},
  {Onaka}, {Shiao}, {Stubbs}, {Tonry}, {White}, {Ba{\~n}ados}, {Bell},
  {Bender}, {Bernard}, {Boegner}, {Boffi}, {Botticella}, {Calamida},
  {Casertano}, {Chen}, {Chen}, {Cole}, {Deacon}, {Frenk}, {Fitzsimmons},
  {Gezari}, {Gibbs}, {Goessl}, {Goggia}, {Gourgue}, {Goldman}, {Grant},
  {Grebel}, {Hambly}, {Hasinger}, {Heavens}, {Heckman}, {Henderson}, {Henning},
  {Holman}, {Hopp}, {Ip}, {Isani}, {Jackson}, {Keyes}, {Koekemoer}, {Kotak},
  {Le}, {Liska}, {Long}, {Lucey}, {Liu}, {Martin}, {Masci}, {McLean}, {Mindel},
  {Misra}, {Morganson}, {Murphy}, {Obaika}, {Narayan}, {Nieto-Santisteban},
  {Norberg}, {Peacock}, {Pier}, {Postman}, {Primak}, {Rae}, {Rai}, {Riess},
  {Riffeser}, {Rix}, {R{\"o}ser}, {Russel}, {Rutz}, {Schilbach}, {Schultz},
  {Scolnic}, {Strolger}, {Szalay}, {Seitz}, {Small}, {Smith}, {Soderblom},
  {Taylor}, {Thomson}, {Taylor}, {Thakar}, {Thiel}, {Thilker}, {Unger},
  {Urata}, {Valenti}, {Wagner}, {Walder}, {Walter}, {Watters}, {Werner},
  {Wood-Vasey}, \& {Wyse}}]{PS1}
{Chambers}, K.~C., {Magnier}, E.~A., {Metcalfe}, N., {et~al.} 2016, arXiv
  e-prints, arXiv:1612.05560.
\newblock \doarXiv{1612.05560}

\bibitem[{{Chandrasekhar}(1931)}]{Chandrasekhar31}
{Chandrasekhar}, S. 1931, \apj, 74, 81

\bibitem[{{Chen} {et~al.}(2019){Chen}, {Dong}, {Katz}, {Kochanek}, {Kollmeier},
  {Maguire}, {Phillips}, {Prieto}, {Shappee}, {Stritzinger}, {Bose}, {Brown},
  {Holoien}, {Galbany}, {Milne}, {Morrell}, {Piro}, {Stanek}, {Thompson}, \&
  {Young}}]{Chen2019ApJ}
{Chen}, P., {Dong}, S., {Katz}, B., {et~al.} 2019, \apj, 880, 35,
  \dodoi{10.3847/1538-4357/ab2630}

\bibitem[{{Cikota} {et~al.}(2019){Cikota}, {Patat}, {Wang}, {Wheeler}, {Bulla},
  {Baade}, {H{\"o}flich}, {Cikota}, {Clocchiatti}, {Maund}, {Stevance}, \&
  {Yang}}]{Cikota19MNRAS}
{Cikota}, A., {Patat}, F., {Wang}, L., {et~al.} 2019, \mnras, 490, 578,
  \dodoi{10.1093/mnras/stz2322}

\bibitem[{{Dan} {et~al.}(2011){Dan}, {Rosswog}, {Guillochon}, \&
  {Ramirez-Ruiz}}]{Dan2011ApJ}
{Dan}, M., {Rosswog}, S., {Guillochon}, J., \& {Ramirez-Ruiz}, E. 2011, \apj,
  737, 89, \dodoi{10.1088/0004-637X/737/2/89}

\bibitem[{{Dan} {et~al.}(2012){Dan}, {Rosswog}, {Guillochon}, \&
  {Ramirez-Ruiz}}]{Dan2012MNRAS}
---. 2012, \mnras, 422, 2417, \dodoi{10.1111/j.1365-2966.2012.20794.x}

\bibitem[{{Dimitriadis} {et~al.}(2017){Dimitriadis}, {Sullivan}, {Kerzendorf},
  {Ruiter}, {Seitenzahl}, {Taubenberger}, {Doran}, {Gal-Yam}, {Laher},
  {Maguire}, {Nugent}, {Ofek}, \& {Surace}}]{Dimitriadis17MNRAS}
{Dimitriadis}, G., {Sullivan}, M., {Kerzendorf}, W., {et~al.} 2017, \mnras,
  468, 3798, \dodoi{10.1093/mnras/stx683}

\bibitem[{{Exelis Visual Information Solutions}(2010)}]{idl}
{Exelis Visual Information Solutions}. 2010, {IDL}

\bibitem[{{Filippenko}(1997)}]{Filippenko97}
{Filippenko}, A.~V. 1997, \araa, 35, 309

\bibitem[{{Fink} {et~al.}(2018){Fink}, {Kromer}, {Hillebrandt}, {R{\"o}pke},
  {Pakmor}, {Seitenzahl}, \& {Sim}}]{Fink18AA}
{Fink}, M., {Kromer}, M., {Hillebrandt}, W., {et~al.} 2018, \aap, 618, A124,
  \dodoi{10.1051/0004-6361/201833475}

\bibitem[{{Fitzpatrick}(1999)}]{Fitzpatrick99}
{Fitzpatrick}, E.~L. 1999, \pasp, 111, 63, \dodoi{10.1086/316293}

\bibitem[{{Foley} {et~al.}(2012){Foley}, {Challis}, {Filippenko},
  {Ganeshalingam}, {Landsman}, {Li}, {Marion}, {Silverman}, {Beaton},
  {Bennert}, {Cenko}, {Childress}, {Guhathakurta}, {Jiang}, {Kalirai},
  {Kirshner}, {Stockton}, {Tollerud}, {Vink{\'o}}, {Wheeler}, \&
  {Woo}}]{Foley12:09ig}
{Foley}, R.~J., {Challis}, P.~J., {Filippenko}, A.~V., {et~al.} 2012, \apj,
  744, 38, \dodoi{10.1088/0004-637X/744/1/38}

\bibitem[{{Freedman}(2021)}]{Freedman21ApJ}
{Freedman}, W.~L. 2021, \apj, 919, 16, \dodoi{10.3847/1538-4357/ac0e95}

\bibitem[{{Freedman} {et~al.}(2019){Freedman}, {Madore}, {Hatt}, {Hoyt},
  {Jang}, {Beaton}, {Burns}, {Lee}, {Monson}, {Neeley}, {Phillips}, {Rich}, \&
  {Seibert}}]{Freedman2019ApJ}
{Freedman}, W.~L., {Madore}, B.~F., {Hatt}, D., {et~al.} 2019, \apj, 882, 34,
  \dodoi{10.3847/1538-4357/ab2f73}

\bibitem[{{Friedman} {et~al.}(2015){Friedman}, {Wood-Vasey}, {Marion},
  {Challis}, {Mandel}, {Bloom}, {Modjaz}, {Narayan}, {Hicken}, {Foley},
  {Klein}, {Starr}, {Morgan}, {Rest}, {Blake}, {Miller}, {Falco}, {Wyatt},
  {Mink}, {Skrutskie}, \& {Kirshner}}]{Friedman2015ApJS}
{Friedman}, A.~S., {Wood-Vasey}, W.~M., {Marion}, G.~H., {et~al.} 2015, \apjs,
  220, 9, \dodoi{10.1088/0067-0049/220/1/9}

\bibitem[{{Gehrels} {et~al.}(2004){Gehrels}, {Chincarini}, {Giommi}, {Mason},
  {Nousek}, {Wells}, {White}, {Barthelmy}, {Burrows}, {Cominsky}, {Hurley},
  {Marshall}, {M{\'e}sz{\'a}ros}, {Roming}, {Angelini}, {Barbier}, {Belloni},
  {Campana}, {Caraveo}, {Chester}, {Citterio}, {Cline}, {Cropper}, {Cummings},
  {Dean}, {Feigelson}, {Fenimore}, {Frail}, {Fruchter}, {Garmire}, {Gendreau},
  {Ghisellini}, {Greiner}, {Hill}, {Hunsberger}, {Krimm}, {Kulkarni}, {Kumar},
  {Lebrun}, {Lloyd-Ronning}, {Markwardt}, {Mattson}, {Mushotzky}, {Norris},
  {Osborne}, {Paczynski}, {Palmer}, {Park}, {Parsons}, {Paul}, {Rees},
  {Reynolds}, {Rhoads}, {Sasseen}, {Schaefer}, {Short}, {Smale}, {Smith},
  {Stella}, {Tagliaferri}, {Takahashi}, {Tashiro}, {Townsley}, {Tueller},
  {Turner}, {Vietri}, {Voges}, {Ward}, {Willingale}, {Zerbi}, \&
  {Zhang}}]{Gehrels04}
{Gehrels}, N., {Chincarini}, G., {Giommi}, P., {et~al.} 2004, \apj, 611, 1005,
  \dodoi{10.1086/422091}

\bibitem[{{Goldstein} \& {Kasen}(2018)}]{Goldstein18ApJ}
{Goldstein}, D.~A., \& {Kasen}, D. 2018, \apjl, 852, L33,
  \dodoi{10.3847/2041-8213/aaa409}

\bibitem[{{Graham} {et~al.}(2017){Graham}, {Kumar}, {Hosseinzadeh},
  {Hiramatsu}, {Arcavi}, {Howell}, {Valenti}, {Sand}, {Parrent}, {McCully}, \&
  {Filippenko}}]{Graham2017MNRAS}
{Graham}, M.~L., {Kumar}, S., {Hosseinzadeh}, G., {et~al.} 2017, \mnras, 472,
  3437, \dodoi{10.1093/mnras/stx2224}

\bibitem[{{Hachinger} {et~al.}(2012){Hachinger}, {Mazzali}, {Taubenberger},
  {Fink}, {Pakmor}, {Hillebrandt}, \& {Seitenzahl}}]{Hachinger12MNRAS}
{Hachinger}, S., {Mazzali}, P.~A., {Taubenberger}, S., {et~al.} 2012, \mnras,
  427, 2057, \dodoi{10.1111/j.1365-2966.2012.22068.x}

\bibitem[{{Hicken} {et~al.}(2007){Hicken}, {Garnavich}, {Prieto}, {Blondin},
  {DePoy}, {Kirshner}, \& {Parrent}}]{Hicken07ApJ}
{Hicken}, M., {Garnavich}, P.~M., {Prieto}, J.~L., {et~al.} 2007, \apjl, 669,
  L17, \dodoi{10.1086/523301}

\bibitem[{{Hicken} {et~al.}(2009){Hicken}, {Challis}, {Jha}, {Kirshner},
  {Matheson}, {Modjaz}, {Rest}, {Wood-Vasey}, {Bakos}, {Barton}, {Berlind},
  {Bragg}, {Brice{\~n}o}, {Brown}, {Caldwell}, {Calkins}, {Cho}, {Ciupik},
  {Contreras}, {Dendy}, {Dosaj}, {Durham}, {Eriksen}, {Esquerdo}, {Everett},
  {Falco}, {Fernandez}, {Gaba}, {Garnavich}, {Graves}, {Green}, {Groner},
  {Hergenrother}, {Holman}, {Hradecky}, {Huchra}, {Hutchison}, {Jerius},
  {Jordan}, {Kilgard}, {Krauss}, {Luhman}, {Macri}, {Marrone}, {McDowell},
  {McIntosh}, {McNamara}, {Megeath}, {Mochejska}, {Munoz}, {Muzerolle},
  {Naranjo}, {Narayan}, {Pahre}, {Peters}, {Peterson}, {Rines}, {Ripman},
  {Roussanova}, {Schild}, {Sicilia-Aguilar}, {Sokoloski}, {Smalley}, {Smith},
  {Spahr}, {Stanek}, {Barmby}, {Blondin}, {Stubbs}, {Szentgyorgyi}, {Torres},
  {Vaz}, {Vikhlinin}, {Wang}, {Westover}, {Woods}, \& {Zhao}}]{Hicken09ApJ}
{Hicken}, M., {Challis}, P., {Jha}, S., {et~al.} 2009, \apj, 700, 331,
  \dodoi{10.1088/0004-637X/700/1/331}

\bibitem[{{Hoeflich} \& {Khokhlov}(1996)}]{Hoeflich96ApJ}
{Hoeflich}, P., \& {Khokhlov}, A. 1996, \apj, 457, 500, \dodoi{10.1086/176748}

\bibitem[{{Hook} {et~al.}(2004){Hook}, {J{\o}rgensen}, {Allington-Smith},
  {Davies}, {Metcalfe}, {Murowinski}, \& {Crampton}}]{GMOS}
{Hook}, I.~M., {J{\o}rgensen}, I., {Allington-Smith}, J.~R., {et~al.} 2004,
  \pasp, 116, 425, \dodoi{10.1086/383624}

\bibitem[{{Howell} {et~al.}(2006){Howell}, {Sullivan}, {Nugent}, {Ellis},
  {Conley}, {Le Borgne}, {Carlberg}, {Guy}, {Balam}, {Basa}, {Fouchez}, {Hook},
  {Hsiao}, {Neill}, {Pain}, {Perrett}, \& {Pritchet}}]{Howell06}
{Howell}, D.~A., {Sullivan}, M., {Nugent}, P.~E., {et~al.} 2006, \nat, 443,
  308, \dodoi{10.1038/nature05103}

\bibitem[{{Hsiao} {et~al.}(2007){Hsiao}, {Conley}, {Howell}, {Sullivan},
  {Pritchet}, {Carlberg}, {Nugent}, \& {Phillips}}]{Hsiao07}
{Hsiao}, E.~Y., {Conley}, A., {Howell}, D.~A., {et~al.} 2007, \apj, 663, 1187,
  \dodoi{10.1086/518232}

\bibitem[{{Hsiao} {et~al.}(2020){Hsiao}, {Hoeflich}, {Ashall}, {Lu},
  {Contreras}, {Burns}, {Phillips}, {Galbany}, {Anderson}, {Baltay}, {Baron},
  {Castell{\'o}n}, {Davis}, {Freedman}, {Gall}, {Gonzalez}, {Graham}, {Hamuy},
  {Holoien}, {Karamehmetoglu}, {Krisciunas}, {Kumar}, {Kuncarayakti},
  {Morrell}, {Moriya}, {Nugent}, {Perlmutter}, {Persson}, {Piro}, {Rabinowitz},
  {Roth}, {Shahbandeh}, {Shappee}, {Stritzinger}, {Suntzeff}, {Taddia}, \&
  {Uddin}}]{Hsiao20ApJ}
{Hsiao}, E.~Y., {Hoeflich}, P., {Ashall}, C., {et~al.} 2020, \apj, 900, 140,
  \dodoi{10.3847/1538-4357/abaf4c}

\bibitem[{{Iben} \& {Tutukov}(1984)}]{Iben84}
{Iben}, I., J., \& {Tutukov}, A.~V. 1984, \apjs, 54, 335,
  \dodoi{10.1086/190932}

\bibitem[{{Ilbert} {et~al.}(2006){Ilbert}, {Arnouts}, {McCracken},
  {Bolzonella}, {Bertin}, {Le F{\`e}vre}, {Mellier}, {Zamorani}, {Pell{\`o}},
  {Iovino}, {Tresse}, {Le Brun}, {Bottini}, {Garilli}, {Maccagni}, {Picat},
  {Scaramella}, {Scodeggio}, {Vettolani}, {Zanichelli}, {Adami}, {Bardelli},
  {Cappi}, {Charlot}, {Ciliegi}, {Contini}, {Cucciati}, {Foucaud}, {Franzetti},
  {Gavignaud}, {Guzzo}, {Marano}, {Marinoni}, {Mazure}, {Meneux}, {Merighi},
  {Paltani}, {Pollo}, {Pozzetti}, {Radovich}, {Zucca}, {Bondi}, {Bongiorno},
  {Busarello}, {de La Torre}, {Gregorini}, {Lamareille}, {Mathez}, {Merluzzi},
  {Ripepi}, {Rizzo}, \& {Vergani}}]{Ilbert2006}
{Ilbert}, O., {Arnouts}, S., {McCracken}, H.~J., {et~al.} 2006, \aap, 457, 841,
  \dodoi{10.1051/0004-6361:20065138}

\bibitem[{{Jeffery}(1999)}]{Jeffery1999}
{Jeffery}, D.~J. 1999, arXiv e-prints, astro.
\newblock \doarXiv{astro-ph/9907015}

\bibitem[{{Jerkstrand} {et~al.}(2020){Jerkstrand}, {Maeda}, \&
  {Kawabata}}]{Jerkstrand2020Sci}
{Jerkstrand}, A., {Maeda}, K., \& {Kawabata}, K.~S. 2020, Science, 367, 415,
  \dodoi{10.1126/science.aaw1469}

\bibitem[{{Jones} {et~al.}(2009){Jones}, {Read}, {Saunders}, {Colless},
  {Jarrett}, {Parker}, {Fairall}, {Mauch}, {Sadler}, {Watson}, {Burton},
  {Campbell}, {Cass}, {Croom}, {Dawe}, {Fiegert}, {Frankcombe}, {Hartley},
  {Huchra}, {James}, {Kirby}, {Lahav}, {Lucey}, {Mamon}, {Moore}, {Peterson},
  {Prior}, {Proust}, {Russell}, {Safouris}, {Wakamatsu}, {Westra}, \&
  {Williams}}]{6df_2}
{Jones}, D.~H., {Read}, M.~A., {Saunders}, W., {et~al.} 2009, \mnras, 399, 683,
  \dodoi{10.1111/j.1365-2966.2009.15338.x}

\bibitem[{{Jones} {et~al.}(2019){Jones}, {Scolnic}, {Foley}, {Rest}, {Kessler},
  {Challis}, {Chambers}, {Coulter}, {Dettman}, {Foley}, {Huber}, {Jha},
  {Johnson}, {Kilpatrick}, {Kirshner}, {Manuel}, {Narayan}, {Pan}, {Riess},
  {Schultz}, {Siebert}, {Berger}, {Chornock}, {Flewelling}, {Magnier},
  {Smartt}, {Smith}, {Wainscoat}, {Waters}, \& {Willman}}]{Jones19}
{Jones}, D.~O., {Scolnic}, D.~M., {Foley}, R.~J., {et~al.} 2019, \apj, 881, 19,
  \dodoi{10.3847/1538-4357/ab2bec}

\bibitem[{{Jones} {et~al.}(2021){Jones}, {Foley}, {Narayan}, {Hjorth}, {Huber},
  {Aleo}, {Alexander}, {Angus}, {Auchettl}, {Baldassare}, {Bruun}, {Chambers},
  {Chatterjee}, {Coppejans}, {Coulter}, {DeMarchi}, {Dimitriadis}, {Drout},
  {Engel}, {French}, {Gagliano}, {Gall}, {Hung}, {Izzo}, {Jacobson-Gal{\'a}n},
  {Kilpatrick}, {Korhonen}, {Margutti}, {Raimundo}, {Ramirez-Ruiz}, {Rest},
  {Rojas-Bravo}, {Siebert}, {Smartt}, {Smith}, {Terreran}, {Wang}, {Wojtak},
  {Agnello}, {Ansari}, {Arendse}, {Baldeschi}, {Blanchard}, {Brethauer},
  {Bright}, {Brown}, {de Boer}, {Dodd}, {Fairlamb}, {Grillo}, {Hajela}, {Hede},
  {Kolborg}, {Law-Smith}, {Lin}, {Magnier}, {Malanchev}, {Matthews}, {Mockler},
  {Muthukrishna}, {Pan}, {Pfister}, {Ramanah}, {Rest}, {Sarangi},
  {Schr{\o}der}, {Stauffer}, {Stroh}, {Taggart}, {Tinyanont}, {Wainscoat}, \&
  {Young Supernova Experiment}}]{Jones21ApJ}
{Jones}, D.~O., {Foley}, R.~J., {Narayan}, G., {et~al.} 2021, \apj, 908, 143,
  \dodoi{10.3847/1538-4357/abd7f5}

\bibitem[{{Kasen} \& {Woosley}(2007)}]{Kasen07:wlr}
{Kasen}, D., \& {Woosley}, S.~E. 2007, \apj, 656, 661, \dodoi{10.1086/510375}

\bibitem[{{Kashi} \& {Soker}(2011)}]{Kashi11MNRAS}
{Kashi}, A., \& {Soker}, N. 2011, \mnras, 417, 1466,
  \dodoi{10.1111/j.1365-2966.2011.19361.x}

\bibitem[{{Kilpatrick} \& {Foley}(2018)}]{Kilpatrick2018}
{Kilpatrick}, C.~D., \& {Foley}, R.~J. 2018, \mnras, 481, 2536,
  \dodoi{10.1093/mnras/sty2435}

\bibitem[{{Lantz} {et~al.}(2004){Lantz}, {Aldering}, {Antilogus}, {Bonnaud},
  {Capoani}, {Castera}, {Copin}, {Dubet}, {Gangler}, {Henault}, {Lemonnier},
  {Pain}, {Pecontal}, {Pecontal}, \& {Smadja}}]{Lantz2004SPIE}
{Lantz}, B., {Aldering}, G., {Antilogus}, P., {et~al.} 2004, in Society of
  Photo-Optical Instrumentation Engineers (SPIE) Conference Series, Vol. 5249,
  Optical Design and Engineering, ed. L.~{Mazuray}, P.~J. {Rogers}, \&
  R.~{Wartmann}, 146--155, \dodoi{10.1117/12.512493}

\bibitem[{{Lu} {et~al.}(2021){Lu}, {Ashall}, {Hsiao}, {Hoeflich}, {Galbany},
  {Baron}, {Phillips}, {Contreras}, {Burns}, {Suntzeff}, {Stritzinger},
  {Anais}, {Anderson}, {Brown}, {Busta}, {Castell{\'o}n}, {Davis}, {Diamond},
  {Falco}, {Gonzalez}, {Hamuy}, {Holmbo}, {Holoien}, {Krisciunas}, {Kirshner},
  {Kumar}, {Kuncarayakti}, {Marion}, {Morrell}, {Persson}, {Piro}, {Prieto},
  {Sand}, {Shahbandeh}, {Shappee}, \& {Taddia}}]{Lu2021ApJ}
{Lu}, J., {Ashall}, C., {Hsiao}, E.~Y., {et~al.} 2021, \apj, 920, 107,
  \dodoi{10.3847/1538-4357/ac1606}

\bibitem[{{Maeda} {et~al.}(2009){Maeda}, {Kawabata}, {Li}, {Tanaka}, {Mazzali},
  {Hattori}, {Nomoto}, \& {Filippenko}}]{Maeda09ApJ}
{Maeda}, K., {Kawabata}, K., {Li}, W., {et~al.} 2009, \apj, 690, 1745,
  \dodoi{10.1088/0004-637X/690/2/1745}

\bibitem[{{Maguire} {et~al.}(2018){Maguire}, {Sim}, {Shingles}, {Spyromilio},
  {Jerkstrand}, {Sullivan}, {Chen}, {Cartier}, {Dimitriadis}, {Frohmaier},
  {Galbany}, {Guti{\'e}rrez}, {Hosseinzadeh}, {Howell}, {Inserra}, {Rudy}, \&
  {Sollerman}}]{Maguire2018MNRAS}
{Maguire}, K., {Sim}, S.~A., {Shingles}, L., {et~al.} 2018, \mnras, 477, 3567,
  \dodoi{10.1093/mnras/sty820}

\bibitem[{{Maoz} {et~al.}(2014){Maoz}, {Mannucci}, \& {Nelemans}}]{Maoz14}
{Maoz}, D., {Mannucci}, F., \& {Nelemans}, G. 2014, \araa, 52, 107,
  \dodoi{10.1146/annurev-astro-082812-141031}

\bibitem[{{Marsh} {et~al.}(2004){Marsh}, {Nelemans}, \&
  {Steeghs}}]{Marsh2004MNRAS}
{Marsh}, T.~R., {Nelemans}, G., \& {Steeghs}, D. 2004, \mnras, 350, 113,
  \dodoi{10.1111/j.1365-2966.2004.07564.x}

\bibitem[{{Matheson} {et~al.}(2012){Matheson}, {Joyce}, {Allen}, {Saha},
  {Silva}, {Wood-Vasey}, {Adams}, {Anderson}, {Beck}, {Bentz}, {Bershady},
  {Binkert}, {Butler}, {Camarata}, {Eigenbrot}, {Everett}, {Gallagher},
  {Garnavich}, {Glikman}, {Harbeck}, {Hargis}, {Herbst}, {Horch}, {Howell},
  {Jha}, {Kaczmarek}, {Knezek}, {Manne-Nicholas}, {Mathieu}, {Meixner},
  {Milliman}, {Power}, {Rajagopal}, {Reetz}, {Rhode}, {Schechtman-Rook},
  {Schwamb}, {Schweiker}, {Simmons}, {Simon}, {Summers}, {Young}, {Weyant},
  {Wilcots}, {Will}, \& {Williams}}]{Matheson2012ApJ}
{Matheson}, T., {Joyce}, R.~R., {Allen}, L.~E., {et~al.} 2012, \apj, 754, 19,
  \dodoi{10.1088/0004-637X/754/1/19}

\bibitem[{{Mazzali} {et~al.}(2011){Mazzali}, {Maurer}, {Stritzinger},
  {Taubenberger}, {Benetti}, \& {Hachinger}}]{Mazzali2011MNRAS}
{Mazzali}, P.~A., {Maurer}, I., {Stritzinger}, M., {et~al.} 2011, \mnras, 416,
  881, \dodoi{10.1111/j.1365-2966.2011.19000.x}

\bibitem[{{Miller} \& {Stone}(1993)}]{KAST}
{Miller}, J.~S., \& {Stone}, R. P.~S. 1993, LOTRM

\bibitem[{{Moll} {et~al.}(2014){Moll}, {Raskin}, {Kasen}, \&
  {Woosley}}]{Moll14}
{Moll}, R., {Raskin}, C., {Kasen}, D., \& {Woosley}, S.~E. 2014, \apj, 785,
  105, \dodoi{10.1088/0004-637X/785/2/105}

\bibitem[{{Mould} {et~al.}(2000){Mould}, {Huchra}, {Freedman}, {Kennicutt},
  {Ferrarese}, {Ford}, {Gibson}, {Graham}, {Hughes}, {Illingworth}, {Kelson},
  {Macri}, {Madore}, {Sakai}, {Sebo}, {Silbermann}, \& {Stetson}}]{Mould00}
{Mould}, J.~R., {Huchra}, J.~P., {Freedman}, W.~L., {et~al.} 2000, \apj, 529,
  786, \dodoi{10.1086/308304}

\bibitem[{{Nicholl} {et~al.}(2017){Nicholl}, {Berger}, {Margutti}, {Blanchard},
  {Guillochon}, {Leja}, \& {Chornock}}]{Nicholl2017ApJ}
{Nicholl}, M., {Berger}, E., {Margutti}, R., {et~al.} 2017, \apjl, 845, L8,
  \dodoi{10.3847/2041-8213/aa82b1}

\bibitem[{{Noebauer} {et~al.}(2016){Noebauer}, {Taubenberger}, {Blinnikov},
  {Sorokina}, \& {Hillebrandt}}]{Noebauer16MNRAS}
{Noebauer}, U.~M., {Taubenberger}, S., {Blinnikov}, S., {Sorokina}, E., \&
  {Hillebrandt}, W. 2016, \mnras, 463, 2972, \dodoi{10.1093/mnras/stw2197}

\bibitem[{{Nugent} {et~al.}(2011){Nugent}, {Sullivan}, {Cenko}, {Thomas},
  {Kasen}, {Howell}, {Bersier}, {Bloom}, {Kulkarni}, {Kandrashoff},
  {Filippenko}, {Silverman}, {Marcy}, {Howard}, {Isaacson}, {Maguire},
  {Suzuki}, {Tarlton}, {Pan}, {Bildsten}, {Fulton}, {Parrent}, {Sand},
  {Podsiadlowski}, {Bianco}, {Dilday}, {Graham}, {Lyman}, {James}, {Kasliwal},
  {Law}, {Quimby}, {Hook}, {Walker}, {Mazzali}, {Pian}, {Ofek}, {Gal-Yam}, \&
  {Poznanski}}]{Nugent11}
{Nugent}, P.~E., {Sullivan}, M., {Cenko}, S.~B., {et~al.} 2011, \nat, 480, 344,
  \dodoi{10.1038/nature10644}

\bibitem[{{Oke} {et~al.}(1995){Oke}, {Cohen}, {Carr}, {Cromer}, {Dingizian},
  {Harris}, {Labrecque}, {Lucinio}, {Schaal}, {Epps}, \& {Miller}}]{LRIS}
{Oke}, J.~B., {Cohen}, J.~G., {Carr}, M., {et~al.} 1995, \pasp, 107, 375,
  \dodoi{10.1086/133562}

\bibitem[{{Parrent} {et~al.}(2014){Parrent}, {Friesen}, \&
  {Parthasarathy}}]{Parrent14}
{Parrent}, J., {Friesen}, B., \& {Parthasarathy}, M. 2014, \apss, 351, 1,
  \dodoi{10.1007/s10509-014-1830-1}

\bibitem[{{Parrent} {et~al.}(2016){Parrent}, {Howell}, {Fesen}, {Parker},
  {Bianco}, {Dilday}, {Sand}, {Valenti}, {Vink{\'o}}, {Berlind}, {Challis},
  {Milisavljevic}, {Sanders}, {Marion}, {Wheeler}, {Brown}, {Calkins},
  {Friesen}, {Kirshner}, {Pritchard}, {Quimby}, \& {Roming}}]{Parrent16MNRAS}
{Parrent}, J.~T., {Howell}, D.~A., {Fesen}, R.~A., {et~al.} 2016, \mnras, 457,
  3702, \dodoi{10.1093/mnras/stw239}

\bibitem[{{Pereira} {et~al.}(2013){Pereira}, {Thomas}, {Aldering}, {Antilogus},
  {Baltay}, {Benitez-Herrera}, {Bongard}, {Buton}, {Canto}, {Cellier-Holzem},
  {Chen}, {Childress}, {Chotard}, {Copin}, {Fakhouri}, {Fink}, {Fouchez},
  {Gangler}, {Guy}, {Hillebrandt}, {Hsiao}, {Kerschhaggl}, {Kowalski},
  {Kromer}, {Nordin}, {Nugent}, {Paech}, {Pain}, {P{\'e}contal}, {Perlmutter},
  {Rabinowitz}, {Rigault}, {Runge}, {Saunders}, {Smadja}, {Tao},
  {Taubenberger}, {Tilquin}, \& {Wu}}]{Pereira13}
{Pereira}, R., {Thomas}, R.~C., {Aldering}, G., {et~al.} 2013, \aap, 554, A27,
  \dodoi{10.1051/0004-6361/201221008}

\bibitem[{{Perlmutter} {et~al.}(1999){Perlmutter}, {Aldering}, {Goldhaber},
  {Knop}, {Nugent}, {Castro}, {Deustua}, {Fabbro}, {Goobar}, {Groom}, {Hook},
  {Kim}, {Kim}, {Lee}, {Nunes}, {Pain}, {Pennypacker}, {Quimby}, {Lidman},
  {Ellis}, {Irwin}, {McMahon}, {Ruiz-Lapuente}, {Walton}, {Schaefer}, {Boyle},
  {Filippenko}, {Matheson}, {Fruchter}, {Panagia}, {Newberg}, \&
  {Couch}}]{Perlmutter99}
{Perlmutter}, S., {Aldering}, G., {Goldhaber}, G., {et~al.} 1999, \apj, 517,
  565, \dodoi{10.1086/307221}

\bibitem[{{Pfannes} {et~al.}(2010{\natexlab{a}}){Pfannes}, {Niemeyer}, \&
  {Schmidt}}]{Pfannes10AA2}
{Pfannes}, J.~M.~M., {Niemeyer}, J.~C., \& {Schmidt}, W. 2010{\natexlab{a}},
  \aap, 509, A75, \dodoi{10.1051/0004-6361/200912033}

\bibitem[{{Pfannes} {et~al.}(2010{\natexlab{b}}){Pfannes}, {Niemeyer},
  {Schmidt}, \& {Klingenberg}}]{Pfannes10AA}
{Pfannes}, J.~M.~M., {Niemeyer}, J.~C., {Schmidt}, W., \& {Klingenberg}, C.
  2010{\natexlab{b}}, \aap, 509, A74, \dodoi{10.1051/0004-6361/200912032}

\bibitem[{{Phillips}(1993)}]{Phillips93}
{Phillips}, M.~M. 1993, \apjl, 413, L105, \dodoi{10.1086/186970}

\bibitem[{{Phillips} {et~al.}(2013){Phillips}, {Simon}, {Morrell}, {Burns},
  {Cox}, {Foley}, {Karakas}, {Patat}, {Sternberg}, {Williams}, {Gal-Yam},
  {Hsiao}, {Leonard}, {Persson}, {Stritzinger}, {Thompson}, {Campillay},
  {Contreras}, {Folatelli}, {Freedman}, {Hamuy}, {Roth}, {Shields}, {Suntzeff},
  {Chomiuk}, {Ivans}, {Madore}, {Penprase}, {Perley}, {Pignata}, {Preston}, \&
  {Soderberg}}]{Phillips2013ApJ}
{Phillips}, M.~M., {Simon}, J.~D., {Morrell}, N., {et~al.} 2013, \apj, 779, 38,
  \dodoi{10.1088/0004-637X/779/1/38}

\bibitem[{{Piro}(2012)}]{Piro12}
{Piro}, A.~L. 2012, \apj, 759, 83, \dodoi{10.1088/0004-637X/759/2/83}

\bibitem[{{Poznanski} {et~al.}(2012){Poznanski}, {Prochaska}, \&
  {Bloom}}]{Poznanski12MNRAS}
{Poznanski}, D., {Prochaska}, J.~X., \& {Bloom}, J.~S. 2012, \mnras, 426, 1465,
  \dodoi{10.1111/j.1365-2966.2012.21796.x}

\bibitem[{{Quimby} {et~al.}(2018){Quimby}, {De Cia}, {Gal-Yam}, {Leloudas},
  {Lunnan}, {Perley}, {Vreeswijk}, {Yan}, {Bloom}, {Cenko}, {Cooke}, {Ellis},
  {Filippenko}, {Kasliwal}, {Kleiser}, {Kulkarni}, {Matheson}, {Nugent}, {Pan},
  {Silverman}, {Sternberg}, {Sullivan}, \& {Yaron}}]{Quimby18}
{Quimby}, R.~M., {De Cia}, A., {Gal-Yam}, A., {et~al.} 2018, \apj, 855, 2,
  \dodoi{10.3847/1538-4357/aaac2f}

\bibitem[{{Raskin} \& {Kasen}(2013)}]{Raskin13}
{Raskin}, C., \& {Kasen}, D. 2013, \apj, 772, 1,
  \dodoi{10.1088/0004-637X/772/1/1}

\bibitem[{{Raskin} {et~al.}(2014){Raskin}, {Kasen}, {Moll}, {Schwab}, \&
  {Woosley}}]{Raskin14}
{Raskin}, C., {Kasen}, D., {Moll}, R., {Schwab}, J., \& {Woosley}, S. 2014,
  \apj, 788, 75, \dodoi{10.1088/0004-637X/788/1/75}

\bibitem[{{Rest} {et~al.}(2014){Rest}, {Scolnic}, {Foley}, {Huber}, {Chornock},
  {Narayan}, {Tonry}, {Berger}, {Soderberg}, {Stubbs}, {Riess}, {Kirshner},
  {Smartt}, {Schlafly}, {Rodney}, {Botticella}, {Brout}, {Challis}, {Czekala},
  {Drout}, {Hudson}, {Kotak}, {Leibler}, {Lunnan}, {Marion}, {McCrum},
  {Milisavljevic}, {Pastorello}, {Sanders}, {Smith}, {Stafford}, {Thilker},
  {Valenti}, {Wood-Vasey}, {Zheng}, {Burgett}, {Chambers}, {Denneau}, {Draper},
  {Flewelling}, {Hodapp}, {Kaiser}, {Kudritzki}, {Magnier}, {Metcalfe},
  {Price}, {Sweeney}, {Wainscoat}, \& {Waters}}]{Rest14}
{Rest}, A., {Scolnic}, D., {Foley}, R.~J., {et~al.} 2014, \apj, 795, 44,
  \dodoi{10.1088/0004-637X/795/1/44}

\bibitem[{{Riess} {et~al.}(1996){Riess}, {Press}, \& {Kirshner}}]{Riess96}
{Riess}, A.~G., {Press}, W.~H., \& {Kirshner}, R.~P. 1996, \apj, 473, 88,
  \dodoi{10.1086/178129}

\bibitem[{{Riess} {et~al.}(1998){Riess}, {Filippenko}, {Challis},
  {Clocchiatti}, {Diercks}, {Garnavich}, {Gilliland}, {Hogan}, {Jha},
  {Kirshner}, {Leibundgut}, {Phillips}, {Reiss}, {Schmidt}, {Schommer},
  {Smith}, {Spyromilio}, {Stubbs}, {Suntzeff}, \& {Tonry}}]{Riess98:Lambda}
{Riess}, A.~G., {Filippenko}, A.~V., {Challis}, P., {et~al.} 1998, \aj, 116,
  1009

\bibitem[{{Riess} {et~al.}(2016){Riess}, {Macri}, {Hoffmann}, {Scolnic},
  {Casertano}, {Filippenko}, {Tucker}, {Reid}, {Jones}, {Silverman},
  {Chornock}, {Challis}, {Yuan}, {Brown}, \& {Foley}}]{Riess16}
{Riess}, A.~G., {Macri}, L.~M., {Hoffmann}, S.~L., {et~al.} 2016, \apj, 826,
  56, \dodoi{10.3847/0004-637X/826/1/56}

\bibitem[{{Roming} {et~al.}(2005){Roming}, {Kennedy}, {Mason}, {Nousek}, {Ahr},
  {Bingham}, {Broos}, {Carter}, {Hancock}, {Huckle}, {Hunsberger}, {Kawakami},
  {Killough}, {Koch}, {McLelland}, {Smith}, {Smith}, {Soto}, {Boyd},
  {Breeveld}, {Holland}, {Ivanushkina}, {Pryzby}, {Still}, \&
  {Stock}}]{Roming05}
{Roming}, P. W.~A., {Kennedy}, T.~E., {Mason}, K.~O., {et~al.} 2005, \ssr, 120,
  95, \dodoi{10.1007/s11214-005-5095-4}

\bibitem[{{Rubin} {et~al.}(2015){Rubin}, {Aldering}, {Barbary}, {Boone},
  {Chappell}, {Currie}, {Deustua}, {Fagrelius}, {Fruchter}, {Hayden}, {Lidman},
  {Nordin}, {Perlmutter}, {Saunders}, {Sofiatti}, \& {Supernova Cosmology
  Project}}]{Rubin2015ApJ}
{Rubin}, D., {Aldering}, G., {Barbary}, K., {et~al.} 2015, \apj, 813, 137,
  \dodoi{10.1088/0004-637X/813/2/137}

\bibitem[{{Saio} \& {Nomoto}(2004)}]{Saio2004ApJ}
{Saio}, H., \& {Nomoto}, K. 2004, \apj, 615, 444, \dodoi{10.1086/423976}

\bibitem[{{Scalzo} {et~al.}(2014){Scalzo}, {Aldering}, {Antilogus}, {Aragon},
  {Bailey}, {Baltay}, {Bongard}, {Buton}, {Cellier-Holzem}, {Childress},
  {Chotard}, {Copin}, {Fakhouri}, {Gangler}, {Guy}, {Kim}, {Kowalski},
  {Kromer}, {Nordin}, {Nugent}, {Paech}, {Pain}, {Pecontal}, {Pereira},
  {Perlmutter}, {Rabinowitz}, {Rigault}, {Runge}, {Saunders}, {Sim}, {Smadja},
  {Tao}, {Taubenberger}, {Thomas}, {Weaver}, \& {Nearby Supernova
  Factory}}]{Scalzo14}
{Scalzo}, R., {Aldering}, G., {Antilogus}, P., {et~al.} 2014, \mnras, 440,
  1498, \dodoi{10.1093/mnras/stu350}

\bibitem[{{Scalzo} {et~al.}(2010){Scalzo}, {Aldering}, {Antilogus}, {Aragon},
  {Bailey}, {Baltay}, {Bongard}, {Buton}, {Childress}, {Chotard}, {Copin},
  {Fakhouri}, {Gal-Yam}, {Gangler}, {Hoyer}, {Kasliwal}, {Loken}, {Nugent},
  {Pain}, {P{\'e}contal}, {Pereira}, {Perlmutter}, {Rabinowitz}, {Rau},
  {Rigaudier}, {Runge}, {Smadja}, {Tao}, {Thomas}, {Weaver}, \&
  {Wu}}]{Scalzo10}
{Scalzo}, R.~A., {Aldering}, G., {Antilogus}, P., {et~al.} 2010, \apj, 713,
  1073, \dodoi{10.1088/0004-637X/713/2/1073}

\bibitem[{{Schlafly} \& {Finkbeiner}(2011)}]{Schlafly11}
{Schlafly}, E.~F., \& {Finkbeiner}, D.~P. 2011, \apj, 737, 103,
  \dodoi{10.1088/0004-637X/737/2/103}

\bibitem[{{Science Software Branch at STScI}(2012)}]{iraf3}
{Science Software Branch at STScI}. 2012, {PyRAF: Python alternative for IRAF}.
\newblock \doeprint{1207.011}

\bibitem[{{Scolnic} {et~al.}(2018){Scolnic}, {Jones}, {Rest}, {Pan},
  {Chornock}, {Foley}, {Huber}, {Kessler}, {Narayan}, {Riess}, {Rodney},
  {Berger}, {Brout}, {Challis}, {Drout}, {Finkbeiner}, {Lunnan}, {Kirshner},
  {Sanders}, {Schlafly}, {Smartt}, {Stubbs}, {Tonry}, {Wood-Vasey}, {Foley},
  {Hand}, {Johnson}, {Burgett}, {Chambers}, {Draper}, {Hodapp}, {Kaiser},
  {Kudritzki}, {Magnier}, {Metcalfe}, {Bresolin}, {Gall}, {Kotak}, {McCrum}, \&
  {Smith}}]{Scolnic18:ps1}
{Scolnic}, D.~M., {Jones}, D.~O., {Rest}, A., {et~al.} 2018, \apj, 859, 101,
  \dodoi{10.3847/1538-4357/aab9bb}

\bibitem[{{Shappee} {et~al.}(2017){Shappee}, {Stanek}, {Kochanek}, \&
  {Garnavich}}]{Shappee17ApJ}
{Shappee}, B.~J., {Stanek}, K.~Z., {Kochanek}, C.~S., \& {Garnavich}, P.~M.
  2017, \apj, 841, 48, \dodoi{10.3847/1538-4357/aa6eab}

\bibitem[{{Shappee} {et~al.}(2014){Shappee}, {Prieto}, {Grupe}, {Kochanek},
  {Stanek}, {De Rosa}, {Mathur}, {Zu}, {Peterson}, {Pogge}, {Komossa}, {Im},
  {Jencson}, {Holoien}, {Basu}, {Beacom}, {Szczygie{\l}}, {Brimacombe},
  {Adams}, {Campillay}, {Choi}, {Contreras}, {Dietrich}, {Dubberley},
  {Elphick}, {Foale}, {Giustini}, {Gonzalez}, {Hawkins}, {Howell}, {Hsiao},
  {Koss}, {Leighly}, {Morrell}, {Mudd}, {Mullins}, {Nugent}, {Parrent},
  {Phillips}, {Pojmanski}, {Rosing}, {Ross}, {Sand}, {Terndrup}, {Valenti},
  {Walker}, \& {Yoon}}]{Shappee14}
{Shappee}, B.~J., {Prieto}, J.~L., {Grupe}, D., {et~al.} 2014, \apj, 788, 48,
  \dodoi{10.1088/0004-637X/788/1/48}

\bibitem[{{Shen} {et~al.}(2012){Shen}, {Bildsten}, {Kasen}, \&
  {Quataert}}]{Shen2012ApJ}
{Shen}, K.~J., {Bildsten}, L., {Kasen}, D., \& {Quataert}, E. 2012, \apj, 748,
  35, \dodoi{10.1088/0004-637X/748/1/35}

\bibitem[{{Siebert} {et~al.}(2020){Siebert}, {Dimitriadis}, {Polin}, \&
  {Foley}}]{Siebert2020ApJ}
{Siebert}, M.~R., {Dimitriadis}, G., {Polin}, A., \& {Foley}, R.~J. 2020,
  \apjl, 900, L27, \dodoi{10.3847/2041-8213/abae6e}

\bibitem[{{Silverman} {et~al.}(2011){Silverman}, {Ganeshalingam}, {Li},
  {Filippenko}, {Miller}, \& {Poznanski}}]{Silverman2011MNRAS}
{Silverman}, J.~M., {Ganeshalingam}, M., {Li}, W., {et~al.} 2011, \mnras, 410,
  585, \dodoi{10.1111/j.1365-2966.2010.17474.x}

\bibitem[{{Silverman} {et~al.}(2012){Silverman}, {Foley}, {Filippenko},
  {Ganeshalingam}, {Barth}, {Chornock}, {Griffith}, {Kong}, {Lee}, {Leonard},
  {Matheson}, {Miller}, {Steele}, {Barris}, {Bloom}, {Cobb}, {Coil},
  {Desroches}, {Gates}, {Ho}, {Jha}, {Kandrashoff}, {Li}, {Mandel}, {Modjaz},
  {Moore}, {Mostardi}, {Papenkova}, {Park}, {Perley}, {Poznanski}, {Reuter},
  {Scala}, {Serduke}, {Shields}, {Swift}, {Tonry}, {Van Dyk}, {Wang}, \&
  {Wong}}]{Silverman2012MNRAS}
{Silverman}, J.~M., {Foley}, R.~J., {Filippenko}, A.~V., {et~al.} 2012, \mnras,
  425, 1789, \dodoi{10.1111/j.1365-2966.2012.21270.x}

\bibitem[{{Silverman} {et~al.}(2013){Silverman}, {Nugent}, {Gal-Yam},
  {Sullivan}, {Howell}, {Filippenko}, {Arcavi}, {Ben-Ami}, {Bloom}, {Cenko},
  {Cao}, {Chornock}, {Clubb}, {Coil}, {Foley}, {Graham}, {Griffith}, {Horesh},
  {Kasliwal}, {Kulkarni}, {Leonard}, {Li}, {Matheson}, {Miller}, {Modjaz},
  {Ofek}, {Pan}, {Perley}, {Poznanski}, {Quimby}, {Steele}, {Sternberg}, {Xu},
  \& {Yaron}}]{Silverman2013ApJS}
{Silverman}, J.~M., {Nugent}, P.~E., {Gal-Yam}, A., {et~al.} 2013, \apjs, 207,
  3, \dodoi{10.1088/0067-0049/207/1/3}

\bibitem[{{Smith} {et~al.}(2008){Smith}, {Foley}, \&
  {Filippenko}}]{Smith2008ApJ}
{Smith}, N., {Foley}, R.~J., \& {Filippenko}, A.~V. 2008, \apj, 680, 568,
  \dodoi{10.1086/587860}

\bibitem[{{Stanek}(2020)}]{Stanek20TNSTR}
{Stanek}, K.~Z. 2020, Transient Name Server Discovery Report, 2020-843, 1

\bibitem[{{Szalai} {et~al.}(2019){Szalai}, {Vink{\'o}}, {K{\"o}nyves-T{\'o}th},
  {Nagy}, {Bostroem}, {S{\'a}rneczky}, {Brown}, {Pejcha}, {B{\'o}di}, {Cseh},
  {Cs{\"o}rnyei}, {Dencs}, {Hanyecz}, {Ign{\'a}cz}, {Kalup}, {Kriskovics},
  {Ordasi}, {P{\'a}l}, {Seli}, {S{\'o}dor}, {Szak{\'a}ts}, {Vida}, {Zsidi},
  {Konkoly Team}, {Arcavi}, {Ashall}, {Burke}, {Galbany}, {Hiramatsu},
  {Hosseinzadeh}, {Hsiao}, {Howell}, {McCully}, {Moran}, {Rho}, {Sand },
  {Shahbandeh}, {Valenti}, {Wang}, {Wheeler}, \& {Supernova
  Project}}]{Szalai19}
{Szalai}, T., {Vink{\'o}}, J., {K{\"o}nyves-T{\'o}th}, R., {et~al.} 2019, \apj,
  876, 19, \dodoi{10.3847/1538-4357/ab12d0}

\bibitem[{{Tanaka} {et~al.}(2010){Tanaka}, {Kawabata}, {Yamanaka}, {Maeda},
  {Hattori}, {Aoki}, {Nomoto}, {Iye}, {Sasaki}, {Mazzali}, \&
  {Pian}}]{Tanaka10ApJ}
{Tanaka}, M., {Kawabata}, K.~S., {Yamanaka}, M., {et~al.} 2010, \apj, 714,
  1209, \dodoi{10.1088/0004-637X/714/2/1209}

\bibitem[{{Taubenberger}(2017)}]{Taubenberger17}
{Taubenberger}, S. 2017, Handbook of Supernovae ({Springer})

\bibitem[{{Taubenberger} {et~al.}(2011){Taubenberger}, {Benetti}, {Childress},
  {Pakmor}, {Hachinger}, {Mazzali}, {Stanishev}, {Elias-Rosa}, {Agnoletto},
  {Bufano}, {Ergon}, {Harutyunyan}, {Inserra}, {Kankare}, {Kromer},
  {Navasardyan}, {Nicolas}, {Pastorello}, {Prosperi}, {Salgado}, {Sollerman},
  {Stritzinger}, {Turatto}, {Valenti}, \& {Hillebrandt}}]{Taubenberger11}
{Taubenberger}, S., {Benetti}, S., {Childress}, M., {et~al.} 2011, \mnras, 412,
  2735, \dodoi{10.1111/j.1365-2966.2010.18107.x}

\bibitem[{{Taubenberger} {et~al.}(2013){Taubenberger}, {Kromer}, {Hachinger},
  {Mazzali}, {Benetti}, {Nugent}, {Scalzo}, {Pakmor}, {Stanishev},
  {Spyromilio}, {Bufano}, {Sim}, {Leibundgut}, \&
  {Hillebrandt}}]{Taubenberger2013MNRAS}
{Taubenberger}, S., {Kromer}, M., {Hachinger}, S., {et~al.} 2013, \mnras, 432,
  3117, \dodoi{10.1093/mnras/stt668}

\bibitem[{{Taubenberger} {et~al.}(2019){Taubenberger}, {Floers}, {Vogl},
  {Kromer}, {Spyromilio}, {Aldering}, {Antilogus}, {Bailey}, {Baltay},
  {Bongard}, {Boone}, {Buton}, {Chotard}, {Copin}, {Dixon}, {Fouchez},
  {Fransson}, {Gangler}, {Gupta}, {Hachinger}, {Hayden}, {Hillebrandt}, {Kim},
  {Kowalski}, {Leget}, {Leibundgut}, {Mazzali}, {Noebauer}, {Nordin}, {Pain},
  {Pakmor}, {Pecontal}, {Pereira}, {Perlmutter}, {Ponder}, {Rabinowitz},
  {Rigault}, {Rubin}, {Runge}, {Saunders}, {Smadja}, {Tao}, \&
  {Thomas}}]{Taubenberger19}
{Taubenberger}, S., {Floers}, A., {Vogl}, C., {et~al.} 2019, \mnras, 488, 5473,
  \dodoi{10.1093/mnras/stz1977}

\bibitem[{{Thomas} {et~al.}(2011){Thomas}, {Nugent}, \& {Meza}}]{SYNAPPS}
{Thomas}, R.~C., {Nugent}, P.~E., \& {Meza}, J.~C. 2011, \pasp, 123, 237,
  \dodoi{10.1086/658673}

\bibitem[{{Tody}(1986)}]{iraf1}
{Tody}, D. 1986, in Society of Photo-Optical Instrumentation Engineers (SPIE)
  Conference Series, Vol. 627, Instrumentation in astronomy VI, ed. D.~L.
  {Crawford}, 733, \dodoi{10.1117/12.968154}

\bibitem[{{Tody}(1993)}]{iraf2}
{Tody}, D. 1993, in Astronomical Society of the Pacific Conference Series,
  Vol.~52, Astronomical Data Analysis Software and Systems II, ed. R.~J.
  {Hanisch}, R.~J.~V. {Brissenden}, \& J.~{Barnes}, 173

\bibitem[{{Tucker} {et~al.}(2020){Tucker}, {Payne}, {Hinkle}, {Do}, {Huber}, \&
  {Shappee}}]{Tucker20TNSCR}
{Tucker}, M.~A., {Payne}, A.~V., {Hinkle}, J., {et~al.} 2020, Transient Name
  Server Classification Report, 2020-861, 1

\bibitem[{{Webbink}(1984)}]{Webbink84}
{Webbink}, R.~F. 1984, \apj, 277, 355, \dodoi{10.1086/161701}

\bibitem[{{Yamanaka} {et~al.}(2009){Yamanaka}, {Kawabata}, {Kinugasa},
  {Tanaka}, {Imada}, {Maeda}, {Nomoto}, {Arai}, {Chiyonobu}, {Fukazawa},
  {Hashimoto}, {Honda}, {Ikejiri}, {Itoh}, {Kamata}, {Kawai}, {Komatsu},
  {Konishi}, {Kuroda}, {Miyamoto}, {Miyazaki}, {Nagae}, {Nakaya}, {Ohsugi},
  {Omodaka}, {Sakai}, {Sasada}, {Suzuki}, {Taguchi}, {Takahashi}, {Tanaka},
  {Uemura}, {Yamashita}, {Yanagisawa}, \& {Yoshida}}]{Yamanaka09ApJ}
{Yamanaka}, M., {Kawabata}, K.~S., {Kinugasa}, K., {et~al.} 2009, \apjl, 707,
  L118, \dodoi{10.1088/0004-637X/707/2/L118}

\bibitem[{{Yamanaka} {et~al.}(2016){Yamanaka}, {Maeda}, {Tanaka}, {Tominaga},
  {Kawabata}, {Takaki}, {Kawabata}, {Nakaoka}, {Ueno}, {Akitaya}, {Nagayama},
  {Takahashi}, {Honda}, {Omodaka}, {Miyanoshita}, {Nagao}, {Watanabe},
  {Isogai}, {Arai}, {Itoh}, {Ui}, {Uemura}, {Yoshida}, {Hanayama}, {Kuroda},
  {Ukita}, {Yanagisawa}, {Izumiura}, {Saito}, {Masumoto}, {Ono}, {Noguchi},
  {Matsumoto}, {Nogami}, {Morokuma}, {Oasa}, \& {Sekiguchi}}]{Yamanaka2016PASJ}
{Yamanaka}, M., {Maeda}, K., {Tanaka}, M., {et~al.} 2016, \pasj, 68, 68,
  \dodoi{10.1093/pasj/psw047}

\bibitem[{{Yoon} \& {Langer}(2005)}]{Yoon05AA}
{Yoon}, S.~C., \& {Langer}, N. 2005, \aap, 435, 967,
  \dodoi{10.1051/0004-6361:20042542}

\bibitem[{{Yuan} {et~al.}(2010){Yuan}, {Quimby}, {Wheeler}, {Vink{\'o}},
  {Chatzopoulos}, {Akerlof}, {Kulkarni}, {Miller}, {McKay}, \&
  {Aharonian}}]{Yuan10ApJ}
{Yuan}, F., {Quimby}, R.~M., {Wheeler}, J.~C., {et~al.} 2010, \apj, 715, 1338,
  \dodoi{10.1088/0004-637X/715/2/1338}

\bibitem[{{Zhang} {et~al.}(2016){Zhang}, {Wang}, {Zhang}, {Zhang},
  {Ganeshalingam}, {Li}, {Filippenko}, {Zhao}, {Zheng}, {Bai}, {Chen}, {Chen},
  {Huang}, {Mo}, {Rui}, {Song}, {Sai}, {Li}, {Wang}, \& {Wu}}]{Zhang16}
{Zhang}, K., {Wang}, X., {Zhang}, J., {et~al.} 2016, \apj, 820, 67,
  \dodoi{10.3847/0004-637X/820/1/67}

\end{thebibliography}
\bibliographystyle{aasjournal}



\end{document}